\documentstyle[aps,prl,preprint,epsf,floats]{revtex}
\tightenlines

\newcommand{\beq}{\begin{equation}}
\newcommand{\eeq}{\end{equation}}
\newcommand{\bqa}{\begin{eqnarray}}
\newcommand{\eqa}{\end{eqnarray}}

\def\square{\vcenter{\vbox{\hrule height.4pt
          \hbox{\vrule width.4pt height8pt
          \kern8pt\vrule width.4pt}\hrule height.4pt}}}

\def\sumint{\hbox{$\sum$}\!\!\!\!\!\!\int}

\def\ranglec{\rangle_{\!\!c}}
\def\ranglex{\rangle_{\!\!x}}
\def\ranglecx{\rangle_{\!\!c,x}}

\begin{document}

\preprint{
\vbox{\halign{&##\hfil\cr
        & hep-ph/0205085 \cr
        & DUKE-TH-02-221 \cr
        & ITF-UU-02/24 \cr
&\today\cr\cr\cr }}}

\title{HTL Perturbation Theory to Two Loops}

\author{Jens O. Andersen}
\address{Institute for Theoretical Physics, University of Utrecht, \\
Leuvenlaan 4, 3584 CC Utrecht, The Netherlands}

\author{Eric Braaten and Emmanuel Petitgirard}
\address{Physics Department, Ohio State University, Columbus OH 43210, USA}

\author{Michael Strickland}
\address{Physics Department, Duke University, Durham NC 27708, USA}
\maketitle

\begin{abstract}
{\footnotesize
We calculate the pressure for pure-glue QCD at high temperature
to two-loop order using hard-thermal-loop (HTL) perturbation theory.
At this order, all the ultraviolet divergences can be absorbed into
renormalizations of the vacuum energy density and the HTL mass parameter.
We determine the HTL mass parameter by a variational prescription.
The resulting predictions for the pressure fail to agree with results from
lattice gauge theory at temperatures for which they are available.
}
\end{abstract}

\newpage

\section{Introduction}

Relativistic heavy-ion collisions allow the experimental study of
hadronic matter at energy densities exceeding
that required to create a quark-gluon plasma.
A quantitative understanding of the properties of a
quark-gluon plasma is essential in order to determine whether it has been
created.  Because QCD is asymptotically free,
its running coupling constant $\alpha_s$ becomes
weaker as the temperature increases.  One  might therefore expect
the behavior of hadronic matter at sufficiently high temperature
to be calculable using perturbative methods.  Unfortunately,
a straightforward perturbative expansion in powers
of $\alpha_s$ does not seem to be of any quantitative use
even at temperatures orders of magnitude higher than those achievable in
heavy-ion collisions.

\begin{figure}[htb]
\hspace{1cm}
\epsfysize=9cm
\centerline{\epsffile{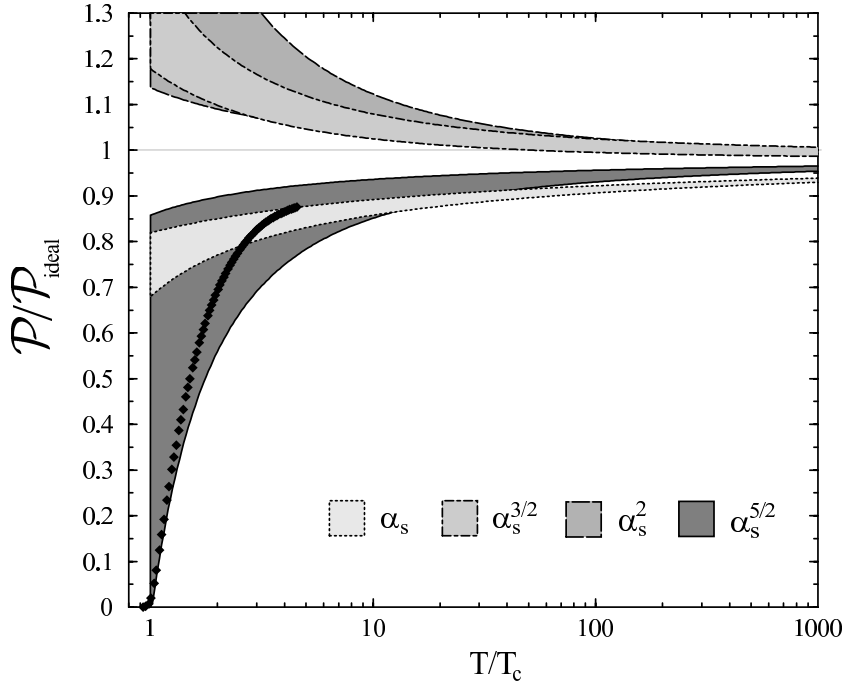}}
\\
\caption[a]{The free energy for pure-glue QCD as a function of $T/T_c$.
The weak-coupling expansions through orders $\alpha_s$,
$\alpha_s^{3/2}$, $\alpha_s^2$, and $\alpha_s^{5/2}$
are shown as bands that correspond to varying
the renormalization scale $\mu$ by a factor of two. The diamonds are
the lattice result from Boyd et al.~\cite{lattice-0}.
The size of the diamonds indicate the typical error bar.}
\label{weakfig}
\end{figure}

The problem is evident in the free energy ${\cal F}$ of the quark-gluon
plasma, whose weak-coupling expansion has been calculated through order
$\alpha_s^{5/2}$ \cite{AZ-95,KZ-96,BN-96}.
For a pure-glue plasma, the first few terms in the weak-coupling expansion
are
\begin{eqnarray}\nonumber
{\cal F}_{\rm QCD} &=& {\cal F}_{\rm ideal}
\Bigg[ 1 \;-\; {15 \over 4} {\alpha_s \over \pi}
\;+\; 30 \left ( {\alpha_s \over \pi} \right )^{3/2} \\
&&\;+\; {135 \over 2} \left( \log {\alpha_s \over \pi}
-{11\over36}\log{\mu\over2\pi T}+ 3.51 \right)
	\left( {\alpha_s \over \pi} \right)^2
\nonumber
\\
&&
+{495\over2}\left(\log{\mu\over2\pi T}-3.23\right) \left ( {\alpha_s \over \pi} \right )^{5/2}
\;+\; {\cal O} (\alpha_s^3\log \alpha_s ) \Bigg],
\label{F1-QCD}
\end{eqnarray}
where ${\cal F}_{\rm ideal}=-(8\pi^2/45)T^4$ is the free energy of an
ideal gas of massless gluons and $\alpha_s=\alpha_s(\mu)$ is the running
coupling constant in the $\overline{\mbox{MS}}$ scheme.
In Fig.~\ref{weakfig}, the free energy is shown as a
function of the temperature $T/T_c$,
 where $T_c$ is the critical temperature
for the deconfinement transition.
The weak-coupling expansions through
orders $\alpha_s$, $\alpha_s^{3/2}$, $\alpha_s^2$, and $\alpha_s^{5/2}$
are shown as bands that correspond to varying the renormalization scale
$\mu$
by a factor of two from the central value $\mu=2\pi T$.
As successive terms in the weak-coupling expansion
are added, the predictions change wildly and the sensitivity to the
renormalization scale grows.
It is clear that a reorganization of the perturbation series
is essential if perturbative calculations are to be of any quantitative
use at temperatures accessible in heavy-ion collisions.

The free energy can also be calculated nonperturbatively using
lattice gauge theory \cite{Karsch}.
The thermodynamic functions for pure-glue QCD have been calculated with
high precision by Boyd et al.~\cite{lattice-0}. There have also been
calculations with $N_f=2$ and 4 flavors of dynamical quarks~\cite{lattice-Nf}.
In Fig.~\ref{weakfig}, the lattice results for the free energy
of pure-glue QCD from Boyd et al.~\cite{lattice-0} are shown as diamonds.
The free energy is very close to zero near $T_c$.
As the temperature increases, the free energy increases
and approaches that of an ideal gas of massless gluons.
We will regard the lattice results as the correct results for the
thermodynamic functions.  One goal of any reorganization
of perturbation theory is to obtain a free energy that agrees
within its domain of validity with the lattice results.

There is of course little to be gained by just reproducing
the results of lattice gauge theory.
A method for reorganizing perturbation theory
is of practical use only if it allows the calculation of quantities
that are not so easily calculated using lattice gauge theory.
There are many observables that are difficult
or even impossible to calculate using lattice gauge theory.
First, lattice gauge theory becomes increasingly inefficient
at higher temperatures,
so some other method is required to extrapolate to high $T$.
Second, calculations with light dynamical quarks require orders of magnitude
more computer power than pure-glue QCD.
Third, the Monte Carlo approach used in lattice gauge theory fails
completely at nonzero baryon number density.
Finally, lattice gauge theory is only effective
for calculating static quantities,
but many of the more promising signatures for a quark-gluon plasma
involve dynamical quantities.

The only rigorous method available for reorganizing
perturbation theory in thermal QCD is {\it dimensional reduction}
to an effective 3-dimensional field theory \cite{KLRS,Kajantie-97}.
The coefficients of the terms in the effective lagrangian are calculated
using perturbation theory, but calculations within the
effective field theory are carried out nonperturbatively
using lattice gauge theory.
Dimensional reduction has the same limitations as ordinary
lattice gauge theory: it can be applied only to static quantities
and only at zero baryon number density.
Unlike in ordinary lattice gauge theory,
light dynamical quarks do not require any additional computer power,
because they only enter through the perturbatively calculated
coefficients in the effective lagrangian.
This method has been applied to the Debye screening mass
for QCD \cite{Kajantie-97} as well as the
pressure \cite{KLRS}.

There are some proposals for reorganizing
perturbation theory in QCD that are essentially just mathematical
manipulations of the weak coupling expansion.
The methods include {\it Pad\'e approximates} \cite{Pade},
{\it Borel resummation} \cite{Parwani},
and {\it self-similar approximates} \cite{Yukalov}.
These methods are used to construct more stable sequences
of successive approximations that agree with the weak-coupling expansion
when expanded in powers of $\alpha_s$.
These methods can only be applied to quantities
for which several orders in the weak-coupling expansion are known,
so they are limited in practice to the thermodynamic functions.

One promising approach for reorganizing perturbation theory
in thermal QCD is to use a variational framework.
The free energy ${\cal F}$ is expressed as the variational
minimum of a thermodynamic potential $\Omega(T,\alpha_s;m^2)$
that depends on one or more variational parameters
that we denote collectively by $m^2$:
\beq
{\cal F}(T,\alpha_s) \;=\; \Omega(T,\alpha_s;m^2)
\bigg|_{\partial\Omega/\partial m^2 = 0} \,.
\label{varf}
\eeq
%
A particularly compelling variational formulation
is the {\it $\Phi$-derivable approximation}, in which the complete propagator
is used as an infinite set of variational parameters \cite{Phi}.
The $\Phi$-derivable thermodynamic potential $\Omega$ is the
2PI effective action, the sum of all diagrams that are
2-particle-irreducible with respect to the complete propagator \cite{CJT-74}.
The $n$-loop $\Phi$-derivable approximations, in which $\Omega$ is the
the sum of 2PI diagrams with up to $n$ loops, form a systematically
improvable sequence of variational approximations.
Until recently, $\Phi$-derivable approximations have proved to be
intractable for relativistic field theories except for simple cases
in which the self-energy is momentum-independent.
However there has been some recent progress in solving the
3-loop $\Phi$-derivable approximation for scalar field theories.
Braaten and Petitgirard have developed an analytic method
for solving the 3-loop $\Phi$-derivable approximation for the
massless $\phi^4$ field theory \cite{BP-01}.
Van Hees and Knoll have developed numerical methods for solving the
3-loop $\Phi$-derivable approximation for the massive $\phi^4$ field theory
\cite{vHK-01}. They have also investigated renormalization issues
associated with the $\Phi$-derivable approximation.

The application of the $\Phi$-derivable approximation to QCD
was first discussed by McLerran and Freedman \cite{FM-77}.
One problem with this approach is that the
thermodynamic potential $\Omega$ is gauge dependent,
and so are the resulting thermodynamic functions.
The gauge dependence is the same order in $\alpha_s$
as the truncation error.
However the most serious problem is that even the 2-loop
$\Phi$-derivable approximation has proved to be intractable.

The 2-loop $\Phi$-derivable approximation for QCD has been used as
the starting point for {\it HTL resummations} of the entropy by
Blaizot, Iancu and Rebhan \cite{BIR-99}
and of the pressure by Peshier \cite{Peshier-00}.
The thermodynamic potential $\Omega_{\rm 2-loop}$ is a functional
of the complete gluon propagator $D_{\mu\nu}(P)$.
The HTL  resummations of Refs.~\cite{BIR-99} and \cite{Peshier-00}
can be derived in 2 steps.
The first step is to replace the 2-loop functional
at its variational point by a 1-loop functional
evaluated at the 2-loop variational point.
In the resummation of the pressure of Ref.~\cite{Peshier-00},
the 2-loop functional is the thermodynamic potential
and this step is a weak-coupling approximation:
\beq
\Omega_{\rm 2-loop}[D_{\mu\nu}]\bigg|_{\delta\Omega_{\rm 2-loop} = 0}
\;\approx\;
\Omega_{\rm 1-loop}[D_{\mu\nu}]\bigg|_{\delta\Omega_{\rm 2-loop} = 0}
\,.
\label{step1}
\eeq
%
In the resummation of the entropy of Ref.~\cite{BIR-99},
the 2-loop functional is the derivative of $\Omega_{\rm 2-loop}$
with respect to $T$ and this step is an exact equality.
The second step exploits the fact that the HTL gluon propagator
$D_{\mu\nu}^{\rm HTL}(P)$ is an approximate solution
to the variational equation $\delta\Omega_{\rm 2-loop} = 0$.
The HTL gluon propagator depends on one parameter $m_D^2$, which can be
interpreted as the Debye screening mass for the gluon.
The HTL gluon propagator satisfies the variational equation
to leading order in $\alpha_s$ provided that $m_D^2$ reduces
in the weak-coupling limit to
\beq
m_D^2 \;=\;
{4 \pi N_c \over 3} \alpha_s(\mu) T^2 \,,
\label{mD-wcl}
\eeq
%
with some appropriate choice for the scale $\mu$ such as $\mu = 2 \pi T$.
Thus we can approximation the solution to the variational equation
in (\ref{step1}) by $D_{\mu\nu}^{\rm HTL}(P)$:
\beq
\Omega_{\rm 1-loop}[D_{\mu\nu}]\bigg|_{\delta\Omega_{\rm 2-loop} = 0}
\;\approx\;
\Omega_{\rm 1-loop}[D^{\rm HTL}_{\mu\nu}]\bigg|_{m_D^2 = 4 \pi \alpha_s T^2} \,.
\label{step2}
\eeq
%
This approximate solution holds when $m_D^2$ is given by (\ref{mD-wcl}), however,
there is some freedom in the choice of the parameter $m_D^2$,
as long as it reduces to (\ref{mD-wcl}) in the weak-coupling limit.
It can not be determined variationally,
because the variational character of the thermodynamic potential was lost
in the first step (\ref{step1}).
With the prescription (\ref{mD-wcl}), the errors in the
thermodynamic functions are of order $\alpha_s^{3/2}$.
The errors can be reduced to order
$\alpha_s^2$ by adding an $\alpha_s^{3/2}$ term to the right side of
(\ref{mD-wcl}).

The intractability of $\Phi$-derivable approximations
motivates the use of simpler variational approximations.
One such strategy that involves a single variational parameter $m$
has been called {\it optimized perturbation theory} \cite{Stevenson-81},
{\it variational perturbation theory} \cite{varpert},
or the {\it linear $\delta$ expansion} \cite{deltaexp}.
This strategy was applied to the thermodynamics of the massless $\phi^4$
field theory by Karsch, Patkos and Petreczky under the name
{\it screened perturbation theory} \cite{KPP-97}.
The method has also been applied to spontaneously broken
field theories at finite temperature \cite{CK-98}.
The calculations of the thermodynamics of the massless $\phi^4$
field theory using screened perturbation theory
have been extended to 3 loops \cite{ABS-01}.
The calculations can be greatly simplified by using a double
expansion in powers of the coupling constant and $m/T$ \cite{AS-01}.

{\it HTL perturbation theory} (HTLpt) is an adaptation of this strategy
to thermal QCD \cite{ABS-99}.
The exactly solvable theory used as the starting point
is one whose propagators are the HTL gluon propagators.
The variational mass parameter $m_D$ can be identified
with the Debye screening mass.
The one-loop free energy in HTLpt was calculated
for pure-glue QCD in Ref.~\cite{ABS-99}
and for QCD with light quarks in Ref.~\cite{ABS-00}.
At this order, the parameter $m_D$ cannot be determined variationally,
so the prescription (\ref{mD-wcl}) was used.
The resulting thermodynamic functions have errors of order $\alpha_s$,
but the terms of order $\alpha_s^{3/2}$ associated with Debye screening
are correct.
A two-loop calculation is required to reduce the errors to order
$\alpha_s^2$.  At two-loop order, it is also possible to determine
$m_D$ using a variational prescription.

One difference between HTLpt and the
HTL resummation methods of Refs.~\cite{BIR-99} and \cite{Peshier-00}
is in how they deal with gauge invariance.
HTLpt is constructed in such a way that physical
observables are gauge invariant order-by-order in perturbation theory.
Gauge invariance arises in the same way as in ordinary perturbation
theory by cancellations between diagrams.
In the HTL resummation methods of Refs.~\cite{BIR-99} and \cite{Peshier-00},
the 2-loop thermodynamic potential $\Omega_{\rm 2-loop}$
that is used as the starting point is gauge dependent.
In the first step (\ref{step1}) of the derivation,
$\Omega_{\rm 2-loop}$ is replaced by a 1-loop functional
$\Omega_{\rm 1-loop}$ that is gauge invariant,
but the variational equation $\delta\Omega_{\rm 2-loop} = 0$
is still gauge dependent.  In the second step (\ref{step2}),
the solution to that variational equation is approximated by
$D_{\mu \nu}^{\rm HTL}$, and it is only at this point that
the gauge dependence disappears.

Another difference between HTLpt
and the HTL resummation methods of Refs.~\cite{BIR-99} and \cite{Peshier-00}
is in the ranges of observables to which they can be applied.
The HTL resummation methods were specifically formulated as approximations
to the thermodynamic functions, so they cannot be easily
applied to other observables.
However, they can be used to calculate the thermodynamic functions
in cases where calculations using
conventional lattice gauge theory are difficult or impossible:
the high-temperature limit of pure-glue QCD, QCD with light quarks,
and QCD with nonzero baryon number density.
In contrast to these methods, HTLpt has the same
wide range of applicability as ordinary perturbation theory.  It can be
used to calculate the thermodynamic functions, but it
can also be applied to all the standard signatures of a quark-gluon plasma.
It has some of the limitations of ordinary perturbation theory.
Calculations can be carried out only up to the order at which
the magnetic screening problem causes diagrammatic methods to break down.

In this paper, we calculate the thermodynamic functions of QCD
to 2-loop order in HTLpt.
We begin with a brief summary of HTLpt
in Section~\ref{HTLpt}.  In Section~\ref{diagrams},
we give the expressions for the one-loop and two-loop
diagrams for the thermodynamic potential.
In Section~\ref{scalarint},  we reduce those diagrams to scalar sum-integrals.
We are unable to compute those sum-integrals, so in
Section~\ref{expand}, we evaluate them approximately by expanding them
in powers of $m_D/T$.  The diagrams are
combined in Section~\ref{thermpot} to obtain the final results for the
two-loop thermodynamic potential up to 5$^{\rm th}$ order in $g$ and $m_D/T$.
In Section~\ref{thermodynamics}, we present our numerical results for the
thermodynamic functions of QCD.
There are several appendices that contain technical details
of the calculations.
In Appendix~\ref{app:rules},
we give the Feynman rules for HTLpt
in Minkowski space to facilitate the application of this formalism
to signatures of the quark-gluon plasma.
The most difficult aspect of these calculations was the evaluation of the
sum-integrals obtained from the expansion in $m_D/T$.
We give the results for these sum-integrals in Appendix~\ref{app:sumint}.
The evaluation of some difficult thermal integrals
that were required to obtain the sum-integrals is described in
Appendix~\ref{app:int}.

\section{HTL Perturbation Theory}
\label{HTLpt}

The lagrangian density that generates the perturbative expansion for
pure-glue QCD can be expressed in the form
\bqa
{\cal L}_{\rm QCD}=-{1\over2}{\rm Tr}\left(G_{\mu\nu}G^{\mu\nu}\right)
+{\cal L}_{\rm gf}+{\cal L}_{\rm ghost}+\Delta{\cal L}_{\rm QCD},
\label{L-QCD}
\eqa
%
where $G_{\mu\nu}=\partial_{\mu}A_{\nu}-\partial_{\nu}A_{\mu}
	-ig[A_{\mu},A_{\nu}]$ is the gluon field strength
and $A_{\mu}$ is the gluon field expressed
as a matrix in the $SU(N_c)$ algebra.
The ghost term ${\cal L}_{\rm ghost}$ depends on the choice of
the gauge-fixing term ${\cal L}_{\rm gf}$.
Two choices for the gauge-fixing term that depend on an arbitrary gauge
parameter $\xi$ are the general covariant gauge
and the general Coulomb gauge:
\bqa
{\cal L}_{\rm gf}&=&
-{1\over\xi}{\rm Tr}\left(\left(\partial^{\mu}A_{\mu}\right)^2\right)
\hspace{1.1cm}\mbox{covariant}\;,
\label{Lgf-cov}
\\
&=&-{1\over\xi}{\rm Tr}\left(\left(\nabla\cdot {\bf A}\right)^2\right)
\hspace{1cm}\mbox{Coulomb}\;.
\label{Lgf-C}
\eqa
%
The perturbative expansion in powers of $g$
generates ultraviolet divergences.
The renormalizability of perturbative QCD guarantees that
all divergences in physical quantities can be removed by
renormalization of the coupling constant $\alpha_s = g^2/4 \pi$.
If we use dimensional regularization with minimal subtraction
as a renormalization  prescription, the renormalization can be
accomplished by substituting $\alpha_s \to \alpha_s + \Delta \alpha_s$,
where the counterterm $\Delta \alpha_s$ is a power series in $\alpha_s$
whose coefficients have only poles in $\epsilon$:
\beq
\Delta \alpha_s  =
-{11 N_c \over 12 \pi \epsilon} \alpha_s^2
+ \left( {121 N_c^2 \over 144 \pi^2 \epsilon^2}
	- {17 N_c^2 \over 48 \pi^2 \epsilon} \right) \alpha_s^3
	+ O(\alpha_s^4)  \,.
\label{d-alphas}
\eeq
%
Renormalized perturbation theory can be implemented
by including among the interactions terms
a counterterm lagrangian $\Delta{\cal L}_{\rm QCD}$
that is given by the change in the first 3 terms
on the right side of (\ref{L-QCD})
upon substituting $g \to g(1+ \Delta \alpha_s)^{1/2}$.

Hard-thermal-loop perturbation theory (HTLpt) is a reorganization
of the perturbation
series for thermal QCD. The lagrangian density is written as
\bqa
{\cal L}= \left({\cal L}_{\rm QCD}
+ {\cal L}_{\rm HTL} \right) \Big|_{g \to \sqrt{\delta} g}
+ \Delta{\cal L}_{\rm HTL}.
\label{L-HTLQCD}
\eqa
%
The HTL improvement term is
\bqa
\label{L-HTL}
{\cal L}_{\rm HTL}=-{1\over2}(1-\delta)m_D^2 {\rm Tr}
\left(G_{\mu\alpha}\left\langle {y^{\alpha}y^{\beta}\over(y\cdot D)^2}
	\right\rangle_{\!\!y}G^{\mu}_{\;\;\beta}\right),
\eqa
%
where
$D_{\mu}$ is the covariant derivative in the adjoint representation,
$y^{\mu}=(1,\hat{{\bf y}})$ is a light-like four-vector,
and $\langle\ldots\rangle_{ y}$
represents the average over the directions
of $\hat{{\bf y}}$.
The term~(\ref{L-HTL}) has the form of the effective lagrangian
that would be induced by
a rotationally invariant ensemble of colored sources with infinitely high
momentum. The parameter $m_D$ can be identified with the
Debye screening mass.
HTLpt is defined by treating
$\delta$ as a formal expansion parameter.
The free lagrangian in general covariant gauge
is obtained by setting $\delta=0$ in~(\ref{L-HTLQCD}):
\bqa
\nonumber
{\cal L}_{\rm free}&=&-{\rm Tr}\left(\partial_{\mu}A_{\nu}
\partial^{\mu}A^{\nu}-\partial_{\mu}A_{\nu}\partial^{\nu}A^{\mu}\right)
-{1\over\xi}{\rm Tr}\left(\left(\partial^{\mu}A_{\mu}\right)^2\right) \\
&&-{1\over2}m_D^2\mbox{Tr}
\left((\partial_{\mu}A_{\alpha}-\partial_{\alpha}A_{\mu})
\left\langle{y^{\alpha} y^{\beta}\over(y\cdot\partial)^2}\right\rangle_{\!\!y}
(\partial^{\mu}A_{\beta}-\partial_{\beta}A^{\mu})\right).
\label{L-free}
\eqa
%
The resulting propagator is the HTL gluon propagator.
The remaining terms in (\ref{L-HTLQCD}) are treated as perturbations.
The Feynman rules for gluon and ghost propagators
and the 3-gluon, ghost-gluon, and 4-gluon vertices
are given in Appendix \ref{app:rules}.

The HTL perturbation expansion generates ultraviolet divergences.
In QCD perturbation theory, renormalizability constrains the ultraviolet
divergences to have a form that can be cancelled by the counterterm
lagrangian $\Delta{\cal L}_{\rm QCD}$.
There is no proof that the HTL perturbation expansion is renormalizable,
so the general structure of the ultraviolet divergences is not known.
The most optimistic possibility is that HTLpt is
renormalizable, so that the ultraviolet divergences in physical
quantities can all be cancelled by renormalization of
the coupling constant $\alpha_s$, the mass parameter $m_D^2$,
and the vacuum energy density ${\cal E}_0$.
If this is the case, the renormalization of a physical quantity can be
accomplished by substituting $\alpha_s \to \alpha_s + \Delta \alpha_s$
and $m_D^2 \to m_D^2 + \Delta m_D^2$, where $\Delta \alpha_s$ and
$\Delta m_D^2$ are counterterms.  In the case of the free energy,
it is also necessary to add a vacuum energy counterterm $\Delta {\cal E}_0$.
If we use dimensional regularization with minimal subtraction
as a renormalization  prescription, the form of the counterterms
for $\delta \alpha_s$, $(1-\delta)m_D^2$, and ${\cal E}_0$
should be the power of $(1-\delta)m_D^2$ required by dimensional analysis
multiplied by a power series in $\delta \alpha_s$ with coefficients
that have only poles in $\epsilon$.
The counterterm for $\delta \alpha_s$ should be identical to that in
ordinary perturbative QCD given in (\ref{d-alphas}) with:
\beq
\delta \Delta \alpha_s  =
-{11 N_c \over 12 \pi \epsilon} \delta^2 \alpha_s^2
+ \left({121 N_c^2 \over 144 \pi^2 \epsilon^2}
	- {17 N_c^2 \over 48 \pi^2 \epsilon} \right) \delta^3 \alpha_s^3
	+ O(\alpha_s^4)  \,.
\label{del-alphas}
\eeq
%
The leading term in the delta expansion of the ${\cal E}_0$ counterterm
$\Delta{\cal E}_0$ was deduced in Ref.~\cite{ABS-99}
by calculating the free energy to leading order in $\delta$.
The ${\cal E}_0$ counterterm $\Delta{\cal E}_0$ must therefore have the form
\beq
\Delta {\cal E}_0 \;=\;
\left( {N_c^2 - 1\over128\pi^2\epsilon}
	+ O(\delta \alpha_s) \right) (1-\delta)^2 m_D^4\,.
\label{del-E0}
\eeq
%
To calculate the free energy to next-to-leading order in $\delta$,
we need the counterterm $\Delta {\cal E}_0$ to order $\delta$
and the counterterm $\Delta m_D^2$ to order $\delta$.
We will show that there is a nontrivial cancellation of the ultraviolet
divergences if the mass counterterm has the form
\beq
\Delta m_D^2 \;=\;
\left( - {11N_c \over 12 \pi \epsilon} \, \delta \alpha_s
	+ O(\delta^2 \alpha_s^2) \right) (1-\delta) m_D^2 \,.
\label{del-mD2}
\eeq
%
Renormalized perturbation theory can be implemented by including
a counterterm lagrangian $\Delta{\cal L}_{\rm HTL}$ among the
interaction terms in (\ref{L-HTLQCD}).

Physical observables are calculated in HTLpt
by expanding them in powers of $\delta$,
truncating at some specified order, and then setting $\delta=1$.
This defines a reorganization of the perturbation series
in which the effects of
the $m_D^2$ term in~(\ref{L-free})
are included to all orders but then systematically subtracted out
at higher orders in perturbation theory
by the $\delta m_D^2$ term in~(\ref{L-HTL}).
If we set $\delta=1$, the lagrangian (\ref{L-HTLQCD})
reduces to the QCD lagrangian (\ref{L-QCD}).
If the expansion in $\delta$ could be calculated to all orders,
all dependence on $m_D$ should disappear when we set $\delta=1$.
However, any truncation of the expansion in $\delta$ produces results
that depend on $m_D$.
Some prescription is required to determine $m_D$
as a function of $T$ and $\alpha_s$.
We choose to treat $m_D$ as a variational parameter that should be
determined by minimizing the free energy.
If we denote the free energy truncated at some order in $\delta$ by
$\Omega(T,\alpha_s,m_D,\delta)$, our prescription is
\beq
{\partial \ \ \over \partial m_D}\Omega(T,\alpha_s,m_D,\delta=1) = 0.
\label{gap}
\eeq
%
Since $\Omega(T,\alpha_s,m_D,\delta=1)$ is a function of a
variational parameter $m_D$, we will refer to it as the
{\it thermodynamic potential}.  We will refer to the variational equation
(\ref{gap}) as the {\it gap equation}.  The free energy ${\cal F}$
is obtained by evaluating the thermodynamic potential at the solution
to the gap equation.  Other thermodynamic functions can then be
obtained by taking appropriate derivatives of ${\cal F}$ with respect to $T$.

\section{Diagrams for the Thermodynamic Potential}
\label{diagrams}

The thermodynamic potential
at leading order in HTL perturbation theory (HTLpt) is
\begin{equation}
\Omega_{\rm LO} \;=\; (N_c^2-1) {\cal F}_g
	\;+\; \Delta_0{\cal E}_0 \;,
\label{Omega-LO:def}
\end{equation}
%
where ${\cal F}_g$ is the contribution to the free energy from
each of the color states of the gluon:
\begin{eqnarray}
{\cal F}_g & = & -{1 \over 2}\sumint_P
\left\{ (d-1) \log [-\Delta_T(P)] \;+\; \log \Delta_L(P) \right\}\;.
\label{Fg-def}
\end{eqnarray}
%
The transverse and longitudinal HTL propagators
$\Delta_T(P)$ and $\Delta_L(P)$
are given in (\ref{Delta-T}) and (\ref{Delta-L}).
We use dimensional regularization with $d = 3-2 \epsilon$ spatial dimensions
to regularize ultraviolet divergences.
The term of order $\delta^0$ in the
vacuum energy counterterm was determined in Ref.~\cite{ABS-99}:
\bqa
\Delta_0{\cal E}_0={N_c^2-1\over128\pi^2\epsilon}m_D^4\;.
\label{delE-0}
\eqa
%

\begin{figure}[htb]
\hspace{1cm}
\epsfysize=3cm
\centerline{\epsffile{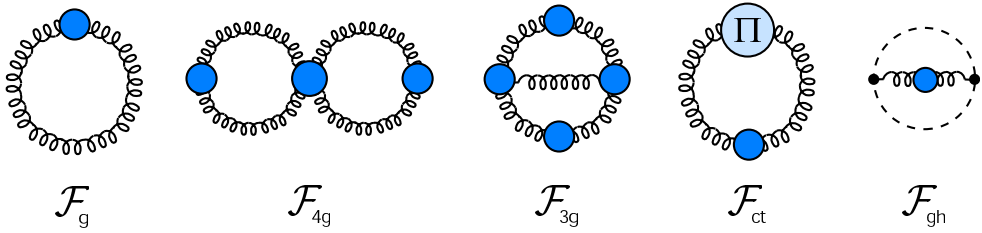}}
\\
\caption[a]{Diagrams contributing through NLO in HTLpt.  Shaded circles
indicate dressed HTL propagators and vertices.}
\label{diagramfig}
\end{figure}

The thermodynamic potential at next-to-leading order in HTLpt
can be written
\bqa
\Omega_{\rm NLO}&=& \Omega_{\rm LO} \;+\; (N_c^2-1)
\left[{\cal F}_{3g}+{\cal F}_{4g}+{\cal F}_{gh}+{\cal F}_{\rm HTL} \right]
+ \Delta_1{\cal E}_0
+ \Delta_1m_D^2 {\partial \ \ \over \partial m_D^2} \Omega_{\rm LO} \;,
\label{F1}
\eqa
%
where $\Delta_1{\cal E}_0$ and $\Delta_1m_D^2$ are the terms of order
$\delta$ in the vacuum energy density and mass counterterms.
The contributions from the two-loop diagrams with the
three-gluon and four-gluon vertices are
\bqa
{\cal F}_{3g}&=&
{N_c\over12}g^2\sumint_{PQ}\Gamma^{\mu\lambda\rho}(P,Q,R)
\Gamma^{\nu\sigma\tau}(P,Q,R)
\Delta^{\mu\nu}(P)\Delta^{\lambda\sigma}(Q)\Delta^{\rho\tau}(R)\;,
\label{F-3g}
\\
{\cal F}_{4g}&=&
{N_c\over8}g^2
\sumint_{PQ}\Gamma^{\mu\nu,\lambda\sigma}(P,-P,Q,-Q)
\Delta^{\mu\nu}(P)\Delta^{\lambda\sigma}(Q)\;.
\label{F-4g}
\eqa
%
Expressions for the gluon propagator tensor $\Delta^{\mu\nu}$,
the 3-gluon vertex tensor $\Gamma^{\mu\lambda\rho}$,
and the 4-gluon vertex tensor $\Gamma^{\mu\nu,\lambda\sigma}$
in Minkowski space are given in (\ref{D-cov}) or (\ref{D-C}),
(\ref{Gam3}), and (\ref{Gam4}).  Prescriptions for translating them
into the Euclidean tensors appropriate for the imaginary time formalism
are given in Appendix \ref{app:ITF}.
The contribution from the ghost diagram depends on the choice of gauge.
The expressions in the covariant and Coulomb gauges are
\bqa
{\cal F}_{gh}&=&
{N_c \over 2} g^2\sumint_{PQ}{1\over Q^2}{1\over R^2}Q^{\mu}R^{\nu}
\Delta^{\mu\nu}(P) \hspace{5.2cm}\mbox{covariant}\;,
\label{F-gh:cov}
\\
&=&
{N_c \over 2} g^2\sumint_{PQ}{1\over q^2}{1\over r^2}
\left(Q^{\mu}-Q\!\cdot\!n\;n^{\mu}\right)\left(R^{\nu}-R\!\cdot\!n\;
n^{\nu}\right)
\Delta^{\mu\nu}(P)
\hspace{1cm}\mbox{Coulomb}\;.
\label{F-gh:C}
\eqa
%
The contribution from the HTL counterterm diagram is
\bqa
{\cal F}_{\rm HTL}&=& {1\over2}\sumint_P\Pi^{\mu\nu}(P)\Delta^{\mu\nu}(P)\;.
\label{F-ct}
\eqa
%
It can also be obtained by substituting $m_D^2 \to (1-\delta) m_D^2$
in the one-loop expression ${\cal F}_g$ in (\ref{Fg-def})
and expanding to first order in
$\delta$:
\bqa
{\cal F}_{\rm HTL} \;=\;
{1 \over 2} \sumint_P \left[ (d-1) \Pi_T(P) \Delta_T(P)
		- \Pi_L(P) \Delta_L(P) \right] \,.
\label{F-ct:scalar}
\eqa
%

Provided that HTLpt is renormalizable,
the ultraviolet divergences at any order in $\delta$
can be cancelled by renormalizations of
the vacuum energy density ${\cal E}_0$,
the HTL mass parameter $m_D^2$, and the coupling constant $\alpha_s$.
Renormalization of the coupling constant
does not enter until order $\delta^2$.
We will calculate the thermodynamic potential as a double expansion
in powers of $g$ and $m_D/T$,
including all terms through 5$^{\rm th}$ order.
The $\delta \alpha_s$ term in $\Delta{\cal E}_0$
does not contribute until 6$^{\rm th}$ order in this expansion,
so the term of order $\delta$
in $\Delta{\cal E}_0$ can be obtained simply by expanding
(\ref{delE-0}) to first order in $\delta$:
\bqa
\Delta_1{\cal E}_0 =
- {N_c^2-1\over 64\pi^2\epsilon} m_D^4 \;.
\label{DelE0:NLO}
\eqa
%
The remaining ultraviolet divergences must be removed by
renormalization of the mass parameter $m_D$.
We will find that there are ultraviolet divergences in the
$\alpha_s m_D^2 T^2$ and $\alpha_s m_D^3 T^3$ terms,
and both are removed by the same counterterm $\Delta_1m_D^2$.
This provides nontrivial evidence for the renormalizability
of HTLpt at this order in $\delta$.

The sum of the 3-gluon, 4-gluon, and ghost contributions in
(\ref{F-3g}), (\ref{F-4g}), and (\ref{F-gh:cov}) or (\ref{F-gh:C})
is gauge-invariant.  By inserting the expression (\ref{D-cov})
or (\ref{D-C}) for the gluon propagator tensor and
using the Ward identities (\ref{ward-3}) and (\ref{ward-4}),
one can easily verify that the sum of these three diagrams is
independent of the
gauge parameter $\xi$ in both covariant gauge and Coulomb gauge.
With more effort, we can verify the equivalence of the
covariant gauge expression with $\xi=0$ (Landau gauge)
and the Coulomb gauge expression with $\xi = 0$.
This involves expanding the tensor $n_P^\mu n_P^\nu$
in the covariant gauge propagator into the sum of terms proportional
to $n^\mu n^\nu$, $P^\mu n^\nu$, $n^\mu P^\nu$, and $P^\mu P^\nu$,
and then applying the Ward identities to the terms
involving $P^\mu$ or $P^\nu$.

\section{Reduction to Scalar Sum-integrals}
\label{scalarint}

The first step in calculating the thermodynamic potential is to reduce
the sum of the diagrams to scalar sum-integrals.
The one-loop diagram in (\ref{Fg-def})
and the HTL counterterm diagram (\ref{F-ct}) are already
expressed in terms of scalar integrals.
We proceed to consider the 3-gluon diagram in (\ref{F-3g}),
the 4-gluon diagram in (\ref{F-4g}),
and the ghost diagram in Landau gauge which is given in (\ref{F-gh:cov}).
The expression for the sum of these three diagrams
is simpler than that of the 3-gluon diagram alone.
We insert the gluon propagator in the form (\ref{gprop-TC}) with $\xi=0$.
It has terms proportional to $\Delta_T$ and $\Delta_X$,
where $\Delta_X$ is the combination of transverse and longitudinal
propagators defined in (\ref{Delta-X}).
When a momentum $P^\mu$ from the gluon propagator tensor
is contracted with a 3-gluon or 4-gluon vertex,
the Ward identities can be used to reduce it ultimately to
expressions involving the inverse propagator (\ref{delta-inv:inf}).
The term $\Delta_T/\Delta_L$ can be eliminated
in favor of $\Delta_X/\Delta_L$ using the
definition (\ref{Delta-X}).
This reduces the sum of the 3-gluon, 4-gluon, and ghost diagrams
to the following form:
\bqa
&& {\cal F}_{3g}+{\cal F}_{4g}+{\cal F}_{gh}
\nonumber
\\
&& \;=\; {N_c \over 12} g^2 \sumint_{PQ} \Bigg\{
\Gamma^{\mu\nu\lambda} \Gamma^{\mu\nu\lambda} \;
	\Delta_T(P) \Delta_T(Q) \Delta_T(R)
- 3 \Gamma^{\mu\nu0} \Gamma^{\mu\nu0} \;
	\Delta_T(P) \Delta_T(Q) \Delta_X(R)
\nonumber
\\
&& \hspace{2.5cm}
+ 3 \Gamma^{\mu00} \Gamma^{\mu00} \;
	\Delta_T(P) \Delta_X(Q) \Delta_X(R)
- \left( \Gamma^{000} \right)^2 \;
	\Delta_X(P) \Delta_X(Q) \Delta_X(R)
\nonumber
\\
&& \hspace{2.5cm}
+ 3 d(d+1) \; \Delta_T(P) \Delta_T(Q)
- 6 d \; \Delta_T(P) \Delta_X(Q)
+ {3\over2} \Gamma^{00,00} \; \Delta_X(P) \Delta_X(Q)
\nonumber
\\
&& \hspace{3cm}
+ 6 \left( { Q \!\cdot\! R \over Q^2 R^2} \;
	\Delta_T(P)
- {n\!\cdot\!Q \; n\!\cdot\! R \over Q^2 R^2}
	\Delta_X(P) \right)
\nonumber
\\
&& \hspace{2.5cm}
- 12 \left( {n \!\cdot\! Q \; n_Q \!\cdot\! R \over q^2 R^2} \;
	\Delta_T(P)
- {n \!\cdot\! Q \; n \!\cdot\! R \over Q^2 R^2}
	\Delta_X(P) \right) {\Delta_X(Q) \over \Delta_L(Q)}
\nonumber
\\
&& \hspace{2.5cm}
+ 6 \left( {n \!\cdot\! Q \; n \!\cdot\! R \; n_Q \!\cdot\! n_R
		\over q^2 r^2} \; \Delta_T(P)
- {n \!\cdot\! Q \; n \!\cdot\! R \over Q^2 R^2} \;
	\Delta_X(P) \right) {\Delta_X(Q) \over \Delta_L(Q)}
			{\Delta_X(R) \over \Delta_L(R)}
\Bigg\} \,.
\label{F1-a}
\eqa
%
In the 3-gluon and 4-gluon vertex tensors,
we have suppressed the momentum arguments:
$\Gamma^{\mu\nu\lambda} = \Gamma^{\mu\nu\lambda}(P,Q,R)$
and $\Gamma^{00,00} = \Gamma^{00,00}(P,-P,Q,-Q)$.

The next step is to insert the
Euclidean analogs of the expressions
(\ref{Gam3}) and (\ref{Gam4}) for the 3-gluon and 4-gluon vertex
tensors.  The combinations of terms that appear in (\ref{F1-a})
can be simplified using the ``Ward identities'' (\ref{ward-t2}),
(\ref{ward-t3}), and (\ref{ward-t4})
satisfied by the HTL correction tensors:
\bqa
\Gamma^{\mu\nu\lambda} \Gamma^{\mu\nu\lambda} &=&
3 d (P^2 + Q^2 + R^2)
+ m_D^4 \; {\cal T}^{\mu\nu\lambda} {\cal T}^{\mu\nu\lambda} \;,
\label{gam3gam3}
\\
\Gamma^{\mu\nu0} \Gamma^{\mu\nu0} &=&
p^2 + q^2 + 4 r^2 + 2d (n\!\cdot\!P)^2 + 2d (n\!\cdot\!Q)^2
- d (n\!\cdot\!R)^2
\nonumber
\\
&& + 2 m_D^2 \left( 2 {\cal T}_R - {\cal T}_P - {\cal T}_Q \right)
+ m_D^4 \; {\cal T}^{\mu\nu0} {\cal T}^{\mu\nu0} \;,
\\
\Gamma^{\mu00} \Gamma^{\mu00} &=&
2 q^2 + 2 r^2 - p^2
\nonumber
\\
&& - 2 m_D^2 \left[ 2 {\cal T}_P - {\cal T}_Q - {\cal T}_R
		+ n\!\cdot\!(Q-R) \; {\cal T}^{000}\right]
+ m_D^4 \; {\cal T}^{\mu00} {\cal T}^{\mu00} \;,
\\
(\Gamma^{000})^2  &=&
m_D^4 \; ({\cal T}^{000})^2 \;,
\\
\Gamma^{00,00} &=&
- m_D^2 \; {\cal T}^{0000} \;.
\label{gam0000}
\eqa
%
In the 3-gluon and 4-gluon HTL correction tensors,
we have suppressed the momentum arguments:
${\cal T}^{\mu\nu\lambda} = {\cal T}^{\mu\nu\lambda}(P,Q,R)$
and ${\cal T}^{0000} = {\cal T}^{0000}(P,Q,-P,-Q)$.
We have also used the short-hand ${\cal T}_P = -{\cal T}^{00}(P,-P)$
for the 2-gluon HTL correction tensor.
Inserting the expressions (\ref{gam3gam3})--(\ref{gam0000}) into
(\ref{F1-a}) and eliminating $1/\Delta_L(P)$ in favor of ${\cal T}_P$,
the reduction to scalar integrals is
\bqa
&& {\cal F}_{3g}+{\cal F}_{4g}+{\cal F}_{gh}
\nonumber
\\
&& \;=\; {N_c\over4} g^2 \sumint_{PQ}
 \Bigg\{
\left[ 3 d R^2
+ {1\over3} m_D^4 \; {\cal T}^{\mu\nu\lambda} {\cal T}^{\mu\nu\lambda} \right]
	\Delta_T(P) \Delta_T(Q) \Delta_T(R)
\nonumber
\\
&& \hspace{3cm}
+ \left[ -2 q^2 - 4 r^2 - 4d (n\!\cdot\!Q)^2 + d (n\!\cdot\!R)^2
\right.
\nonumber
\\
&& \hspace{4cm} \left.
- 4 m_D^2 \left( {\cal T}_R - {\cal T}_Q \right)
- m_D^4 \; {\cal T}^{\mu\nu0} {\cal T}^{\mu\nu0} \right]
	\Delta_T(P) \Delta_T(Q) \Delta_X(R)
\nonumber
\\
&& \hspace{3cm}
+ \left[ -p^2 + 4 r^2
- 2 m_D^2 \, n \!\cdot\! (Q-R) \; {\cal T}^{000}
\right.
\nonumber
\\
&& \hspace{4cm} \left.
- 4 m_D^2 \left( {\cal T}_P - {\cal T}_R \right)
+ m_D^4 \; {\cal T}^{\mu00} {\cal T}^{\mu00} \right]
	\Delta_T(P) \Delta_X(Q) \Delta_X(R)
\nonumber
\\
&& \hspace{3cm}
- {1\over3} m_D^4 \; ({\cal T}^{000})^2 \;
	\Delta_X(P) \Delta_X(Q) \Delta_X(R)
\nonumber
\\
&&\hspace{3cm}
+ d(d+1) \; \Delta_T(P) \Delta_T(Q)
- 2 d \;  \Delta_T(P) \Delta_X(Q)
\nonumber
\\
&& \hspace{3cm}
- {1\over2}  m_D^2 \; {\cal T}^{0000} \; \Delta_X(P) \Delta_X(Q)
\nonumber
\\
&& \hspace{3cm}
+ 2 {Q \!\cdot\! R \over Q^2 R^2} \Delta_T(P)
	\left( 1 - [q^2 + m_D^2 (1-{\cal T}_Q)] \Delta_X(Q) \right)
\nonumber
\\
&& \hspace{6cm}
	\times \left( 1 - [r^2 + m_D^2 (1-{\cal T}_R)] \Delta_X(R) \right)
\nonumber
\\
&& \hspace{3cm}
- 2 {n\!\cdot\!Q \; n\!\cdot\!R \over Q^2 R^2} \Delta_X(P)
	\left( 1 - [q^2 + m_D^2 (1-{\cal T}_Q)] \Delta_X(Q) \right)
\nonumber
\\
&& \hspace{6cm}
	\times \left( 1 - [r^2 + m_D^2 (1-{\cal T}_R)] \Delta_X(R) \right)
\nonumber
\\
&& \hspace{3cm}
+ 4 {{\bf q}\cdot{\bf r} \over q^2 R^2} \;
	[q^2 + m_D^2 (1-{\cal T}_Q)] \; \Delta_T(P) \Delta_X(Q)
\nonumber
\\
&& \hspace{3cm}
- 2 { (2 n_Q^2 - 1){\bf q}\cdot{\bf r}
		\over q^2 r^2 } \;
	[q^2 + m_D^2 (1-{\cal T}_Q)] [r^2 + m_D^2 (1-{\cal T}_R)] \;
\nonumber
\\
&& \hspace{6cm}
	\times \Delta_T(P) \Delta_X(Q) \Delta_X(R)
\Bigg\} \,.
\label{F1-b}
\eqa
%

\section{Expansion in the mass parameter}
\label{expand}

The thermodynamic potential has been reduced to scalar sum-integrals.
In Ref.~\cite{ABS-99}, the sum-integrals for the one-loop free energy were
evaluated exactly by replacing the sums by contour integrals,
extracting the poles in $\epsilon$, and then reducing the momentum
integrals to integrals that were at most two-dimensional
and could therefore be easily evaluated numerically.
It was also shown that the sum-integrals could be expanded
in powers of $m_D/T$, and that the first few terms in the expansion
gave a surprisingly accurate approximation to the exact result.

If we tried to evaluate the 2-loop HTL free energy exactly,
there are terms such as those involving
${\cal T}^{\mu\nu\lambda} {\cal T}^{\mu\nu\lambda}$
that could at best be reduced to 5-dimensional integrals
that would have to be evaluated numerically.
We will therefore evaluate the sum-integrals approximately
by expanding them in powers of $m_D/T$.  We will carry out the
$m_D/T$ expansion to high enough order  to include all terms
through  order $g^5$ if $m_D/T$ is taken to be of order $g$.

\subsection{1-loop sum-integrals}

The one-loop sum-integrals include the leading order free energy
given by the sum-integrals (\ref{Fg-def})
and the HTL counterterm given by (\ref{F-ct:scalar}).
The leading order free energy must be expanded to order $(m_D/T)^5$
in order to include all terms through order $g^5$.
The HTL counterterm has an explicit factor of
$m_D^2$, so the sum-integral for the HTL counterterm diagram
need only to be expanded
to order $(m_D/T)^3$ to include all terms through order $g^5$.

The sum-integrals over $P$ involve two momentum scales: $m_D$ and $T$.
In order to expand them in powers of $m_D/T$,
we separate them into contributions from hard loop momentum,
for which some of the components of $P$ are of order $T$,
and soft loop momenta, for which all the components of $P$
are of order $m_D$.
We will denote these regions by $(h)$ and $(s)$.
Since the Euclidean energy $P_0$ is an integer multiple of
$2 \pi T$, the soft region requires $P_0=0$.

\subsubsection{Hard contributions}

If $P$ is hard, the denominators $P^2 + \Pi_T$ and $p^2 +\Pi_L$
in the propagators are of order $T$, but the self-energy functions
$\Pi_T$ and $\Pi_L$ are of order $m_D^2$.  The $m_D/T$ expansion
can therefore be obtained by expanding in powers of
$\Pi_T$ and $\Pi_L$.

For one-loop free energy, we need to expand to
second order in $m_D^2$:
\bqa
{\cal F}_g^{(h)}
&=& {d-1 \over 2} \sumint_P \log(P^2)
+ {1 \over 2} m_D^2 \sumint_P {1 \over P^2}
\nonumber
\\
&&\;-\;
{1\over4(d-1)} m_D^4 \sumint_P
\left[ {1 \over (P^2)^2} - 2 \, {1 \over p^2 P^2}
	- 2  d \, {1 \over p^4} {\cal T}_P
	+ 2 \, {1 \over p^2 P^2} {\cal T}_P
	+ d \, {1 \over p^4} ({\cal T}_P)^2 \right]
\,.
\eqa
%
Note that the function ${\cal T}_P$ cancels from the $m_D^2$ term
because of the identity (\ref{PiTL-id}).
The values of the sum-integrals are given in Appendix \ref{app:sumint}.
Inserting those expressions, the hard contributions to the
leading-order free energy reduce to
\bqa
{\cal F}_g^{(h)}
&=& - {\pi^2 \over 45} T^4
+ {1 \over 24} \left[ 1
	+ \left( 2 + 2{\zeta'(-1) \over \zeta(-1)} \right) \epsilon \right]
\left( {\mu \over 4 \pi T} \right)^{2\epsilon} m_D^2 T^2
\nonumber
\\
&& - {1 \over 128 \pi^2}
\left( {1 \over \epsilon} - 7 + 2 \gamma + {2 \pi^2\over 3} \right)
\left( {\mu \over 4 \pi T} \right)^{2\epsilon} m_D^4 \,.
\label{Flo-h}
\eqa
%
Note that the pole in the $m_D^4$ term is cancelled by the
counterterm (\ref{delE-0}).

The HTL counterterm diagram has an explicit factor of
$m_D^2$, so we need only to expand the sum-integral to
first order in $m_D^2$.  Eliminating
$\Pi_T(P)$ and $\Pi_L(P)$ in favor of the function ${\cal T}_P$,
the result is
\bqa
{\cal F}_{ct}^{(h)}
&=& - {1 \over 2} m_D^2 \sumint_P {1 \over P^2}
\nonumber
\\
&&\;+\;
{1\over 2(d-1)} m_D^4 \sumint_P
\left[ {1 \over (P^2)^2} - 2 \, {1 \over p^2 P^2}
	- 2  d \, {1 \over p^4} {\cal T}_P
	+ 2 \, {1 \over p^2 P^2} {\cal T}_P
	+ d \, {1 \over p^4} ({\cal T}_P)^2 \right]
\,.
\eqa
%
The values of the sum-integrals are given in Appendix \ref{app:sumint}.
Inserting those expressions, the hard contributions to the
HTL counterterm in the free energy reduce to
\bqa
{\cal F}_{ct}^{(h)}
&=& - {1\over24} m_D^2 T^2
+ {1 \over 64 \pi^2}
\left( {1 \over \epsilon} - 7 + 2 \gamma + {2 \pi^2\over 3} \right)
\left( {\mu \over 4 \pi T} \right)^{2\epsilon} m_D^4 \,.
\label{Fct-h}
\eqa
%
Note that the first term in (\ref{Fct-h}) cancels the
order-$\epsilon^0$ term
in the coefficient of $m_D^2 T^2$ in (\ref{Flo-h}).
We have kept the order-$\epsilon$ term in the
coefficient of $m_D^2 T^2$ in (\ref{Flo-h}),
because it will contribute to the final result
through the mass counterterm.

\subsubsection{Soft contributions}

The soft contribution comes from the $P_0=0$ term in the sum-integral.
At soft momentum $P=(0,{\bf p})$, the HTL self-energy functions
reduce to $\Pi_T(P) = 0$ and $\Pi_L(P) = m_D^2$.
The transverse term vanishes in dimensional regularization
because there is no momentum scale in the integral over ${\bf p}$.
Thus the soft contribution comes from the longitudinal term only.

The soft contribution to the leading order free energy is
\bqa
{\cal F}_g^{(s)}
&=& {1 \over 2} T \int_{\bf p} \log(p^2 + m_D^2) \,.
\eqa
%
Using the expression for the integral in Appendix \ref{app:int}, we obtain
\bqa
{\cal F}_g^{(s)}
&=& - {1 \over 12 \pi}
\left[ 1 + {8 \over 3}\epsilon \right]
\left( {\mu \over 2 m_D} \right)^{2 \epsilon}
m_D^3 T \,.
\label{Flo-s}
\eqa
%

The soft contribution to the HTL counterterm  is
\bqa
{\cal F}_{ct}^{(s)}
&=& - {1 \over 2} m_D^2 T \int_{\bf p} {1 \over p^2 + m_D^2} \,.
\eqa
%
Using the expression for the integral in Appendix \ref{app:int}, we obtain
\bqa
{\cal F}_{ct}^{(s)}
&=& {1 \over 8 \pi}  m_D^3 T \,.
\label{Fct-s}
\eqa
%

\subsection{2-loop sum-integrals}

The sum of the two-loop sum-integrals is given in (\ref{F1-b}).
Since these integrals have an explicit factor of $g^2$,
we need only expand the sum-integrals
to order $(m_D/T)^3$ to include all terms through order $g^5$.

The sum-integrals involve two momentum scales: $m_D$ and $T$.
In order to expand them in powers of $m_D/T$,
we separate them into contributions from hard loop momenta
and soft loop momenta.  This gives 3 separate regions
which we will denote $(hh)$, $(hs)$, and $(ss)$.
In the $(hh)$ region, all 3 momenta $P$, $Q$, $R$ are hard.
In the $(hs)$ region, two of the 3 momenta are hard
and the other is soft.
In the $(ss)$ region, all 3 momenta are soft.

\subsubsection{Contributions from $(hh)$ region}

If $P$, $Q$, $R$ are all hard, we can obtain the $m_D/T$ expansion simply
by expanding in powers of $m_D^2$.
To obtain the expansion through order $m_D^3/T^3$,
we need only expand to first order in $m_D^2$,
with $\Delta_X$ and $\Pi_T$ taken to be of order $m_D^2$:
%
%
\bqa
{\cal F}_{3g+4g+gh}^{(hh)}
&=& {N_c\over4} g^2 \sumint_{PQ}
 \Bigg\{
3 d R^2 \; \Delta_T(P) \Delta_T(Q) \Delta_T(R)
\nonumber
\\
&& \hspace{2cm}
+ \left[ -2 q^2 - 4 r^2 - 4d (n\!\cdot\!Q)^2 + d (n\!\cdot\!R)^2
	\right] \Delta_T(P) \Delta_T(Q) \Delta_X(R)
\nonumber
\\
&&\hspace{2cm}
+ d(d+1) \; \Delta_T(P) \Delta_T(Q)
- 2 d \;  \Delta_T(P) \Delta_X(Q)
\nonumber
\\
&& \hspace{2cm}
+ 2 {Q \!\cdot\! R \over Q^2 R^2} \Delta_T(P)
	\left( 1 - 2 q^2 \Delta_X(Q) \right)
\nonumber
\\
&& \hspace{2cm}
- 2 {n\!\cdot\!Q \; n\!\cdot\!R \over Q^2 R^2} \Delta_X(P)
+ 4 {{\bf q}\!\cdot\!{\bf r} \over R^2 } \; \Delta_T(P) \Delta_X(Q)
\Bigg\}  \,.
\label{Fhh-1}
\eqa
%
For hard momenta, the self-energies are suppressed by $m_D^2/T^2$
relative to the propagators, so they can be expanded in powers of
$\Pi_T$ and $\Pi_L$.
Expanding all terms to first order in $m_D^2$, and
using (\ref{pit2}) and (\ref{pil2}) to
eliminate $\Pi_T(P)$ and $\Pi_L(P)$ in favor of ${\cal T}_P$,
we obtain
\bqa
{\cal F}_{3g+4g+gh}^{(hh)}
&=& {N_c\over4} g^2 \sumint_{PQ}
 \Bigg\{
(d-1)^2 \, {1 \over P^2} \, {1 \over Q^2}
\Bigg\}
\nonumber
\\
&&\;+\;
{N_c\over4} g^2 m_D^2 \sumint_{PQ}
 \Bigg\{ -2(d-1) \, {1 \over P^2} \, {1 \over (Q^2)^2}
+ 2(d-2)  \, {1 \over P^2} \, {1 \over q^2 Q^2}
\nonumber
\\
&& \hspace{4cm}
+ 2 \, {1 \over P^2 Q^2 R^2}
+ (d+2) \, {1 \over P^2 Q^2 r^2}
\nonumber
\\
&& \hspace{4cm}
- 2 d \, {P\!\cdot\!Q \over P^2 Q^2 (r^2)^2}
- 4d \, {q^2 \over P^2 Q^2 (r^2)^2}
+ 4 \, {q^2 \over P^2 Q^2 r^2 R^2}
\nonumber
\\
&& \hspace{4cm}
- 2(d-1) \, {1 \over P^2} \, {1 \over q^2 Q^2}{\cal T}_Q
- (d+1) {1 \over P^2 Q^2 r^2} {\cal T}_R
\nonumber
\\
&& \hspace{4cm}
+ 4d \, {q^2 \over P^2 Q^2 (r^2)^2}{\cal T}_R
+ 2d \,{P\!\cdot\!Q \over P^2 Q^2 (r^2)^2} {\cal T}_R
\Bigg\}  \,.
\label{Fhh-3}
\eqa
%
Inserting the sum-integrals from Appendix \ref{app:sumint}, this reduces to
\bqa
{\cal F}_{3g+4g+gh}^{(hh)}
&=& {\pi^2 \over 12} {N_c \alpha_s \over 3 \pi} T^4
\;-\; {7 \over 96} \left[ {1\over\epsilon} + 4.621 \right]
{N_c \alpha_s \over 3 \pi}
\left( {\mu \over 4 \pi T} \right)^{4\epsilon}
m_D^2 T^2 \,.
\label{F2loop-hh}
\eqa
%

\subsubsection{The $(hs)$ contributions}

In the $(hs)$ region, the soft momentum can be any one of the
three momenta $P$, $Q$, or $R$.  However we can always permute
the momenta so that the soft momentum is $P = (0,{\bf p})$.
The function that multiplies the soft propagator $\Delta_T(0,{\bf p})$
or $\Delta_X(0,{\bf p})$ can be expanded in powers of
the soft momentum ${\bf p}$.  In the case of $\Delta_T(0,{\bf p})$,
the resulting integrals over ${\bf p}$ have no scale
and therefore vanish in dimensional regularization.
The integration measure $\int_{\bf p}$ scales like $m_D^3$,
the soft propagator $\Delta_X(0,{\bf p})$ scales like $1/m_D^2$,
and every power of $p$ in the numerator scales like $m_D$.
The only terms that contribute through order $g^2 m_D^3 T$ are
%
%
\bqa
&&{\cal F}_{3g+4g+gh}^{(hs)}
\nonumber
\\
&& \;=\; {N_c\over4} g^2 T \int_{\bf p} \Delta_X(0,{\bf p}) \sumint_Q
\Bigg\{
\left[ - 2 q^2 - 4 p^2 - 4d (n\!\cdot\!Q)^2 + 4 m_D^2 {\cal T}_Q \right]
		\Delta_T(Q) \Delta_T(R)
\nonumber
\\
&& \hspace{5cm}
+ \left[ 4 r^2 - 2 q^2 + 4 p^2 \right]
		\Delta_T(Q) \Delta_X(R)
\nonumber
\\
&& \hspace{5cm}
- 2 d \Delta_T(Q)
+ 2 {(n\!\cdot\!Q)^2 \over Q^2 R^2}
\left(1 - 2 q^2 \Delta_X(Q)  \right)
\Bigg\} \,.
\label{Fhs-1}
\eqa
%
In the terms that are already of order $g^2 m_D^3 T$, we can set $R=-Q$.
In the terms of order $g^2 m_D T^3$, we must expand the sum-integrand
to second order in ${\bf p}$.  After averaging over angles of ${\bf p}$,
the linear terms in ${\bf p}$ vanish and quadratic terms
of the form $p^i p^j$ are replaced by $p^2 \delta^{ij}/d$.
We can set $p^2 = -m_D^2$, because any factor
proportional to $p^2 + m_D^2$ will cancel the denominator of the
integral over ${\bf p}$, leaving an integral with no scale.
Our expression for the $(hs)$ contribution reduces to
\bqa
{\cal F}_{3g+4g+gh}^{(hs)}
&=& {N_c\over2} g^2 T \int_{\bf p} {1 \over p^2 + m_D^2} \sumint_Q
\Bigg\{
-(d-1) \, {1 \over Q^2}
+ 2(d-1) \, {q^2 \over (Q^2)^2}
\Bigg\}
\nonumber
\\
&&\;+\; N_c g^2 m_D^2 T \int_{\bf p} {1 \over p^2 + m_D^2}
	\sumint_Q
\Bigg\{
- (d-4) {1 \over (Q^2)^2}
+ {(d-1)(d+2) \over d} \, {q^2 \over (Q^2)^3}
\nonumber
\\
&& \hspace{6cm}
- {4(d-1) \over d} \, {q^4 \over (Q^2)^4}
\Bigg\} \,.
\label{Fhs-3}
\eqa
%
Inserting the sum-integrals from Appendix \ref{app:sumint}
and the integrals from Appendix \ref{app:int}, this reduces to
\bqa
{\cal F}_{3g+4g+gh}^{(hs)}
&=& - {\pi \over 2} {N_c \alpha_s \over 3 \pi} m_D T^3
\nonumber
\\
&& \;-\; {11 \over 32 \pi}
\left( {1 \over \epsilon} + {27 \over 11} + 2 \gamma \right)
{N_c \alpha_s \over 3 \pi}
\left( {\mu \over 4 \pi T} \right)^{2\epsilon}
\left( {\mu \over 2m_D} \right)^{2\epsilon}
	m_D^3 T \,.
\label{F2loop-hs}
\eqa
%

\subsubsection{The $(ss)$ contributions}

The $(ss)$ contributions comes from the zero-frequency modes
of the sum-integrals.
The HTL correction functions ${\cal T}_P$,  ${\cal T}^{000}$,
and ${\cal T}^{0000}$ vanish when all the external frequencies
are zero.  The self-energy functions at zero-frequency are
$\Pi_T(0,{\bf p}) = 0$ and $\Pi_L(0,{\bf p}) = m_D^2$ .
The only scales in the integrals come from the longitudinal
propagators $\Delta_L(0,{\bf p}) = 1/(p^2 +m_D^2)$.
Therefore in dimensional regularization, at least one such propagator
is required in order for the integral to be nonzero.
The only terms in (\ref{F1-b}) that give nonzero contributions are
\bqa
{\cal F}_{3g+4g+gh}^{(ss)}
&=& {N_c\over4} g^2 T^2 \int_{\bf pq}
\Bigg\{
\left[ - 2 q^2 - 4 r^2 \right]
	\Delta_T(0,{\bf p}) \Delta_T(0,{\bf q}) \Delta_X(0,{\bf r})
\nonumber
\\
&& \hspace{2cm}
+ \left[ - p^2 + 4 r^2 \right]
	\Delta_T(0,{\bf p}) \Delta_X(0,{\bf q}) \Delta_X(0,{\bf r})
\Bigg\} \,.
\label{Fss-1}
\eqa
%
After simplifying the integral by dropping terms that vanish in
dimensional regularization, it reduces to
\bqa
{\cal F}_{3g+4g+gh}^{(ss)}
&=& {N_c\over4} g^2 T^2 \int_{\bf pq}
{p^2 + 4 m_D^2 \over p^2 (q^2 + m_D^2)(r^2 + m_D^2)}\,.
\label{Fss-2}
\eqa
%
Inserting the integrals from Appendix \ref{app:int}, this reduces to
\bqa
{\cal F}_{3g+4g+gh}^{(ss)}
&=& {3 \over 16} \left[ {1 \over \epsilon} + 3 \right]
{N_c \alpha_s \over 3 \pi}
\left( {\mu \over 2 m_D} \right)^{4 \epsilon}
m_D^2 T^2 \,.
\label{F2loop-ss}
\eqa
%

\section{Thermodynamic potential}
\label{thermpot}

In this section, we calculate the thermodynamic potential
$\Omega(T,\alpha_s,m_D,\delta=1)$ explicitly,
first to leading order in the $\delta$ expansion and then to
next-to-leading order.

\subsection{Leading order}

The complete expression for the leading order thermodynamic potential
is the sum of the contributions from 1-loop diagrams
and the leading term (\ref{delE-0})
in the vacuum energy counterterm.
The contributions from the 1-loop diagrams,
including all terms through order $g^5$, is the sum of
(\ref{Flo-h}) and (\ref{Flo-s}):
\bqa
\Omega_{1-{\rm loop}} &=& {\cal F}_{\rm ideal}
\left\{ 1 - {15 \over 2} \hat m_D^2
+ 30 \hat m_D^3
+ {45 \over 8}
\left( {1\over \epsilon} + 2 \log {\hat \mu \over 2}
	- 7 + 2 \gamma + {2 \pi^2\over 3} \right)
	\hat m_D^4  \right\} \;,
\label{Omega-1loop}
\eqa
%
where ${\cal F}_{\rm ideal}$ is the free energy of an ideal gas of
$N_c^2 -1$ massless spin-one bosons,
\bqa
{\cal F}_{\rm ideal} =
(N_c^2 - 1) \left( - {\pi^2 \over 45} T^4 \right) \;,
\eqa
%
and $\hat m_D$ and $\hat \mu$ are dimensionless variables:
\bqa
\hat m_D &=& {m_D \over 2 \pi T}  \;,
\\
\hat \mu &=& {\mu \over 2 \pi T}  \;.
\eqa
%
Adding the counterterm (\ref{delE-0}),
we obtain the thermodynamic potential at leading order in the
delta expansion:
\bqa
\Omega_{\rm LO} &=& {\cal F}_{\rm ideal}
\left\{ 1 - {15 \over 2} \hat m_D^2
+ 30 \hat m_D^3
+ {45 \over 4}
\left( \log {\hat \mu \over 2}
	- {7 \over 2} + \gamma + {\pi^2\over 3} \right)
	\hat m_D^4  \right\} \;.
\label{Omega-LO}
\eqa
%

The coefficient of $\hat m_D^4$ in  (\ref{Omega-LO})
differs from the result calculated previously in Ref.~\cite{ABS-99}.
In that paper, the constant under the logarithm of $\hat \mu /2$
was $-{3\over2} + \gamma + \log 2$ instead of
$-{7 \over 2} + \gamma + {1\over 3}\pi^2$.
The reason for the difference is that the sum-integral ${\cal F}_g$
was calculated
in Ref.~\cite{ABS-99} using dimensional regularization to regularize
the integral, but using the 3-dimensional expressions
for the HTL propagators $\Delta_T$ and $\Delta_L$.
At leading order, the difference can be absorbed into the
definition of the scale $\mu$.
For calculations beyond leading order, it is essential
for consistency to use the $d$-dimensional expressions for these
propagators.\footnote{We thank E.~Iancu and A.~Rebhan for first bringing this problem
			to our attention.}

\subsection{Next-to-leading order}

The complete expression for the next-to-leading order correction
to the thermodynamic potential
is the sum of the contributions from the 2-loop diagrams,
the HTL counterterms, and renormalization counterterms.
The contributions from the 2-loop diagrams,
including all terms through order $g^5$, is the sum of
(\ref{F2loop-hh}), (\ref{F2loop-hs}), and (\ref{F2loop-ss}):
\bqa
\Omega_{2-{\rm loop}} &=& {\cal F}_{\rm ideal} \, {N_c \alpha_s \over 3 \pi}
\Bigg\{ - {15\over4} + 45 \, \hat m_D
 - {165 \over 8}
\left[ {1\over\epsilon} + 4 \log{\hat \mu \over 2}
			- {72\over 11} \log{\hat m_D}
			+ 1.969 \right] \hat m_D^2
\nonumber
\\
&& \hspace{3cm}
+ {495\over 4}
\left[ {1\over\epsilon} + 4 \log{\hat \mu \over 2} - 2 \log{\hat m_D}
	+ {27\over 11} + 2 \gamma \right] \hat m_D^3
\Bigg\} \;.
\label{Omega-2loop}
\eqa
%
The HTL counterterm contribution is the sum of
(\ref{Fct-h}) and (\ref{Fct-s}):
\bqa
\Omega_{\rm HTL} &=& {\cal F}_{\rm ideal}
\left\{ {15 \over 2} \, \hat m_D^2
- 45  \, \hat m_D^3
- {45 \over 4}
\left( {1 \over \epsilon} + 2 \log {\hat \mu \over 2}
	- 7 + 2 \gamma + {2 \pi^2\over 3} \right)
	\hat m_D^4  \right\} \;,
\label{Omega-HTL}
\eqa
%
The ultraviolet divergences that remain after these 3 terms are added
can be removed by renormalization of the vacuum energy density ${\cal E}_0$
and the HTL mass parameter $m_D$.
The renormalization contributions at first order in $\delta$ are
\bqa
\Delta \Omega \;=\; \Delta_1{\cal E}_0
+ \Delta_1m_D^2 {\partial \ \ \over \partial m_D^2} \Omega_{\rm LO} \;,
\label{Omega-renorm}
\eqa
%
where $\Delta_1{\cal E}_0$ and $\Delta_1m_D^2$ are the terms
of order $\delta$ in the vacuum energy counterterm
and the mass counterterm.
The expression for $\Delta_1{\cal E}_0$ is given in (\ref{DelE0:NLO}).
It cancels the poles in $\epsilon$ proportional to $m_D^4$
in (\ref{Omega-1loop}) and (\ref{Omega-HTL}).
The remaining ultraviolet divergences are poles in $\epsilon$
proportional to $m_D^2$ and $m_D^3$ in (\ref{Omega-2loop}).
If HTL perturbation theory (HTLpt) is renormalizable,
both divergences must be removed by the same mass counterterm.
This requires a remarkable coincidence between the coefficients
of the two poles, and provides a nontrivial test of renormalizability.
The value of the counterterm required is
\bqa
\Delta_1m_D^2 \;=\;
- {11 \over 4\epsilon} \, {N_c \alpha_s \over 3 \pi}
	\, m_D^2 \;.
\eqa
%
The complete contribution from the counterterms
through first order in $\delta$ is
\bqa
\Delta\Omega &=& {\cal F}_{\rm ideal} \,
\Bigg\{ {45 \over 4 \epsilon} \, \hat m_D^4
+ {165 \over 8}
\left[ {1\over\epsilon} + 2 \log{\hat \mu \over 2}
			+ 2 + 2 {\zeta'(-1) \over \zeta(-1)} \right]
	{N_c \alpha_s \over 3 \pi} \, \hat m_D^2
\nonumber
\\
&& \hspace{2cm}
- {495\over 4}
\left[ {1\over\epsilon} + 2\log{\hat \mu \over 2} - 2 \log{\hat m_D}
	+ 2 \right]
	{N_c \alpha_s \over 3 \pi} \, \hat m_D^3
\Bigg\} \;.
\label{Omega-ren}
\eqa
%

Adding the contributions from the
two-loop diagrams in (\ref{Omega-2loop}),
the HTL counterterm in (\ref{Omega-HTL}),
and the renormalization counterterms in (\ref{Omega-renorm})
and adding them to the leading order thermodynamic potential
in (\ref{Omega-LO}),
we obtain the complete expression for the thermodynamic potential
at next-to-leading order in HTLpt:
\bqa
\Omega_{\rm NLO} &=& {\cal F}_{\rm ideal}
\Bigg\{ 1 - 15 \hat m_D^3
- {45 \over 4} \left( \log{\hat \mu \over 2}
		- {7\over 2} + \gamma + {\pi^2\over3} \right)
		\hat m_D^4
\nonumber
\\
&&  \hspace{1.5cm}
+ {N_c \alpha_s \over 3 \pi}
\Bigg[ - {15 \over 4} + 45 \hat m_D
 - {165 \over 4}
\left( \log{\hat \mu \over 2 }
	- {36 \over 11} \log \hat m_D - 2.001 \right) \hat m_D^2
\nonumber
\\
&& \hspace{3.5cm}
+ {495\over 2}
\left( \log{\hat \mu \over 2} + {5\over22} + \gamma \right) \hat m_D^3 \Bigg]
\Bigg\} \;.
\label{Omega-NLO}
\eqa
%

\subsection{Gap Equation}

\begin{figure}[htb]
\hspace{1cm}
\epsfysize=9cm
\centerline{\epsffile{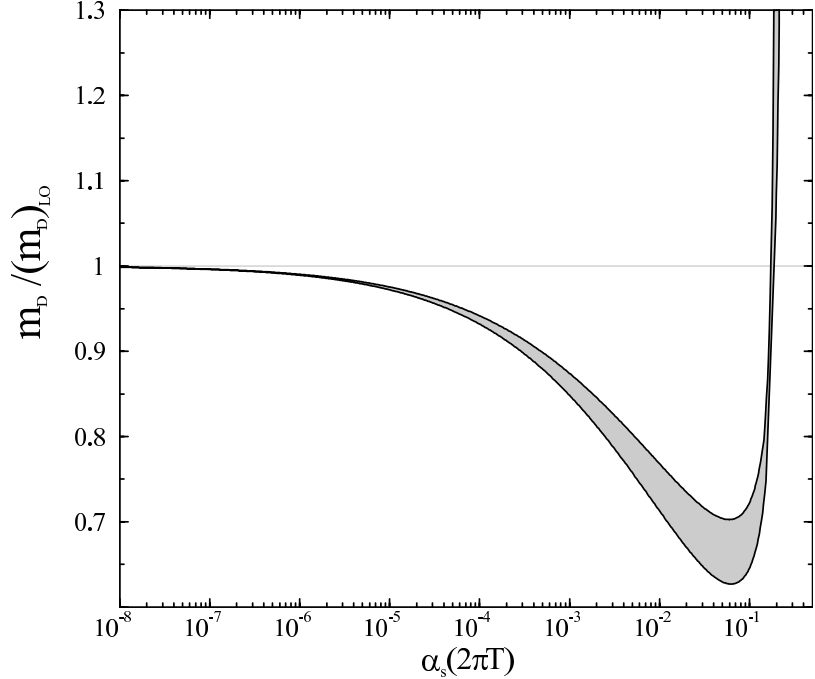}}
\\
\caption[a]{Solution to the gap equation (\ref{gap-NLO}) as a function of 
$\alpha_s(2 \pi T)$.  The shaded band corresponds to variation
of the renormalization scale $\mu$ by a factor of two around $\mu=2\pi T$.
}
\label{gapfig}
\end{figure}

The gap equation which determines $m_D$ is obtained by
differentiating (\ref{Omega-NLO}) with respect to $m_D$ and setting this
derivative equal to zero yielding:
\bqa
\hat m_D^2 \left[ 1
	+ \left( \log{\hat \mu \over 2}
		- {7\over 2} + \gamma + {\pi^2\over3} \right)
		\hat m_D \right]
&=&
{N_c \alpha_s \over 3 \pi}
\Bigg[ 1 - {11 \over 6}
\left( \log{\hat \mu \over 2 }
	- {36 \over 11} \log \hat m_D - 3.637 \right) \hat m_D
\nonumber
\\
&& \hspace{2cm}
\;+\; {33\over 2}
\left( \log{\hat \mu \over 2} + {5\over22} + \gamma \right) \hat m_D^2 \Bigg] \;.
\label{gap-NLO}
\eqa
%
In Fig.~\ref{gapfig}, we have plotted the solution to this gap equation
normalized to the leading-order perturbative result in (\ref{mD-wcl})
as a function of $\alpha_s(2 \pi T)$.  The shaded band indicates the range
resulting from varying the renormalization scale $\mu$ by a factor of two
around $\mu=2\pi T$.  From this plot, we see that the gap equation solution
matches nicely onto the perturbative result as $\alpha_s \rightarrow 0$.
The solution decreases with $\alpha_s(2 \pi T)$ out to about $\alpha_s \approx 0.06$
and then begins to increase.  It exceeds the perturbative result at
around $\alpha_s \approx 0.18$,
and then quickly diverges to $+\infty$.

\section{Thermodynamic functions}
\label{thermodynamics}

In this section, we compare the thermodynamic functions
calculated at next-to-leading order in HTL perturbation theory (HTLpt)
with those calculated using lattice gauge theory.

\subsection{Pressure}

\begin{figure}[htb]
\hspace{1cm}
\epsfysize=9cm
\centerline{\epsffile{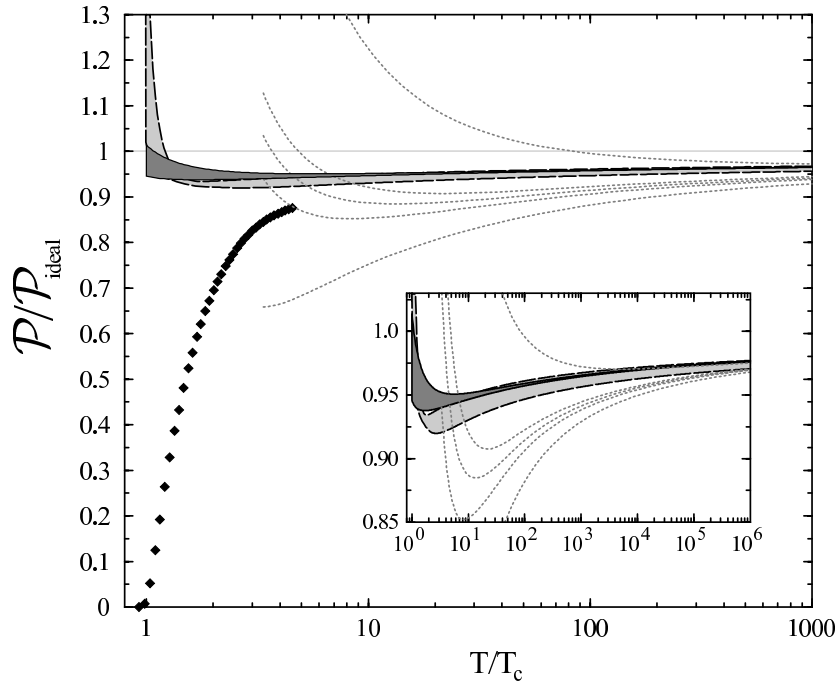}}
\\
\caption[a]{
The LO and NLO results for the pressure in HTLpt
compared with 4-d lattice results (diamonds)
and 3-d lattice results (dotted lines) for various values of an unknown
coefficient in the 3-d effective Lagrangian.
The LO HTLpt result is shown as a light-shaded band outlined by a dashed line.
The NLO HTLpt result is shown as a dark-shaded band outlined by a solid line.
The shaded bands correspond to variations
of the renormalization scale $\mu$ by a factor of two around $\mu=2\pi T$.
}
\label{NLOfig}
\end{figure}

The final results for the LO and NLO HTLpt predictions for the free energy of 
pure-glue QCD are obtained by evaluating the thermodynamic potentials (\ref{Omega-LO}) and
(\ref{Omega-NLO}) at the solution to the gap equation (\ref{gap-NLO}).  
Once the free energy ${\cal F}(T)$ is given as a function of $T$, all
other thermodynamic functions are determined. In particular, the pressure
${\cal P}$ and the energy density ${\cal E}$ are
\bqa
{\cal P}&=&-{\cal F},\\
{\cal E}&=&{\cal F}-T{d{\cal F}\over d T}.
\eqa
In Figure \ref{NLOfig},
we have plotted the LO and NLO HTLpt predictions for
the pressure of pure-glue QCD
as a function of $T/T_c$, where $T_c$ is the deconfinement transition
temperature.  To translate $\alpha_s(2 \pi T)$ into a value of $T/T_c$,
we use the two-loop running formula for pure-glue QCD with
$\Lambda_{\rm \overline{MS}} = 0.65\,T_c$.
Thus $\alpha_s(2 \pi T)= 0.06$ and 0.2 translate into
$T/T_c = 415$ and 0.906, respectively.
The LO and NLO HTLpt results are shown in Fig.~\ref{NLOfig} as
a light-shaded band outlined by a dashed line and
a dark-shaded band outlined by a solid line, respectively.
The LO and NLO bands overlap all the way down to $T=T_c$,
and the bands are very narrow compared to the corresponding bands
for the weak-coupling predictions in Fig.~\ref{weakfig}.
Thus the convergence of HTLpt
seems to be dramatically improved over naive perturbation theory
and the final result is extremely insensitive to the scale $\mu$.

In Fig.~\ref{NLOfig},
we have also included the 4-dimensional lattice gauge theory
results of Boyd et al. \cite{lattice-0} and the 3-dimensional
lattice gauge theory results of Kajantie et al. \cite{KLRS}.
The LO and NLO HTLpt predictions differ significantly from the
4-d lattice results of Ref.~\cite{lattice-0},
even at the highest temperatures for which they are available.
At $T=5\,T_c$, the HTLpt prediction for the deviation of the pressure
from that of the ideal gas is only 45\%
of the 4-d lattice result.
In the high temperature limit, the HTLpt prediction approaches that
of the ideal gas very slowly, in qualitative agreement with the results of
the 3-d lattice calculations of Ref.~\cite{KLRS}.
However the quantitative agreement is not very good.
The results of  Ref.~\cite{KLRS} depend on an unknown coefficient
in the effective lagrangian for the dimensionally reduced theory.
The 5 dotted lines in Fig.~\ref{NLOfig} correspond to 5 possible
values for that coefficient.  We assume that the coefficient
is such that the 3-d results match on reasonably well to the 4-d results,
such as one of the middle 3 of the 5 dotted lines.
In that case, the HTLpt prediction for the deviation from the ideal gas
at $T=10^3\,T_c$ is only about 59\%
of the 3-d lattice result.
We conclude that HTLpt at this order does not describe the pressure
for pure-glue QCD.


\subsection{Trace Anomaly}

The combination ${\cal E}-3{\cal P}$
can be written as
\bqa
\label{em3p}
{\cal E}-3{\cal P}=-T^5{d \ \over d T}\left({{\cal F}\over T^4}\right).
\eqa
This combination is  proportional to the trace of the energy-momentum
tensor. In QCD with massless quarks, it is nonzero only because scale
invariance is broken by renormalization effects. We will call it the trace
anomaly density. It of course vanishes for an ideal gas of massless particles.
However, it also vanishes for a gas of quasiparticles whose masses are
linear in $T$ and whose interactions are governed by a dimensionless
coupling constant that is independent of $T$.

\begin{figure}[htb]
\hspace{1cm}
\epsfysize=9cm
\centerline{\epsffile{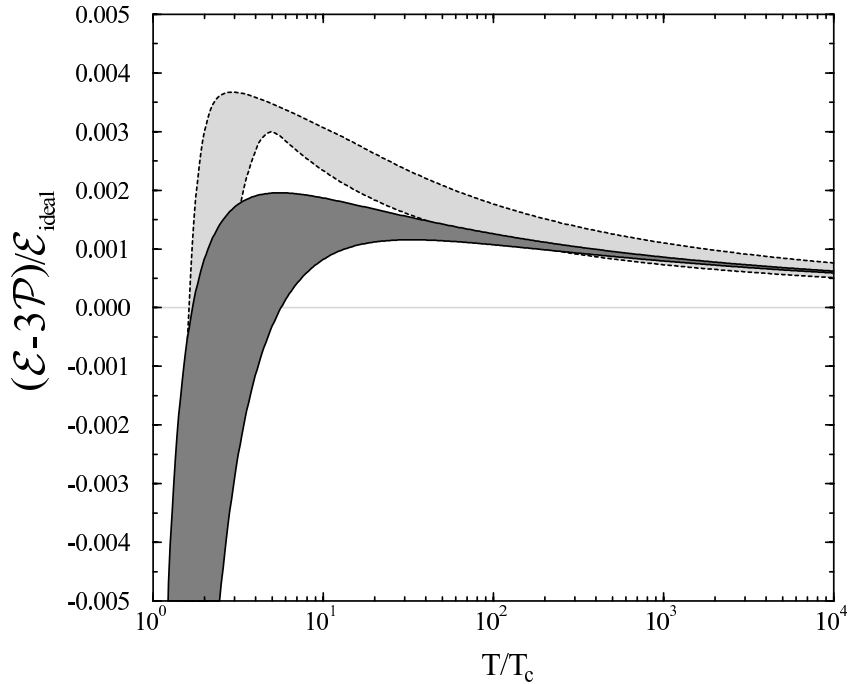}}
\\
\caption[a]{
The LO and NLO results for the trace anomaly in HTLpt.
The LO HTLpt result is shown as a light-shaded band outlined by a dashed line.
The NLO HTLpt result is shown as a dark-shaded band outlined by a solid line.
The shaded bands correspond to variations
of the renormalization scale $\mu$ by a factor of two around $\mu=2\pi T$.
}
\label{tracefig}
\end{figure}

In Fig.~\ref{tracefig}, we have plotted the LO and NLO HTLpt predictions for
the trace anomaly density as a function of $T/T_c$.
At large $T$, the HTLpt prediction is very small and positive.
As $T$ decreases, the NLO prediction for ${\cal E}-3{\cal P}$ increases
to its maximum value around $10 T_c$ and then begins decreasing
and quickly turns negative.
The maximum value is less than about 0.2\% of the energy density
${\cal E}_{\rm ideal}$ of the ideal gas.
In contrast, the 4-d lattice result increases to a
maximum of about 70\% of ${\cal E}_{\rm ideal}$
at a temperature that is very close to $T_c$ and then
decreases rapidly to 0 \cite{lattice-0}.

\section{Conclusions}

We have calculated the free energy of pure-glue QCD at high temperature to
2-loop order using HTL perturbation theory (HTLpt).
The gauge invariance of the 2-loop expression was verified explicitly in
generalized covariant gauge and generalized Coulomb gauge.
The expression was reduced to a relatively compact form involving only
scalar sum-integrals.
The numerical evaluation of the scalar sum-integrals would have been
extremely difficult.
We chose instead to approximate them by expanding in powers of $m_D/T$,
keeping all terms through 5$^{\rm th}$ order in $g$ and $m_D/T$.
The ultraviolet divergences in the resulting expression for the
thermodynamic potential can be removed by renormalization
of the vacuum energy density and the HTL mass parameter $m_D$.
This provides a nontrivial test of the renormalizability of HTL
perturbation theory to this order.

The two-loop order of HTLpt is the first order
at which $m_D$ can be determined by a variational prescription.
The condition that $m_D$ be a stationary point of the thermodynamic
potential provides a ``gap equation'' for $m_D$.
The only ambiguity in the free energy then resides in the scale
$\mu$ associated with renormalizations of the vacuum energy density
and $m_D$.  The predictions for the thermodynamic functions are extremely
insensitive to the choice of $\mu$.

The quantitative predictions for the pressure in 2-loop HTLpt
are disappointing.  In the range $2 T_c < T < 20 T_c$,
the pressure is predicted to be nearly constant with a value of about
95\% of that of an ideal gas of gluons.
The HTLpt prediction for the deviation from the ideal gas is
about 45\%
of the result from 4-dimensional lattice gauge theory at $T= 5 T_c$,
the highest temperature for which the lattice result is available.
At very high temperature, the approach to the ideal gas
limit is extremely slow,
in qualitative agreement with the results of 3-d lattice
gauge theory calculations.
However, assuming that the 3-d results match on reasonably well
to the 4-d results, the HTLpt prediction for the deviation from the ideal gas
at $T=10^3\,T_c$ is only about 59\%
of the 3-d lattice result.

A possible conclusion is that HTLpt at two-loop order is simply not a useful
approximation for thermal QCD.  Another possibility is that the
problem lies not with HTLpt but with our use of the $m_D/T$ expansion
to approximate the scalar sum-integrals.
The sum-integrals that were encountered at 4$^{\rm th}$ and 5$^{\rm th}$
order in $m_D/T$ were so difficult to evaluate that it seems hopeless
to try to expand to higher order. However it is possible that the
scalar sum-integrals could be evaluated numerically.
Part of the difficulty is that it is necessary to isolate the
infrared divergent and ultraviolet divergent terms analytically
before evaluating the remaining terms numerically.
Our $m_D/T$ expansions of the sum-integrals might be useful for
generating the necessary subtractions that would allow the
scalar sum-integrals to be evaluated numerically.

Our calculations required the development of new methods
for evaluating sum-integrals.  The most difficult were two-loop sum-integrals 
that also involved an HTL angular average.  These sum-integrals 
may be useful in other applications, such as solving the 2-loop 
$\Phi$-derivable approximation for QCD.

\section{Acknowledgements}
E.B. and E.P. were supported in part by Department of Energy grant
DE-FG02-91-ER4069.  J.O.A. was supported by the 
Stichting voor Fundamenteel Onderzoek der Materie (FOM), which is 
supported by the Nederlandse Organisatie voor Wetenschapplijk 
Onderzoek (NWO).  M.S. was supported by US DOE Grants DE-FG02-96ER40945
and DE-FG03-97ER41014.  

\newpage

\appendix
\renewcommand{\theequation}{\thesection.\arabic{equation}}

\section{HTL Feynman Rules}
\label{app:rules}

In this appendix, we present Feynman rules for HTL perturbation theory
in pure-glue QCD.  We give explicit expressions for the propagators
and for the 3-particle and 4-particle vertices.
The Feynman rules are given in Minkowski space to facilitate
applications to real-time processes.
A Minkowski momentum is denoted $p = (p_0, {\bf p})$,
and the inner product is $p \cdot q = p_0 q_0 - {\bf p} \cdot {\bf q}$.
The vector that specifies the thermal rest frame
is $n = (1,{\bf 0})$.

\subsection{Gluon Self-energy}

The HTL gluon self-energy tensor for a gluon of momentum $p$ is
\bqa
\label{a1}
\Pi^{\mu\nu}(p)=m_D^2\left[
{\cal T}^{\mu\nu}(p,-p)-n^{\mu}n^{\nu}
\right]\;.
\eqa
%
The tensor ${\cal T}^{\mu\nu}(p,q)$, which is defined only for momenta
that satisfy $p+q=0$, is
\bqa
{\cal T}^{\mu\nu}(p,-p)=
\left \langle y^{\mu}y^{\nu}{p\!\cdot\!n\over p\!\cdot\!y}
\right\rangle_{\bf\hat{y}} \;.
\label{T2-def}
\eqa
%
The angular brackets indicate averaging
over the spatial directions of the light-like vector $y=(1,\hat{\bf y})$.
The tensor ${\cal T}^{\mu\nu}$ is symmetric in $\mu$ and $\nu$
and satisfies the ``Ward identity''
\bqa
p_{\mu}{\cal T}^{\mu\nu}(p,-p)=
p\!\cdot\!n\;n^{\nu}\;.
\label{ward-t2}
\eqa
%
The self-energy tensor $\Pi^{\mu\nu}$ is therefore also
symmetric in $\mu$ and $\nu$ and satisfies
\bqa
p_{\mu}\Pi^{\mu\nu}(p)&=&0\;,\\
\label{contr}
g_{\mu\nu}\Pi^{\mu\nu}(p)&=&-m_D^2\;.
\eqa
%

The gluon self-energy tensor can be expressed in terms of two scalar functions,
the transverse and longitudinal self-energies $\Pi_T$ and $\Pi_L$,
defined by
\bqa
\label{pit2}
\Pi_T(p)&=&{1\over d-1}\left(
\delta^{ij}-\hat{p}^i\hat{p}^j
\right)\Pi^{ij}(p)\;, \\
\label{pil2}
\Pi_L(p)&=&-\Pi^{00}(p)\;,
\eqa
%
where ${\bf \hat p}$ is the unit vector
in the direction of ${\bf p}$.
In terms of these functions, the self-energy tensor is
\bqa
\label{pi-def}
\Pi^{\mu\nu}(p) \;=\; - \Pi_T(p) T_p^{\mu\nu}
- {1\over n_p^2} \Pi_L(p) L_p^{\mu\nu}\;,
\eqa
%
where the tensors $T_p$ and $L_p$ are
\bqa
T_p^{\mu\nu}&=&g^{\mu\nu} - {p^{\mu}p^{\nu} \over p^2}
-{n_p^{\mu}n_p^{\nu}\over n_p^2}\;,\\
L_p^{\mu\nu}&=&{n_p^{\mu}n_p^{\nu} \over n_p^2}\;.
\eqa
%
The four-vector $n_p^{\mu}$ is
\bqa
n_p^{\mu} \;=\; n^{\mu} - {n\!\cdot\!p\over p^2} p^{\mu}
\eqa
%
and satisfies $p\!\cdot\!n_p=0$ and $n^2_p = 1 - (n\!\cdot\!p)^2/p^2$.
The equation~(\ref{contr}) reduces to the identity
\bqa
(d-1)\Pi_T(p)+{1\over n^2_p}\Pi_L(p) \;=\; m_D^2 \;.
\label{PiTL-id}
\eqa
%
We can express both self-energy functions in terms of the function
${\cal T}^{00}$ defined by (\ref{T2-def}):
\bqa
\Pi_T(p)&=& {m_D^2 \over (d-1) n_p^2}
\left[ {\cal T}^{00}(p,-p) - 1 + n_p^2  \right] \;,
\label{PiT-T}
\\
\Pi_L(p)&=& m_D^2
\left[ 1- {\cal T}^{00}(p,-p) \right]\;,
\label{PiT-L}
\eqa
%

In the tensor ${\cal T}^{\mu \nu}(p,-p)$ defined in~(\ref{T2-def}),
the angular brackets indicate the angular average over
the unit vector $\hat{\bf y}$.
In almost all previous work, the angular average in~(\ref{T2-def}) has been
taken in $d=3$ dimensions. For consistency of higher order radiative
corrections, it is essential to take the angular average in $d=3-2\epsilon$
dimensions and analytically continue to $d=3$ only after all poles in
$\epsilon$ have been cancelled.
Expressing the angular average as an integral over the cosine of an angle,
the expression for the $00$ component of the tensor is
\bqa
{\cal T}^{00}(p,-p) &=& {w(\epsilon)\over2}
\int_{-1}^1dc\;(1-c^2)^{-\epsilon}{p_0\over p_0-|{\bf p}|c} \;,
\label{T00-int}
\eqa
%
where the weight function $w(\epsilon)$ is
\bqa
w(\epsilon)={\Gamma(2-2\epsilon)\over\Gamma^2(1-\epsilon)}2^{2\epsilon}
= {\Gamma({3\over2}-\epsilon)
	\over \Gamma({3\over2}) \Gamma(1-\epsilon)} \;.
\label{weight}
\eqa
%
The integral in (\ref{T00-int}) must be defined so that it is analytic
at $p_0=\infty$.
It then has a branch cut running from $p_0=-|{\bf p}|$ to $p_0=+|{\bf p}|$.
If we take the limit $\epsilon\rightarrow 0$, it reduces to
\begin{eqnarray}
{\cal T}^{00}(p,-p) &=&
{p_0 \over 2|{\bf p}|}
		\log {p_0 +|{\bf p}| \over p_0-|{\bf p}|}\;,
\end{eqnarray}
%
which is the expression that
appears in the usual HTL self-energy functions.

\subsection{Gluon Propagator}
\label{app:prop}

The Feynman rule for the gluon propagator is
\bqa
i \delta^{a b} \Delta_{\mu\nu}(p) \;,
\eqa
%
where the gluon propagator tensor $\Delta_{\mu\nu}$
depends on the choice of gauge fixing.
We consider two possibilities that introduce an arbitrary
gauge parameter $\xi$:  general covariant gauge and
general Coulomb gauge.
In both cases, the inverse propagator reduces in the
limit $\xi\rightarrow\infty$ to
\bqa
\Delta^{-1}_{\infty}(p)^{\mu\nu}&=&
-p^2 g^{\mu \nu} + p^\mu p^\nu - \Pi^{\mu\nu}(p)\;.
\label{delta-inv:inf0}
\eqa
%
This can also be written
\bqa
\Delta^{-1}_{\infty}(p)^{\mu\nu}&=&
- {1 \over \Delta_T(p)}       T_p^{\mu\nu}
+ {1 \over n_p^2 \Delta_L(p)} L_p^{\mu\nu}\;,
\label{delta-inv:inf}
\eqa
%
where $\Delta_T$ and $\Delta_L$ are the transverse and longitudinal
propagators:
\bqa
\Delta_T(p)&=&{1 \over p^2-\Pi_T(p)}\;,
\label{Delta-T:M}
\\
\Delta_L(p)&=&{1 \over - n_p^2 p^2+\Pi_L(p)}\;.
\label{Delta-L:M}
\eqa
%
The inverse propagator for general $\xi$ is
\bqa
\Delta^{-1}(p)^{\mu\nu}&=&
\Delta^{-1}_{\infty}(p)^{\mu\nu}-{1\over\xi}
p^{\mu}p^{\nu}\hspace{5.3cm}\mbox{covariant}\;,
\label{Delinv:cov}
\\
&=&\Delta^{-1}_{\infty}(p)^{\mu\nu}-{1\over\xi}
\left(p^{\mu}-p\!\cdot\!n\;n^{\mu}\right)
\left(p^{\nu}-p\!\cdot\!n\;n^{\nu}\right)
\hspace{1cm}\mbox{Coulomb}\;.
\label{Delinv:C}
\eqa
%
The propagators obtained by inverting the tensors in (\ref{Delinv:C})
and  (\ref{Delinv:cov}) are
\bqa
\Delta^{\mu\nu}(p)&=&-\Delta_T(p)T_p^{\mu\nu}
+\Delta_L(p)n_p^{\mu}n_p^{\nu}
- \xi {p^{\mu}p^{\nu} \over (p^2)^2}
\hspace{1.5cm}\mbox{covariant}\;,
\label{D-cov}
\\
&=&-\Delta_T(p)T_p^{\mu\nu}
+\Delta_L(p)n^{\mu}n^{\nu}-\xi{p^{\mu}p^{\nu}\over\left(n_p^2p^2\right)^2}
\hspace{1cm}\mbox{Coulomb}\;.
\label{D-C}
\eqa
%

It is convenient to define the following combination of propagators:
\bqa
\Delta_X(p) &=& \Delta_L(p)+{1\over n_p^2}\Delta_T(p) \;.
\label{Delta-X}
\eqa
%
Using (\ref{PiTL-id}), (\ref{Delta-T:M}), and (\ref{Delta-L:M}),
it can be expressed in the alternative form
\bqa
\Delta_X(p) &=&
\left[ m_D^2 - d \, \Pi_T(p) \right] \Delta_L(p) \Delta_T(p) \;,
\label{Delta-X:2}
\eqa
%
which shows that it vanishes in the limit $m_D \to 0$.
In the covariant gauge, the propagator tensor can be written
\bqa
\Delta^{\mu\nu}(p) &=&
\left[ - \Delta_T(p) g^{\mu \nu} + \Delta_X(p) n^\mu n^\nu \right]
- {n \!\cdot\! p \over p^2} \Delta_X(p)
	\left( p^\mu n^\nu  + n^\mu p^\nu \right)
\nonumber
\\
&&
+ \left[ \Delta_T(p) + {(n \!\cdot\! p)^2 \over p^2} \Delta_X(p)
	- {\xi \over p^2} \right] {p^\mu p^\nu \over p^2} \;.
\label{gprop-TC}
\eqa
%
This decomposition of the propagator into three terms
has proved to be particularly convenient for explicit calculations.
For example, the first term satisfies the identity
\bqa
\left[- \Delta_T(p) g_{\mu \nu} + \Delta_X(p) n_\mu n_\nu \right]
\Delta^{-1}_{\infty}(p)^{\nu\lambda}  &=&
{g_\mu}^\lambda - {p_\mu p^\lambda \over p^2}
+ {n \!\cdot\! p \over n_p^2 p^2} {\Delta_X(p) \over \Delta_L(p)}
	p_\mu n_p^\lambda \;.
\label{propid:2}
\eqa
%

\subsection{Three-gluon Vertex}
\label{app:3gluon}

The three-gluon vertex
for gluons with outgoing momenta $p$, $q$, and $r$,
Lorentz indices $\mu$, $\nu$, and $\lambda$,
and color indices $a$, $b$, and $c$ is
\bqa
i\Gamma_{abc}^{\mu\nu\lambda}(p,q,r)=-gf_{abc}
\Gamma^{\mu\nu\lambda}(p,q,r)\;,
\eqa
%
where the three-gluon vertex tensor is
\bqa
\Gamma^{\mu\nu\lambda}(p,q,r)=
g^{\mu\nu}(p-q)^{\lambda}+
g^{\nu\lambda}(q-r)^{\mu}+
g^{\lambda\mu}(r-p)^{\nu}
-m_D^2{\cal T}^{\mu\nu\lambda}(p,q,r)\;.
\label{Gam3}
\eqa
%
The tensor ${\cal T}^{\mu\nu\lambda}$ in the HTL correction term
is defined only for $p+q+r=0$:
\bqa
{\cal T}^{\mu\nu\lambda}(p,q,r) \;=\;
 - \Bigg\langle y^{\mu} y^{\nu} y^{\lambda}
\left( {p\!\cdot\!n\over p\!\cdot\!y\;q\!\cdot\!y}
	- {r\!\cdot\!n\over\!r\cdot\!y\;q\!\cdot\!y} \right)
	\Bigg\rangle\;.
\label{T3-def}
\eqa
%
This tensor is totally symmetric in its three indices and traceless in any
pair of indices: $g_{\mu\nu}{\cal T}^{\mu\nu\lambda}=0$.
It is odd (even) under odd (even) permutations of the momenta $p$, $q$, and
$r$. It satisfies the ``Ward identity''
\bqa
q_{\mu}{\cal T}^{\mu\nu\lambda}(p,q,r) \;=\;
{\cal T}^{\nu\lambda}(p+q,r)-
{\cal T}^{\nu\lambda}(p,r+q)\;.
\label{ward-t3}
\eqa
%
The three-gluon vertex tensor therefore satisfies the Ward identity
\bqa
p_{\mu}\Gamma^{\mu\nu\lambda}(p,q,r) \;=\;
\Delta_{\infty}^{-1}(q)^{\nu\lambda}-\Delta_{\infty}^{-1}(r)^{\nu\lambda}\;.
\label{ward-3}
\eqa
%

\subsection{Four-gluon Vertex}
\label{app:4gluon}

The four-gluon vertex
for gluons with outgoing momenta $p$, $q$, $r$, and $s$,
Lorentz indices $\mu$, $\nu$, $\lambda$, and $\sigma$,
and color indices $a$, $b$, $c$, and $d$ is
\bqa
i\Gamma^{\mu\nu\lambda\sigma}_{abcd}(p,q,r,s) &=&
- ig^2\Bigg\{ f_{abx}f_{xcd} \left(g^{\mu\lambda}g^{\nu\sigma}
				-g^{\mu\sigma}g^{\nu\lambda}\right)
\nonumber
\\
&&
\hspace{.67cm}
+2m_D^2\mbox{tr}\left[T^a\left(T^bT^cT^d+T^dT^cT^b
\right)\right]{\cal T}^{\mu\nu\lambda\sigma}(p,q,r,s)
\Bigg\}
\nonumber
\\
&&+ \; 2 \; \mbox{cyclic permutations}\;,
\eqa
%
where the cyclic permutations are of
$(q,\nu,b)$, $(r,\lambda,c)$, and $(s,\sigma,d)$.
The matrices $T^a$ are the fundamental representation
of the $SU(3)$ algebra with the standard normalization
${\rm tr}(T^a T^b) = {1 \over 2} \delta^{ab}$.
The tensor ${\cal T}^{\mu\nu\lambda\sigma}$
in the HTL correction term is defined only for $p+q+r+s=0$:
\bqa
{\cal T}^{\mu\nu\lambda\sigma}(p,q,r,s) &=&
\Bigg\langle y^{\mu} y^{\nu} y^{\lambda} y^{\sigma}
\left( {p\!\cdot\!n \over p\!\cdot\!y \; q\!\cdot\!y \; (q+r)\!\cdot\!y}
\right.
\nonumber
\\
&&
\left.
+{(p+q)\!\cdot\!n\over q\!\cdot\!y\;r\!\cdot\!y\;(r+s)\!\cdot\!y}
+{(p+q+r)\!\cdot\!n\over r\!\cdot\!y\;s\!\cdot\!y\;(s+p)\!\cdot\!y}\right)
\Bigg\rangle\;.
\label{T4-def}
\eqa
%
This tensor is totally symmetric in its four indices and traceless in any
pair of indices: $g_{\mu\nu}{\cal T}^{\mu\nu\lambda\sigma}=0$.
It is even under cyclic or anti-cyclic
permutations of the momenta $p$, $q$, $r$, and $s$.
It satisfies the ``Ward identity''
\bqa
q_{\mu}{\cal T}^{\mu\nu\lambda\sigma}(p,q,r,s)=
{\cal T}^{\nu\lambda\sigma}(p+q,r,s)
-{\cal T}^{\nu\lambda\sigma}(p,r+q,s)
\label{ward-t4}
\eqa
%
and the ``Bianchi identity''
\bqa
{\cal T}^{\mu\nu\lambda\sigma}(p,q,r,s)
+ {\cal T}^{\mu\nu\lambda\sigma}(p,r,s,q)+
{\cal T}^{\mu\nu\lambda\sigma}(p,s,q,r)=0\;.
\label{Bianchi}
\eqa
%

When its color indices are traced in pairs, the four-gluon vertex becomes
particularly simple:
\bqa
\delta^{ab} \delta^{cd} i \Gamma_{abcd}^{\mu\nu\lambda\sigma}(p,q,r,s)
= -i g^2 N_c (N_c^2-1) \Gamma^{\mu\nu,\lambda\sigma}(p,q,r,s) \;,
\eqa
%
where the color-traced four-gluon vertex tensor is
\bqa
\Gamma^{\mu\nu,\lambda\sigma}(p,q,r,s)=
2g^{\mu\nu}g^{\lambda\sigma}
-g^{\mu\lambda}g^{\nu\sigma}
-g^{\mu\sigma}g^{\nu\lambda}
-m_D^2{\cal T}^{\mu\nu\lambda\sigma}(p,s,q,r)\;.
\label{Gam4}
\eqa
%
Note the ordering of the momenta in the arguments of the tensor
${\cal T}^{\mu\nu\lambda\sigma}$, which comes from the use of the
Bianchi identity (\ref{Bianchi}).
The tensor (\ref{Gam4}) is symmetric
under the interchange of $\mu$ and $\nu$,
under the interchange of $\lambda$ and $\sigma$,
and under the interchange of $(\mu,\nu)$ and $(\lambda,\sigma)$.
It is also symmetric under the interchange of $p$ and $q$,
under the interchange of $r$ and $s$,
and under the interchange of $(p,q)$ and $(r,s)$.
It satisfies the Ward identity
\bqa
p_{\mu}\Gamma^{\mu\nu,\lambda\sigma}(p,q,r,s)
=\Gamma^{\nu\lambda\sigma}(q,r+p,s)
-\Gamma^{\nu\lambda\sigma}(q,r,s+p)\;.
\label{ward-4}
\eqa
%

\subsection{Ghost propagator and vertex}
\label{app:ghost}

The ghost propagator and the ghost-gluon vertex depend on the gauge.
The Feynman rule for the ghost propagator is
\bqa
&&{i\over p^2}\delta^{ab}      \hspace{2.5cm}\mbox{covariant}\;,
\\
&&{i\over n_p^2 p^2}\delta^{ab} \hspace{2cm}\mbox{Coulomb}\;.
\eqa
%
The Feynman rule for the vertex in which a gluon with
indices $\mu$ and $a$ interacts with an outgoing ghost
with outgoing momentum $r$ and color index $c$ is
\bqa
&&-gf^{abc}r^{\mu} \hspace{3.3cm}\mbox{covariant}\;,
\\
&&-gf^{abc}\left(r^{\mu}-r\!\cdot\!n\;n^{\mu}\right)
\hspace{1cm}\mbox{Coulomb}\;.
\eqa
%
Every closed ghost loop requires a multiplicative factor of $-1$.

\subsection{HTL Counterterm}
\label{app:HTLct}

The Feynman rule for the insertion of an HTL counterterm into a gluon
propagator is
\bqa
-i\delta^{ab}\Pi^{\mu\nu}(p)\;,
\eqa
%
where $\Pi^{\mu\nu}(p)$ is the HTL gluon self-energy tensor given
in~(\ref{pi-def}).

\subsection{Imaginary-time formalism}
\label{app:ITF}

In the imaginary-time formalism,
Minkoswski energies have discrete imaginary values
$p_0 = i (2 \pi n T)$
and integrals over Minkowski space are replaced by sum-integrals over
Euclidean vectors $(2 \pi n T, {\bf p})$.
We will use the notation $P=(P_0,{\bf p})$ for Euclidean momenta.
The magnitude of the spatial momentum will be denoted $p = |{\bf p}|$,
and should not be confused with a Minkowski vector.
The inner product of two Euclidean vectors is
$P \cdot Q = P_0 Q_0 + {\bf p} \cdot {\bf q}$.
The vector that specifies the thermal rest frame
remains $n = (1,{\bf 0})$.

The Feynman rules for Minkowski space given above can be easily
adapted to Euclidean space.  The Euclidean tensor in a given
Feynman rule is obtained from the corresponding Minkowski tensor
with raised indices by replacing each Minkowski energy $p_0$
by $iP_0$, where $P_0$ is the corresponding Euclidean energy,
and multipying by $-i$ for every $0$ index.
This prescription transforms $p=(p_0,{\bf p})$ into $P=(P_0,{\bf p})$,
$g^{\mu \nu}$ into $- \delta^{\mu \nu}$,
and $p\!\cdot\!q$ into $-P\!\cdot\!Q$.
The effect on the HTL tensors defined in (\ref{T2-def}),
(\ref{T3-def}), and (\ref{T4-def}) is equivalent to
substituting $p\!\cdot\!n \to - P\!\cdot\!N$ where $N = (-i,{\bf 0})$,
$p\!\cdot\!y \to -P\!\cdot\!Y$ where $Y = (-i,{\bf \hat y})$,
and $y^\mu \to Y^\mu$.
For example, the Euclidean tensor corresponding to (\ref{T2-def}) is
\bqa
{\cal T}^{\mu\nu}(P,-P)=
\left \langle Y^{\mu}Y^{\nu}{P\!\cdot\!N \over P\!\cdot\!Y}
\right\rangle \;.
\label{T2E-def}
\eqa
%
The average is taken over the directions of the unit vector ${\bf \hat y}$.

Alternatively, one can calculate a diagram
by using the Feynman rules for Minkowski momenta,
reducing the expressions for diagrams to scalars,
and then make the appropriate substitutions,
such as $p^2 \to -P^2$, $p \cdot q \to - P \cdot Q$,
and $n \cdot p \to i n \cdot P$.
For example, the propagator functions (\ref{Delta-T:M})
and (\ref{Delta-L:M}) become
\bqa
\Delta_T(P)&=&{-1 \over P^2 + \Pi_T(P)}\;,
\label{Delta-T}
\\
\Delta_L(P)&=&{1 \over p^2+\Pi_L(P)}\;.
\label{Delta-L}
\eqa
%
The expressions for the HTL self-energy functions $\Pi_T(P)$
and $\Pi_L(P)$ are given by
(\ref{PiT-T}) and (\ref{PiT-L}) with $n_p^2$ replaced by
$n_P^2 = p^2/P^2$ and ${\cal T}^{00}(p,-p)$ replaced by
\bqa
{\cal T}_P &=& {w(\epsilon)\over2}
	\int_{-1}^1dc\;(1-c^2)^{-\epsilon}{iP_0\over iP_0-pc} \;.
\label{TP-def}
\eqa
%
Note that this function differs by a sign from the 00 component
${\cal T}^{00}(P,-P)$ of the Euclidean tensor corresponding
to~(\ref{T2-def}):
\bqa
{\cal T}^{00}(P,-P) = - {\cal T}^{00}(p,-p)\bigg|_{p_0 \to iP_0}
		    = - {\cal T}_P \;.
\eqa
%
A more convenient form for calculating sum-integrals
that involve the function ${\cal T}_P$ is
\bqa
{\cal T}_P &=&
	\left\langle {P_0^2 \over P_0^2 + p^2c^2} \right\ranglec \, ,
\label{TP-int}
\eqa
%
%
where the angular brackets represent an average over $c$ defined by
\begin{equation}
\left\langle f(c) \right\rangle_{\!c} \equiv w(\epsilon) \int_0^1 dc \,
(1-c^2)^{-\epsilon} f(c)
\label{c-average}
\end{equation}
%
and $w(\epsilon)$ is given in (\ref{weight}).

\section{Sum-integrals}
\label{app:sumint}
\setcounter{equation}{0}

In the imaginary-time formalism for thermal field theory, a boson has
Euclidean 4-momentum $P=(P_0,{\bf p})$, with $P^2=P_0^2+{\bf p}^2$.
The Euclidean energy $P_0$ has discrete values: $P_0=2\pi nT$,
where $n$ is an integer.
Loop diagrams involve sums over $P_0$ and integrals over ${\bf p}$.
With dimensional regularization, the integral is generalized
to $d = 3-2 \epsilon$ spatial dimensions.
We define the dimensionally regularized sum-integral by
\begin{equation}
  \hbox{$\sum$}\!\!\!\!\!\!\int_P \;\equiv\;
  \left(\frac{e^\gamma\mu^2}{4\pi}\right)^\epsilon\;
  T\sum_{P_0}\:\int {d^{3-2\epsilon}p \over (2 \pi)^{3-2\epsilon}}\,,
\label{sumint-def}
\end{equation}
%
where $d=3-2\epsilon$ is the dimension of space
and $\mu$ is an arbitrary
momentum scale. The factor $(e^\gamma/4\pi)^\epsilon$
is introduced so that, after minimal subtraction
of the poles in $\epsilon$
due to ultraviolet divergences, $\mu$ coincides
with the renormalization
scale of the $\overline{\rm MS}$ renormalization scheme.

\subsection{Simple one-loop sum-integrals}

The simple one-loop sum-integrals required in our calculations
are
\begin{eqnarray}
\sumint_P \log P^2 &=& -{\pi^2 \over 45} T^4 \,,
\\
\sumint_P {1 \over P^2} \hspace{0.5cm} &=&
T^2 \left({\mu\over4\pi T}\right)^{2\epsilon}
{1 \over 12} \left[ 1
	+ \left( 2 + 2{\zeta'(-1) \over \zeta(-1)} \right) \epsilon \right] \,,
\label{sumint:2}
\\
\sumint_P {p^2 \over (P^2)^2} &=& {1 \over 8} T^2 \,,
\\
\sumint_P {1 \over p^2 P^2} &=&
{1 \over (4\pi)^2} \left({\mu\over4\pi T}\right)^{2\epsilon} \,
2 \left[ {1\over\epsilon} + 2 \gamma + 2
	+ \left( 4 + 4 \gamma + {\pi^2 \over 4} - 4 \gamma_1 \right)
		\epsilon \right]\,,
\label{sumint:4}
\\
\sumint_P {1 \over (P^2)^2} &=&
{1 \over (4\pi)^2} \left({\mu\over4\pi T}\right)^{2\epsilon}
\left[ {1 \over \epsilon} + 2 \gamma
	+ \left( {\pi^2 \over 4} - 4 \gamma_1 \right) \epsilon \right] \,,
\\
\sumint_P {p^2 \over (P^2)^3} &=&
{1 \over (4\pi)^2}\left({\mu\over4\pi T}\right)^{2\epsilon}
{3\over4} \left[ {1\over\epsilon}+2\gamma-{2\over3} \right] \,,
\\
\sumint_P {p^4 \over (P^2)^4} &=&
{1 \over (4\pi)^2}\left({\mu\over4\pi T}\right)^{2\epsilon}
{5\over8} \left[ {1\over\epsilon}+2\gamma-{16\over15} \right] \,,
\\
\sumint_P {1 \over (P^2)^3} &=&
{2 \zeta(3) \over (4\pi)^4}{1 \over T^2} \,.
\end{eqnarray}
%
The calculation of these sum-integrals is standard.
The errors are all one order higher in $\epsilon$
than the smallest term shown.
The number $\gamma_1$ is the first Stieltjes gamma constant
defined by the equation
\begin{equation}
\zeta(1+z) = {1 \over z} + \gamma - \gamma_1 z + O(z^2) \, .
\label{zeta}
\end{equation}
%

\subsection{One-loop HTL sum-integrals}

The one-loop sum-integrals involving
the HTL function ${\cal T}_P$ defined in (\ref{TP-def}) are
\begin{eqnarray}
\sumint_P {1 \over P^2} {\cal T}_P &=&
T^2 \left({\mu\over4\pi T}\right)^{2\epsilon}
\left(- {1 \over 24} \right)
\left[ {1 \over \epsilon} + 2 {\zeta'(-1) \over \zeta(-1)} \right] \,,
\label{sumint-T:1}
\\
\sumint_P {1 \over p^4} {\cal T}_P &=&
{1 \over (4 \pi)^2} \left({\mu\over4\pi T}\right)^{2\epsilon}
(-1) \left[ {1 \over \epsilon} + 2 \gamma + 2 \log 2 \right] \,,
\label{sumint-T:2}
\\
\sumint_P {1 \over p^2 P^2} {\cal T}_P &=&
{1 \over (4\pi)^2} \left({\mu\over4\pi T}\right)^{2\epsilon}
\left[ 2 \log2 \left({1 \over \epsilon} + 2 \gamma \right)
	+ 2 \log^2 2 + {\pi^2 \over 3}  \right] \,,
\\
\sumint_P {1 \over (P^2)^2} {\cal T}_P &=&
{1 \over (4\pi)^2} \left({\mu\over4\pi T}\right)^{2\epsilon}
{1 \over 2}
\left[ {1 \over \epsilon} + 2 \gamma + 1 \right] \,,
\\
\sumint_P {1 \over p^4} ({\cal T}_P)^2 &=&
{1 \over (4\pi)^2} \left({\mu\over4\pi T}\right)^{2\epsilon}
\left( - {2 \over 3} \right)
\left[ (1+ 2 \log 2) \left( {1 \over \epsilon} + 2 \gamma \right)
\right.
\nonumber
\\
&& \hspace{5cm} \left.
	- {4 \over 3} + {22 \over 3} \log 2 + 2 \log^2 2 \right] \, .
\label{sumint-T:5}
\end{eqnarray}
%
The errors are all of order $\epsilon$.

It is straightforward to calculate the sum-integrals
(\ref{sumint-T:1})--(\ref{sumint-T:5}) using the
representation (\ref{TP-int}) of the function ${\cal T}_P$.
For example, the sum-integral (\ref{sumint-T:1}) can be written
\begin{equation}
\sumint_P {1 \over P^2} {\cal T}_P =
\sumint_P {1 \over P_0^2 + p^2}
	\left\langle {P_0^2\over P_0^2 + p^2c^2} \right\ranglec \, ,
\end{equation}
%
%
where the angular brackets denote an average over $c$
as defined in (\ref{c-average}).
Using the factor of $P_0^2$ in the numerator to cancel denominators,
this becomes
%
%
\begin{equation}
\sumint_P {1 \over P^2} {\cal T}_P =
\left\langle {1 \over 1-c^2} \,
\sumint_P \left( {1 \over P^2}
- {c^2 \over P_0^2 + p^2c^2} \right) \right\ranglec \, .
\end{equation}
%
%
After rescaling the momentum by ${\bf p} \to {\bf p}/c$,
the second sum-integral on the right side becomes the same as the first
sum-integral, and the expression reduces to
%
%
%
\begin{equation}
\sumint_P {1 \over P^2} {\cal T}_P =
\left\langle{1-c^{-1+2\epsilon}\over 1 - c^2}\right\ranglec
\sumint_P {1 \over P^2} \,.
\end{equation}
%
%
Evaluating the average over $c$,
using the expression (\ref{sumint:2}) for the sum-integral,
and expanding in powers of $\epsilon$, we obtain the result
(\ref{sumint-T:1}).
Following the same strategy, all the sum-integrals
(\ref{sumint-T:1})--(\ref{sumint-T:5}) can be reduced to linear
combinations of the simple sum-integrals (\ref{sumint:2}) and
(\ref{sumint:4}) with coefficients that are averages over $c$.
The only difficult integral is the double average over $c$
that arises from (\ref{sumint-T:5}):
%
%
%
\begin{eqnarray}
\left\langle {c_1^{3+2\epsilon} - c_2^{3+2\epsilon}
	\over c_1^2 - c_2^2} \right\rangle_{\!\!c_1,c_2}
= {1 + 2 \log 2 \over 3}  +
\left( - {10 \over 9} + {10 \over 9} \log 2 + {2 \over 3} \log^2 2
	\right) \epsilon  \,.
\end{eqnarray}
%

\subsection{Simple two-loop sum-integrals}

The simple two-loop sum-integrals that are needed are
\begin{eqnarray}
\sumint_{PQ} {1\over P^2 Q^2 R^2} &=& 0 \,,
\\
\sumint_{PQ} {1 \over P^2 Q^2 r^2} &=&
{T^2 \over (4 \pi)^2} \left({\mu\over4\pi T}\right)^{4\epsilon}
{1 \over 12}
\left[ {1 \over \epsilon} + 10 - 12 \log 2
	+ 4 {\zeta'(-1) \over \zeta(-1)} \right]    \,,
\label{sumint2:2}
\\
\sumint_{PQ} {q^2 \over P^2 Q^2 r^4} &=&
{T^2 \over (4 \pi)^2} \left({\mu\over4\pi T}\right)^{4\epsilon}
{1 \over 6}
\left[ {1 \over \epsilon} + {8 \over 3} + 2 \gamma
	+ 2 {\zeta'(-1) \over \zeta(-1)} \right]    \,,
\label{sumint2:3}
\\
\sumint_{PQ} {q^2 \over P^2 Q^2 r^2 R^2} &=&
{T^2 \over (4 \pi)^2} \left({\mu\over4\pi T}\right)^{4\epsilon}
{1 \over 9}
\left[ {1 \over \epsilon} + 7.521 \right]    \,,
\label{sumint2:4}
\\
\sumint_{PQ} {P\!\cdot\!Q \over P^2 Q^2 r^4} &=&
{T^2 \over (4 \pi)^2} \left({\mu\over4\pi T}\right)^{4\epsilon}
\left( -{1 \over 8} \right)
\left[ {1 \over \epsilon} + {2 \over 9} + 4 \log 2 + {8\over3} \gamma
	+ {4\over3} {\zeta'(-1) \over \zeta(-1)} \right]     \,,
\label{sumint2:5}
\end{eqnarray}
%
where $R = -(P+Q)$ and $r=|{\bf p} + {\bf q}|$.
The errors are all of order $\epsilon$.

To motivate the integration formula we will use to evaluate the two-loop
sum-integrals, we first present the analogous integration formula
for one-loop sum-integrals.  In a one-loop sum-integral,
the sum over $P_0$ can be replaced by a contour integral in $p_0 = -i P_0$:
\begin{eqnarray}
\sumint_P F(P) &=& \lim_{\eta \to 0^+}
\int {d p_0 \over 2 \pi i} \int_{\bf p}
\left[ F(-i p_0,{\bf p})  - F(0,{\bf p}) \right]
e^{\eta p_0} n(p_0) \,,
\end{eqnarray}
where $n(p_0) = 1/(e^{\beta p_0} - 1)$ is the Bose-Einstein thermal
distribution and the contour runs
from $-\infty$ to $+\infty$ above the real axis and
from  $+\infty$ to $-\infty$ below the real axis.
This formula can be expressed in a more convenient form by
collapsing the contour onto the real axis
and separating out those terms  with the exponential convergence
factor $n(|p_0|)$.  The remaining terms run along contours from
$-\infty \pm i \varepsilon$ to 0 and have the convergence factor
$e^{\eta p_0}$.  This allows the contours to be deformed so that
they run from 0 to $\pm i \infty$ along the imaginary $p_0$ axis,
which corresponds to real values of $P_0 = -i p_0$.
Assuming that $F(-i p_0,{\bf p})$ is a real function of $p_0$,
i.e. that it satisfies
$F(-i p_0^*,{\bf p})= F(-i p_0,{\bf p})^*$,
the resulting formula for the sum-integral is
\begin{eqnarray}
\sumint_P F(P) &=&
\int_P F(P)
+ \int_p \epsilon(p_0) n(|p_0|) \,
	2 {\rm Im} F(-i p_0+ \varepsilon,{\bf p}) \,,
\label{int-1loop}
\end{eqnarray}
%
where $\epsilon(p_0)$ is the sign of $p_0$.
The first integral on the right side is over the $(d+1)$-dimensional
Euclidean vector $P = (P_0,{\bf p})$
and the second is over the $(d+1)$-dimensional
Minkowskian vector $p = (p_0,{\bf p})$.

The two-loop sum-integrals can be evaluated by using a
generalization of the one-loop formula (\ref{int-1loop}):
\begin{eqnarray}
&& \sumint_{PQ} F(P) G(Q) H(R) \;=\;
{1 \over 3} \int_{PQ} F(P) G(Q) H(R)
\nonumber
\\
&& \hspace{1cm}
\;+\; \int_p \epsilon(p_0) n(|p_0|) \,
	2 {\rm Im} F(-i p_0+ \varepsilon,{\bf p}) \,
	{\rm Re} \int_Q G(Q) H(R)\bigg|_{P_0 = -ip_0 + \varepsilon}
\nonumber
\\
&&  \hspace{1cm}
\;+\; \int_p \epsilon(p_0) n(|p_0|) \,
		2 {\rm Im} F(-i p_0+ \varepsilon,{\bf p}) \,
	\int_q \epsilon(q_0) n(|q_0|) \,
		2 {\rm Im} G(-i q_0+ \varepsilon,{\bf q}) \,
\nonumber
\\
&&  \hspace{3cm}
	\times {\rm Re} H(R)\bigg|_{R_0 = i (p_0 + q_0)+ \varepsilon}
\;+\; ({\rm cyclic \; permutations \; of \;} F, \, G, \, H) \,.
\label{int-2loop}
\end{eqnarray}
The sum over cyclic permutations multiplies the first term on the
right side by 3, so there are a total of 7 terms.
This formula can be derived in 3 steps.
First, express the sum over $P_0$ as the sum of two contour integrals over
$p_0$, one that encloses the real axis ${\rm Im}\, p_0 = 0$
and another that encloses the line ${\rm Im} \, p_0= - {\rm Im} \, q_0$.
Second, express the sum over $q_0$ as a contour integral
that encloses the real-$q_0$ axis.
Third, symmetrize the resulting expression under the 6 permutations
of $F$, $G$, and $H$.  The resulting terms can be combined into the expression
(\ref{int-2loop}).
The integrals of the imaginary parts that enter into our calculation can be
reduced to
\begin{eqnarray}
&&\int_p \epsilon(p_0) n(|p_0|) \, 2 {\rm Im}
	{1 \over P^2}\bigg|_{P_0=-i p_0+ \varepsilon}
	f(-i p_0 + \varepsilon,{\bf p})
	= \int_{\bf p} {n(p)\over p} {1 \over 2} \sum_{\pm}
	f(\pm i p + \varepsilon,{\bf p}) \label {impart1} \, , \\
&&\int_p \epsilon(p_0) n(|p_0|) \,  2 {\rm Im}
	{\cal T}_P\bigg|_{P_0=-i p_0+ \varepsilon}
	f(-i p_0 + \varepsilon,{\bf p}) \nonumber \\
	&& \hspace {6cm} =  - \int_{\bf p} p\,n(p) {1 \over 2} \sum_{\pm}
	\left\langle c^{-3+2\epsilon}
	f(\pm i p + \varepsilon,{\bf p}/c) \right\ranglec \, .
\end{eqnarray}
%
The latter equation is obtained by inserting the expression (\ref{TP-int})
for ${\cal T}_P$, using (\ref{impart1}), and then making the change of
variable ${\bf p} \rightarrow {\bf p}/c$ to put the thermal integral into
a standard form.

As a simple illustration, we apply the formula (\ref{int-2loop})
to the sum-integral (\ref{sumint2:2}).  The nonvanishing terms are
\begin{eqnarray}
\sumint_{PQ} {1 \over P^2 Q^2 r^2} &=&
2 \int_p n(|p_0|) \, 2 \pi \delta(p_0^2 - p^2) \int_Q {1 \over Q^2 r^2}
\nonumber
\\
&& \;+\; \int_p n(|p_0|) \, 2 \pi \delta(p_0^2 - p^2)
	 \int_q n(|q_0|) \, 2 \pi \delta(q_0^2 - q^2) {1 \over r^2}   \,.
\end{eqnarray}
%
The delta functions can be used to evaluate the integrals over
$p_0$ and $q_0$.  The integral over $Q$ is given in (\ref{int4:1})
up to corrections of order $\epsilon$.
This reduces the sum-integral to
\begin{eqnarray}
\sumint_{PQ} {1 \over P^2 Q^2 r^2} &=&
{4 \over (4 \pi)^2} \left[ {1 \over \epsilon} + 4 - 2 \log 2 \right]
\mu^{2 \epsilon} \int_{\bf p} {n(p) \over p} p^{-2 \epsilon}
\;+\; \int_{\bf p q} {n(p) n(q) \over p q} {1 \over r^2} \,.
\end{eqnarray}
%
The momentum integrals are evaluated in (\ref{int-th:1}) and
(\ref{int-th:2}).  Keeping all terms that contribute through
order $\epsilon^0$, we get the result (\ref{sumint2:2}).
The sum-integral (\ref{sumint2:3}) can be evaluated in the same way:
\begin{eqnarray}
\sumint_{PQ} {q^2 \over P^2 Q^2 r^4} &=&
{2 \over (4 \pi)^2 }
\left[ {1 \over \epsilon} - 2 \log 2 \right]
\mu^{2 \epsilon} \int_{\bf p} {n(p) \over p} p^{-2 \epsilon}
\;+\; \int_{\bf p q} {n(p) n(q) \over p q} {q^2 \over r^4} \,.
\end{eqnarray}
%
The sum-integral (\ref{sumint2:5}) can be reduced to a linear combination of
(\ref{sumint2:2}) and (\ref{sumint2:3}) by expressing the numerator
in the form $P\!\cdot\!Q = P_0 Q_0 + (r^2 - p^2 - q^2)/2$
and noting that the $P_0 Q_0$ term vanishes upon summing over $P_0$ or
$Q_0$.

The sum-integral (\ref{sumint2:4}) is a little more difficult.
After applying the formula (\ref{int-2loop}) and using the delta
functions to integrate over $p_0$, $q_0$, and $r_0$,
it can be reduced to
\begin{eqnarray}
\sumint_{PQ} {q^2\over P^2 Q^2 r^2 R^2} &=&
\int_{\bf p} {n(p) \over p} \int_Q {1 \over Q^2 R^2}
	\left( {p^2 \over r^2} + {q^2 \over r^2} + {q^2 \over p^2} \right)
	\bigg|_{P_0 = -i p}
\nonumber
\\
&& \;+\; \int_{\bf p q} {n(p) n(q) \over p q}
	\left( {p^2 \over r^2} + {p^2 \over q^2} + {r^2 \over q^2} \right)
	{r^2 - p^2 - q^2 \over \Delta(p,q,r)} \,,
\label{sumint:q/PQrR}
\end{eqnarray}
%
where $\Delta(p,q,r)$ is the triangle function that is negative
when $p$, $q$, and $r$ are the lengths of 3 sides of a triangle:
\begin{equation}
\Delta(p,q,r) = p^4 + q^4 + r^4 - 2 (p^2 q^2 + q^2 r^2 + r^2 p^2) \,.
\label{triangle}
\end{equation}
%
After using (\ref{int4:4})--(\ref{int4:6}) to integrate over $Q$,
the first term on the right side of (\ref{sumint:q/PQrR})
is evaluated using (\ref{int-th:1}).
The 2-loop thermal integrals on the right side of (\ref{sumint:q/PQrR})
are given in (\ref{int-thT:1})--(\ref{int-thT:4}).
Adding together all the terms, we get the final result (\ref{sumint2:4}).

\subsection{Two-loop HTL sum-integrals}

The two-loop sum-integrals involving
the HTL function ${\cal T}_P$ defined in (\ref{TP-def}) are
\begin{eqnarray}
\sumint_{PQ} {1 \over P^2 Q^2 r^2} {\cal T}_R &=&
{T^2 \over (4 \pi)^2} \left({\mu\over4\pi T}\right)^{4\epsilon}
\left(- {1\over 48} \right)
\left[ {1 \over \epsilon^2}
\right.
\nonumber
\\
&& \hspace{1cm} \left.
\;+\; \left( 2 - 12 \log2 + 4 {\zeta'(-1) \over \zeta(-1)} \right)
	{1 \over \epsilon}
- 19.83 \right]    \,, 
\label{sumint2:6}
\\
\sumint_{PQ} {q^2 \over P^2 Q^2 r^4} {\cal T}_R &=&
{T^2 \over (4 \pi)^2} \left({\mu\over4\pi T}\right)^{4\epsilon}
\left(- {1\over 576} \right)
\left[ {1 \over \epsilon^2}
\right.
\nonumber
\\
&& \hspace{1cm} \left.
\;+\; \left( {26\over3} - {24 \over \pi^2} - 92 \log2
	+ 4 {\zeta'(-1) \over \zeta(-1)} \right) {1 \over \epsilon}
- 477.7 \right] \,,  
\label{sumint2:7}
\\
\sumint_{PQ} {P\!\cdot\!Q \over P^2 Q^2 r^4} {\cal T}_R &=&
{T^2 \over (4 \pi)^2} \left({\mu\over4\pi T}\right)^{4\epsilon}
\left(- {1\over 96} \right)
\left[ {1 \over \epsilon^2}
\right.
\nonumber
\\
&& \hspace{1cm} \left.
\;+\; \left( {8 \over \pi^2} + 4 \log2 + 4 {\zeta'(-1) \over \zeta(-1)} \right)
	{1 \over \epsilon}
+ 59.66 \right] \,.
\label{sumint2:8}
\end{eqnarray}
%

To calculate the sum-integral (\ref{sumint2:6}), we begin by using the
representation (\ref{TP-int}) of the function ${\cal T}_R$:
\begin{equation}
\sumint_{PQ} {1 \over P^2 Q^2 r^2} {\cal T}_R =
	\sumint_{PQ} {1 \over P^2 Q^2 r^2}
	- \sumint_{PQ} {1 \over P^2 Q^2 } \left\langle
	{c^2 \over R_0^2 + r^2 c^2} \right\ranglec \, .
\label{sumint2:6a}
\end{equation}
%
The first sum-integral on the right side is given by (\ref{sumint2:2}).
To evaluate the second sum-integral, we apply the sum-integral formula
(\ref{int-2loop}):
\begin{eqnarray}
\sumint_{PQ} {1 \over P^2 Q^2 (R_0^2 + r^2 c^2)} &&
\nonumber
\\
	&& \hspace{-3cm} = \int_{\bf p} {n(p) \over p}
	\left(2 {\rm Re} \int_Q {1 \over Q^2 (R_0^2+r^2c^2)}
	\bigg|_{P_0 = - i p + \varepsilon}
	+ c^{-3+2\epsilon}
	\int_Q {1 \over Q^2 R^2} \bigg|_{P \to (-i p,{\bf p}/c)}
	\, \right)
\nonumber
\\
&& \hspace{-2cm} + \int_{\bf pq} {n(p) n(q) \over p q}
	\left( {\rm Re} {r^2 c^2 - p^2 - q^2 \over
	\Delta(p+i\varepsilon,q,r c)}
	+ 2 c^{-3+2\epsilon} \, {\rm Re}
	{ r_c^2 - p^2 - q^2 \over \Delta(p+i\varepsilon,q,r_c)} \,
	\right) \, ,
\label{sumint2:6b}
\end{eqnarray}
%
where $r_c = |{\bf p} + {\bf q}/c|$.
In the terms on the right side with a single thermal integral,
the appropriate averages over $c$ of the integrals over $Q$
are given in (\ref{int4HTL:1}) and (\ref{int4:8.1}).
\begin{eqnarray}
&&
\left\langle c^2 \left(2 {\rm Re}
\int_Q {1 \over Q^2 (R_0^2+r^2c^2)} \bigg|_{P_0 = - i p + \varepsilon}
+ c^{-3+2\epsilon}
	\int_Q {1 \over Q^2 R^2} \bigg|_{P \to (-i p,{\bf p}/c)}
	\, \right) \right\ranglec
\nonumber
\\
&& \;=\;  {1 \over (4 \pi)^2} \mu^{2\epsilon} p^{-2 \epsilon}
\left[ {1\over4\epsilon^2}+\left( 4-{7\over2}\log 2\right){1\over\epsilon}
+16-{13\pi^2\over16}-8\log 2+{17\over2}\log^2 2 \right] \, .
\label{intave:1a}
\end{eqnarray}
%
The subsequent integral over ${\bf p}$ is a special case of
(\ref{int-th:1}):
\begin{equation}
\int_{\bf p} n(p) \, p^{-1-2\epsilon} =
2^{8 \epsilon}
{(1)_{-4\epsilon} ({1\over2})_{2\epsilon}
	\over (1)_{-2\epsilon} ({3\over2})_{-\epsilon}} \,
{\zeta(-1+4\epsilon) \over \zeta(-1)}
(e^\gamma \mu^2)^\epsilon (4 \pi T)^{-4\epsilon} \, {T^2 \over 12} \, ,
\label{int-th:-1}
\end{equation}
%
where $(a)_b=\Gamma(a+b)/\Gamma(a)$ is Pochhammer's symbol.
Combining this with (\ref{intave:1a}), we obtain
\begin{eqnarray}
&& \int_{\bf p} {n(p) \over p}
\left( 2 \, {\rm Re} \int_Q {1 \over Q^2}
	\left\langle {c^2 \over R_0^2 + r^2 c^2} \right\ranglec
		\bigg|_{P_0 = - i p + \varepsilon}
+ \left\langle c^{-1+2\epsilon}
	\int_Q {1 \over Q^2 R^2} \bigg|_{P \to (-i p,{\bf p}/c)}
	\right\ranglec\, \right)
\nonumber
\\
&& \;=\; {T^2 \over (4 \pi)^2} \left({\mu\over4\pi T}\right)^{4\epsilon}
{1 \over 48}
\left[ {1 \over \epsilon^2}
+ \left( 18 - 12 \log 2 + 4 {\zeta'(-1) \over \zeta(-1)} \right)
	{1 \over \epsilon}
+ 173.30233 \right] \, .
\label{intave:1}
\end{eqnarray}
%
For the two terms in (\ref{sumint2:6b}) with a double thermal integral,
the averages weighted by $c^2$ are given in (\ref{intHTL:1}) and
(\ref{intHTL:4}).
Adding them to (\ref{intave:1}), the final result is
\begin{eqnarray}
\sumint_{PQ} {1 \over P^2 Q^2}
	\left\langle {c^2 \over R_0^2 + r^2 c^2} \right\ranglec
&=&  {T^2 \over (4 \pi)^2} \left({\mu\over4\pi T}\right)^{4\epsilon}
{1\over 48}
\left[ {1 \over \epsilon^2}
\right.
\nonumber
\\
&& \hspace{2cm} \left.
\;+\; \left( 6 - 12 \log2 + 4 {\zeta'(-1) \over \zeta(-1)} \right)
	{1 \over \epsilon}
+ 18.66 \right] \, .
\label{sumintave:2}
\end{eqnarray}
%
Inserting this into (\ref{sumint2:6a}),
we obtain the final result (\ref{sumint2:6}).

The sum-integral (\ref{sumint2:7}) is evaluated in a similar way to
(\ref{sumint2:6}).
Using the representation (\ref{TP-int}) for ${\cal T}_R$, we get
\begin{equation}
\sumint_{PQ} {q^2 \over P^2 Q^2 r^4} {\cal T}_R =
	\sumint_{PQ} {q^2 \over P^2 Q^2 r^4}
	- \sumint_{PQ} {q^2 \over P^2 Q^2 r^2} \left\langle
	{c^2 \over R_0^2 + r^2 c^2} \right\ranglec \, .
\label{sumint2:7a}
\end{equation}
%
The first sum-integral on the right hand side is given by (\ref{sumint2:3}).
To evaluate the second sum-integral, we apply the sum-integral
formula (\ref{int-2loop}):
\begin{eqnarray}
\sumint_{PQ} {q^2 \over P^2 Q^2 r^2 (R_0^2+r^2c^2)} &&
\nonumber
\\
&& \hspace{-3.5cm}
= \int_{\bf p} {n(p) \over p} \left( {\rm Re}
	\int_Q {p^2+q^2 \over Q^2 r^2 (R_0^2+r^2c^2)}
	\bigg|_{P_0 = -i p + \varepsilon}
	+ {1 \over p^2} c^{-1+2\epsilon}
	\int_Q {q^2 \over Q^2 R^2}\bigg|_{P \to (-i p,{\bf p}/c)} \right)
\nonumber
\\
&& \hspace{-3cm}
+ \int_{\bf pq} {n(p) n(q) \over p q}
	\left({q^2 \over r^2} \, {\rm Re}
	{r^2 c^2 - p^2 - q^2 \over \Delta(p+i\varepsilon,q,r c)}
	+ c^{-1 + 2\epsilon} {p^2 + r_c^2 \over q^2} \,
	{\rm Re} { r_c^2-p^2-q^2
	\over \Delta(p+i\varepsilon,q,r_c)} \right) \, .
\label{sumint3:6b}
\end{eqnarray}
%
In the terms on the right side with a single thermal integral,
the weighted averages over $c$ of the integrals over $Q$
are given in (\ref{int4HTL:3}),
(\ref{int4HTL:4}), and (\ref{int4:7a}):
\begin{eqnarray}
&&
\left\langle c^2 \left(
{\rm Re} \int_Q {p^2+q^2 \over Q^2 r^2 (R_0^2+r^2c^2)}
	\bigg|_{P_0 = -i p + \varepsilon}
+ {1 \over p^2} c^{-1+2\epsilon} \int_Q {q^2 \over Q^2 R^2}
	\bigg|_{P \to (-i p,{\bf p}/c)} \right) \right\ranglec
\nonumber
\\
&& \;=\;  {1 \over (4 \pi)^2} \mu^{2\epsilon} p^{-2 \epsilon}
\left[ {1\over 48\epsilon^2}+\left({35\over36}-{31\over24}\log 2 \right)
{1\over\epsilon}+{313\over108}-{247\pi^2\over576}-{17\over18}\log 2
+{65\over24}\log^2 2\right] \, ,
\end{eqnarray}
%
After using (\ref{int-th:-1}) to evaluate the thermal integral, we obtain
\begin{eqnarray}
&& \int_{\bf p} {n(p) \over p}
\left( {\rm Re} \int_Q {p^2+q^2 \over Q^2 r^2}
	\left\langle {c^2 \over R_0^2+r^2c^2} \right\ranglec
	\bigg|_{P_0 = -i p + \varepsilon}
+ {1 \over p^2} \left\langle c^{1+2\epsilon}
	\int_Q {q^2 \over Q^2 R^2} \bigg|_{P \to (-i p,{\bf p}/c)}
	\right\ranglec \right)
\nonumber
\\
&& \;=\;   {T^2 \over (4 \pi)^2} \left({\mu\over4\pi T}\right)^{4\epsilon}
{1\over576}\left[ {1\over\epsilon^2}+\left({146\over3}-60\log 2+
4 {\zeta'(-1) \over \zeta(-1)}\right){1\over\epsilon}+84.72308\right]\, ,
\label{intave:2}
\end{eqnarray}
%
For the two terms in (\ref{sumint3:6b}) with a double thermal integral,
the averages weighted by $c^2$ are given in (\ref{intHTL:3}),
(\ref{intHTL:6}), and (\ref{intHTL:7}).
Adding them to (\ref{intave:2}), the final result is
\begin{eqnarray}
\sumint_{PQ} {q^2 \over P^2 Q^2 r^2}
	\left\langle {c^2 \over R_0^2 + r^2 c^2} \right\ranglec
&=&  {T^2 \over (4 \pi)^2} \left({\mu\over4\pi T}\right)^{4\epsilon}
{1\over 576}
\left[ {1 \over \epsilon^2}
\right.
\nonumber
\\
&&  \left.
\;+\;
\left( {314\over3} -{24 \over \pi^2} - 92 \log2
	+ 4 {\zeta'(-1) \over \zeta(-1)} \right) {1 \over \epsilon}
+ 270.2 \right] \, .
\label{sumintHTL:c2}
\end{eqnarray}
%
Inserting this into (\ref{sumint2:7a}),
we obtain the final result (\ref{sumint2:7}).

To evaluate (\ref{sumint2:8}), we use the expression (\ref{TP-int})
for ${\cal T}_R$ and the identity $P\!\cdot\!Q = (R^2-P^2-Q^2)/2$
to write it in the form
\begin{eqnarray}
\sumint_{PQ} {P\!\cdot\!Q \over P^2 Q^2 r^4} {\cal T}_R &=&
\sumint_{PQ} {P\!\cdot\!Q \over P^2 Q^2 r^4}
- \sumint_P {1\over P^2} \sumint_R {1 \over r^4} {\cal T}_R
\nonumber
\\
&&
- {1\over2} \langle c^2 \rangle_c \sumint_{PQ} {1 \over P^2 Q^2 r^2}
- {1\over2} \sumint_{PQ} {1 \over P^2 Q^2}
	\left\langle {c^2(1-c^2) \over R_0^2+r^2c^2} \right\ranglec \,.
\label{sumint2:8:2}
\end{eqnarray}
%
The sum-integrals in the first 3 terms on the right side of
(\ref{sumint2:8:2}) are given in (\ref{sumint:2}), (\ref{sumint-T:2}),
(\ref{sumint2:2}), and (\ref{sumint2:5}).  The last sum-integral
before the average weighted by $c$ is given in (\ref{sumint2:6a}).
The average weighted by $c^2$ is given in (\ref{sumintave:2}).
The average weighted by $c^4$ can be computed in the same way.
In the integrand of the single thermal integral,
the weighted averages over $c$ of the integrals over $Q$ are given in
(\ref{int4HTL:2}) and (\ref{int4:8.2}):
\begin{eqnarray}
&&
\left\langle c^4 \left(
2 {\rm Re} \int_Q {1 \over Q^2 (R_0^2+r^2c^2)}
	\bigg|_{P_0 = - i p + \varepsilon}
+ c^{-3+2\epsilon} \int_Q {1 \over Q^2 R^2}
	\bigg|_{P \to (-i p,{\bf p}/c)} \right) \right\ranglec
\nonumber
\\
&& \;=\;  {1 \over (4 \pi)^2} \mu^{2\epsilon} p^{-2 \epsilon}
\left[ \left({23\over6}-4\log 2\right){1\over\epsilon}+{104\over9}-\pi^2
-3\log 2+8\log^2 2 \right] \, ,
\end{eqnarray}
%
After using (\ref{int-th:-1}) to evaluate the thermal integral, we obtain
\begin{eqnarray}
&&
\int_{\bf p} {n(p) \over p}
\left( 2 {\rm Re} \int_Q {1 \over Q^2}
	\left\langle {c^4 \over R_0^2 + r^2 c^2} \right\ranglec
		\bigg|_{P_0 = - i p + \varepsilon}
+ \left\langle c^{1+2\epsilon}
	\int_Q {1 \over Q^2 R^2} \bigg|_{P \to (-i p,{\bf p}/c)}
	\right\ranglec \right)
\nonumber
\\
&& \;=\;  {T^2 \over (4 \pi)^2} \left({\mu\over4\pi T}\right)^{4\epsilon}
\left[\left({23\over72}-{1\over3}\log 2\right){1\over\epsilon}+1.28872
\right]\, ,
\label{intave:3}
\end{eqnarray}
%
For the two terms with a double thermal integral,
the averages weighted by $c^4$ are given in (\ref{intHTL:2}) and
(\ref{intHTL:5}).
Adding them to (\ref{intave:3}), we obtain
\begin{eqnarray}
\sumint_{PQ} {1 \over P^2 Q^2}
\left\langle {c^4 \over R_0^2 + r^2 c^2} \right\ranglec
&=&
{T^2 \over (4 \pi)^2} \left({\mu\over4\pi T}\right)^{4\epsilon}
\left[ \left( {17\over 72} - {1 \over 6 \pi^2} - {1\over 3} \log2 \right)
	{1 \over \epsilon}
	- 0.1917 \right] \,.
\label{sumintHTL:c4}
\end{eqnarray}
%
Inserting this into (\ref{sumint2:8:2}) along with (\ref{sumintave:2}),
we get the final result (\ref{sumint2:8}).

\section{Integrals}
\label{app:int}
\setcounter{equation}{0}

Dimensional regularization can be used to
regularize both the ultraviolet divergences and infrared divergences
in 3-dimensional integrals over momenta.
The spatial dimension is generalized to  $d = 3-2\epsilon$ dimensions.
Integrals are evaluated at a value of $d$ for which they converge and then
analytically continued to $d=3$.
We use the integration measure
\begin{equation}
  \int_{\bf p} \;\equiv\;
  \left(\frac{e^\gamma\mu^2}{4\pi}\right)^\epsilon\,
  \int {d^{3-2\epsilon}p \over (2 \pi)^{3-2\epsilon}}\,.
\end{equation}
%

\subsection{3-dimensional integrals}

We require several integrals that do not involve the
Bose-Einstein distribution function.
The momentum scale in these integrals is set by the mass
parameter $m_D$.
The one-loop integrals are
\begin{eqnarray}
\int_{\bf p} \log(p^2+m^2) & = &
- {m^3\over 6\pi}  \left( {\mu \over 2 m} \right)^{2 \epsilon}
\left[ 1 + {8 \over 3}\epsilon \right]  \,,
\label{int-3:1}
\\
\int_{\bf p} {1 \over p^2+m^2} & = &
- {m\over 4\pi} \left( {\mu \over 2 m} \right)^{2 \epsilon}
\left[1 + 2 \epsilon  \right] \,.
\label{bi3}
\end{eqnarray}
%
We also require a two-loop integral:
\begin{eqnarray}
\int_{\bf pq} {1 \over p^2 (q^2+m^2)(r^2+m^2)} & = &
{1\over (4\pi)^2} \left( {\mu \over 2 m} \right)^{4 \epsilon}
{1 \over 4} \left[ {1\over\epsilon} + 2 \right] \,.
\label{int-3:3}
\end{eqnarray}
%
The errors in (\ref{int-3:1})--(\ref{int-3:3})
are all one order higher in $\epsilon$ than the smallest term shown.

\subsection{Thermal integrals}

The thermal integrals involve the
Bose-Einstein distribution $n(p) = 1/(e^{\beta p} - 1)$.
The one-loop integrals can all be obtained from the general
formula
\begin{eqnarray}
\int_{\bf p} {n(p) \over p} p^{2 \alpha} &=&
{\zeta(2+2\alpha-2\epsilon) \over  4 \pi^2}
{\Gamma(2+2\alpha-2\epsilon) \Gamma({1\over2})
	\over  \Gamma({3\over2}-\epsilon)}
\left( e^\gamma \mu^2 \right)^\epsilon T^{2+2\alpha-2\epsilon} \,.
\label{int-th:1}
\end{eqnarray}
%
The simple two-loop thermal integrals that we need are
\begin{eqnarray}
\int_{\bf pq} {n(p) n(q) \over p q} \, {1 \over r^2} &=&
{T^2 \over (4\pi)^2} \left({\mu\over4\pi T}\right)^{4\epsilon}
\left(- {1 \over 4} \right)
\left[ {1\over\epsilon} + {14\over3} + 4 \log 2
	+ 4 {\zeta'(-1) \over \zeta(-1)} \right] \,,
\label{int-th:2}
\\
\int_{\bf pq} {n(p) n(q) \over p q} \, {p^2 \over r^4} &=&
{T^2 \over (4\pi)^2}  \left({\mu\over4\pi T}\right)^{4\epsilon}
\left[ {1\over9} + {1\over3} \gamma
	- {1\over3} {\zeta'(-1) \over \zeta(-1)}
	\;-\; 4.855 \, \epsilon \right] \,,
\label{int-th:3}
\end{eqnarray}
%
We also need some more complicated 2-loop thermal integrals
that involve the triangle function defined in (\ref{triangle}):
\begin{eqnarray}
\int_{\bf pq} {n(p) n(q) \over p q} \,
	{r^4 \over q^2 \Delta(p,q,r)} &=&
{T^2 \over (4\pi)^2} \left({\mu\over4\pi T}\right)^{4\epsilon}
\; {7 \over 48}
\left[ {1\over\epsilon^2}
	+ \left( {22\over 7} + 2 \gamma + 2 {\zeta'(-1)\over\zeta(-1)}
			- {\zeta(3)\over 35} \right) {1\over\epsilon}
\right.
\nonumber
\\
&& \hspace{4cm} \left.
	+\;40.3896 \right] \,,
\label{int-thT:1}
\\
\int_{\bf pq} {n(p) n(q) \over p q} \,
	{r^2 \over \Delta(p,q,r)} &=&
{T^2 \over (4\pi)^2} \left({\mu\over4\pi T}\right)^{4\epsilon}
\; {1 \over 24}
\left[ {1\over\epsilon^2}
	+ 2 \left( 1 + \gamma + {\zeta'(-1)\over\zeta(-1)} \right)
		{1\over\epsilon}
\right. \nonumber
\\
&& \hspace{0cm} \left.
	 + 4 +\;4 \gamma + {\pi^2 \over 2} - 4 \gamma_1
	+ 4 (1 + \gamma) {\zeta'(-1)\over\zeta(-1)}
	+2 {\zeta''(-1)\over\zeta(-1)} \right] \,,
\label{int-thT:2}
\\
\int_{\bf pq} {n(p) n(q) \over p q} \,
	{p^4 \over q^2 \Delta(p,q,r)} &=&
{T^2 \over (4\pi)^2} \left({\mu\over4\pi T}\right)^{4\epsilon}
\left(- {\zeta(3) \over 240} \right)
\left[ {1\over\epsilon} + 2 + 2 {\zeta'(-3)\over\zeta(-3)}
	+ 2 {\zeta'(3)\over\zeta(3)}   \right] ,
\label{int-thT:3}
\\
\int_{\bf pq} {n(p) n(q) \over p q} \, {p^2 \over r^2} \,
	{p^2+q^2 \over \Delta(p,q,r)} &=&
{T^2 \over (4\pi)^2} \left({\mu\over4\pi T}\right)^{4\epsilon}
\; {1 \over 48}
\left[ {1\over\epsilon^2}
	+ \left( {14 \over 3} + 10 \gamma - 6 {\zeta'(-1)\over\zeta(-1)} \right)
		{1\over\epsilon}
\right.
\nonumber
\\
&& \hspace{4cm} \left.
	-\; 86.46  \right]  \,.
\label{int-thT:4}
\end{eqnarray}
%
The most difficult thermal integrals to evaluate involve both the
triangle function and the HTL average defined in
(\ref{c-average}). There are 2 sets of these integrals.
The first set is
\begin{eqnarray}
&& \int_{\bf pq} {n(p) n(q) \over p q}
{\rm Re} \left\langle c^2 {r^2 c^2 - p^2 - q^2 \over
	\Delta(p+i\varepsilon,q,r c)}  \right\ranglec \;=\;
{T^2 \over (4\pi)^2}  \left[ \, 0.138727 \, \right] \, ,
\label{intHTL:1}
\\
&& \int_{\bf pq} {n(p) n(q) \over p q}
{\rm Re} \left\langle c^4 {r^2 c^2 - p^2 - q^2 \over
	\Delta(p+i\varepsilon,q,r c)}  \right\ranglec \;=\;
{T^2 \over (4\pi)^2} \left({\mu\over4\pi T}\right)^{4\epsilon}
\left(- {1 \over 6\pi^2} \right)
\left[ {1\over\epsilon} +  6.8343 \right]\, ,
\label{intHTL:2}
\\
&& \int_{\bf pq} {n(p) n(q) \over p q}{q^2 \over r^2}
{\rm Re} \left\langle c^2
	{r^2 c^2 - p^2 - q^2 \over \Delta(p+i\varepsilon,q,r c)}
	\right\ranglec \;=\;
{T^2 \over (4\pi)^2} \left({\mu\over4\pi T}\right)^{4\epsilon}
{\pi^2 -1 \over 24 \pi^2} \left[ {1\over\epsilon} + 15.3782 \right] \, .
\label{intHTL:3}
\end{eqnarray}
%
The second set of these integrals involve the variable
$r_c = |{\bf p} + {\bf q}/c|$:
\begin{eqnarray}
&& \int_{\bf pq} {n(p) n(q) \over p q}
{\rm Re} \left\langle c^{-1+2\epsilon}
	{r_c^2 - p^2 - q^2 \over \Delta(p+i\varepsilon,q,r_c)} \,
	\right\ranglec \;=\;
{T^2 \over (4\pi)^2} \left({\mu\over4\pi T}\right)^{4\epsilon}
\left(- {1 \over 8} \right) \left[ {1\over\epsilon} + 13.442 \right] \, ,
\label{intHTL:4}
\\
&& \int_{\bf pq} {n(p) n(q) \over p q}
{\rm Re} \left\langle c^{1+2\epsilon}
	{r_c^2 - p^2 - q^2 \over \Delta(p+i\varepsilon,q,r_c)} \,
	\right\ranglec \;=\;
{T^2 \over (4\pi)^2} \left({\mu\over4\pi T}\right)^{4\epsilon}
\left(- {1 \over 24} \right) \left[ {1\over\epsilon} +  16.381 \right] \, ,
\label{intHTL:5}
\\
&& \int_{\bf pq} {n(p) n(q) \over p q}{p^2 \over q^2}
{\rm Re} \left\langle c^{1+2\epsilon}
	{r_c^2 - p^2 - q^2 \over \Delta(p+i\varepsilon,q,r_c)} \,
	\right\ranglec \;=\;
{T^2 \over (4\pi)^2} \left({\mu\over4\pi T}\right)^{4\epsilon}
{1 \over 48} \left[ {1\over\epsilon} + 6.1227 \right] \, ,
\label{intHTL:6}
\\
&& \int_{\bf pq} {n(p) n(q) \over p q}
{\rm Re} \left\langle c^{1+2\epsilon} {r_c^2 \over q^2}
	{r_c^2 - p^2 - q^2\over \Delta(p+i\varepsilon,q,r_c)} \,
	\right\ranglec \;=\;
{T^2 \over (4\pi)^2} \left({\mu\over4\pi T}\right)^{4\epsilon}
\nonumber
\\
&& \hspace{9cm} \times {5-8\log 2 \over 144}
\left[ {1\over\epsilon} +  100.73  \right] \, .
\label{intHTL:7}
\end{eqnarray}
%
The errors in (\ref{int-th:2})--(\ref{intHTL:7})
are all one order higher in $\epsilon$ than the smallest term shown.
The numerical constant in (\ref{int-thT:1}) can be expressed
analytically in terms of the transcendental numbers appearing in
(\ref{int-thT:2}) and (\ref{int-thT:3}).
We do not know how to calculate the numerical constants in
(\ref{int-th:3}), (\ref{int-thT:4}),
(\ref{intHTL:1})--(\ref{intHTL:7})  analytically.

The simplest way to evaluate integrals like (\ref{int-th:2})
and (\ref{int-th:3}) whose integrands factor into
separate functions of $p$, $q$, and $r$  is to Fourier transform
to coordinate space where they reduce to an integral over a single
coordinate ${\bf R}$:
\begin{eqnarray}
\int_{\bf pq} f(p) \, g(q) \, h(r) &=&
\int_{\bf R} \tilde f(R) \, \tilde g(R) \, \tilde h(R) \,.
\label{int-fgh}
\end{eqnarray}
%
The Fourier transform is
\begin{eqnarray}
\tilde f(R) &=&
\int_{\bf p} f(p) e^{i {\bf p} \cdot {\bf R}} \, ,
\end{eqnarray}
%
and the dimensionally regularized coordinate integral is
\begin{eqnarray}
\int_{\bf R} &=&
\left( {e^\gamma \mu^2 \over 4 \pi} \right)^{-\epsilon}
\int d^{3-2 \epsilon}R \,.
\end{eqnarray}
%
The Fourier transforms we need are
\begin{eqnarray}
\int_{\bf p} p^{2 \alpha} \,  e^{i {\bf p} \cdot {\bf R}} &=&
{1 \over 8\pi}
{\Gamma({3\over2} + \alpha - \epsilon)
	\over \Gamma({1\over2}) \Gamma(-\alpha)}
\left( e^\gamma \mu^2 \right)^\epsilon
\left( {2 \over R} \right)^{3 + 2 \alpha - 2\epsilon} \,,
\\
\int_{\bf p} {n(p) \over p} \,  p^{2 \alpha} \,
	e^{i {\bf p} \cdot {\bf R}} &=&
{1 \over 4\pi} {1 \over \Gamma({1\over2})}
\left( e^\gamma \mu^2 \right)^\epsilon
\left( {2 \over R} \right)^{{1\over2} - \epsilon}
\int_0^\infty dp \, p^{2 \alpha + {1\over2} - \epsilon} n(p)
	J_{{1\over2}-\epsilon}(pR) \,.
\label{fourier-n}
\end{eqnarray}
%
If $\alpha$ is an even integer, the Fourier transform (\ref{fourier-n})
is particularly simple in the limit $d \to 3$:
\begin{eqnarray}
\int_{\bf p} {n(p) \over p} \,  e^{i {\bf p} \cdot {\bf R}} &\longrightarrow&
{T \over 4 \pi R}
\left( \coth x - {1 \over x} \right) \,,
\\
\int_{\bf p} {n(p) \over p} \, p^2 \,  e^{i {\bf p} \cdot {\bf R}}
	&\longrightarrow&
- {\pi T^3 \over 2 R}
\left( \coth^3x - \coth x - {1 \over x^3} \right) \,,
\end{eqnarray}
%
where $x = \pi R T$.
We can use these simple expressions only if the integral
over the coordinate ${\bf R}$ in (\ref{int-fgh})
converges for $d=3$.  Otherwise, we must first make subtractions
on the integrand to make the integral convergent.

The integral (\ref{int-th:3}) can be evaluated directly by applying
the Fourier transform formula (\ref{int-fgh}) in the limit $\epsilon \to 0$.
The integral (\ref{int-th:2}) however requires subtractions.
It can be written
\begin{eqnarray}
\int_{\bf pq} {n(p) n(q) \over p q} \, {1 \over r^2} &=&
\int_{\bf pq} {n(p) \over p} \,
	\left( {n(q) \over q} - {T \over q^2} \right) \, {1 \over r^2}
+ T \int_{\bf p} {n(p) \over p} \int_{\bf q} {1 \over q^2 r^2} \,.
\end{eqnarray}
%
In the second term on the right side, the integral over ${\bf q}$
is proportional to $p^{-1-2 \epsilon}$, so the integral over ${\bf p}$
can be evaluated using (\ref{int-th:1}).  This first term on the right side
is convergent for $d=3$ so it can be evaluated easily
using the Fourier transform formula (\ref{int-fgh}).
The integral over ${\bf R}$ reduces to a sum of integrals of the form
$\int_0^\infty dx \, x^m \coth ^n x$.  Although the sum of the integrals
converges, each of the individual integrals diverges either as
$x \to 0$ or as $x \to \infty$.  A convenient way to evaluate these
integrals is to use the strategy in Appendix C of Ref.~\cite{AZ-95}.
The integrals are regularized by using the substitution
\begin{eqnarray}
\int_0^\infty dx \, x^m \coth^n x &\longrightarrow&
{\Gamma(1+\delta) \over 2^\delta}
\int_0^\infty dx \, x^{m+\delta} \coth^n x \,.
\end{eqnarray}
%
The divergences appear as poles in $\delta$ that cancel upon adding
a convergent combination of these integrals.

The integrals (\ref{int-thT:1})--(\ref{int-thT:3}) can be evaluated
by first averaging over angles.  The triangle function can be expressed as
\begin{equation}
\Delta(p,q,r) = - 4 p^2 q^2 (1 - \cos^2 \theta) \,,
\label{triangle-theta}
\end{equation}
%
where $\theta$ is the angle between ${\bf p}$ and ${\bf q}$.
For example, the angle average for (\ref{int-thT:1}) is
\begin{equation}
\left\langle {r^4 \over \Delta(p,q,r)}
\right\rangle_{\!\!{\bf \hat p}\cdot{\bf \hat q}}
= -{w(\epsilon) \over 8} \int_{-1}^{+1} dx \, (1-x^2)^{-1-\epsilon} \,
	{(p^2 + q^2 + 2 p q x)^2 \over p^2 q^2} \, .
\label{ang-ave:1}
\end{equation}
%
After integrating over $x$ and inserting the result into
(\ref{int-thT:1}), the integral reduces to
\begin{equation}
\int_{\bf pq} {n(p) n(q) \over p q} \,
	{r^4 \over q^2\Delta(p,q,r)} \;=\;
\int_{\bf pq} {n(p) n(q) \over p q}
\left( {1 - 2 \epsilon \over 8 \epsilon} \, {p^2 \over q^4}
	+ {7 - 6 \epsilon \over 8 \epsilon} \, {1 \over q^2} \right) \,.
\end{equation}
%
The integrals over ${\bf p}$ and ${\bf q}$ factor into separate integrals
that can be evaluated using (\ref{int-th:1}).
After averaging over angles, the integrals
(\ref{int-thT:2}) and (\ref{int-thT:3}) reduce to
\begin{eqnarray}
\int_{\bf pq} {n(p) n(q) \over p q} \,
	{r^2 \over \Delta(p,q,r)}
&=& {1 - 2 \epsilon \over 4 \epsilon}
\int_{\bf p} {n(p) \over p}
\int_{\bf q} {n(q) \over q} \, {1 \over  q^2} \,,
\\
\int_{\bf pq} {n(p) n(q) \over p q} \,
	{p^4 \over q^2 \Delta(p,q,r)}
&=& {1 - 2 \epsilon \over 8 \epsilon}
\int_{\bf p} {n(p) \over p} \, p^2
\int_{\bf q} {n(q) \over q} \, {1 \over  q^4} \,.
\end{eqnarray}
%

The integral (\ref{int-thT:4}) can be evaluated by using the remarkable
identity
\begin{equation}
\left\langle {p^2+q^2 \over r^2\Delta(p,q,r)}
\right\rangle_{\!\!{\bf \hat p}\cdot{\bf \hat q}}
= {1 \over 2 \epsilon} \left\langle {1 \over r^4}
\right\rangle_{\!\!{\bf \hat p}\cdot{\bf \hat q}}
+ {1-2\epsilon \over 8\epsilon} {1 \over p^2 q^2} \, .
\label{ang-ave:2}
\end{equation}
%
The identity can be proved by expressing the angular averages
in terms of integrals over the cosine of the angle between
${\bf p}$ and ${\bf q}$ as in (\ref{ang-ave:1}),
and then integrating by parts. Inserting the identity (\ref{ang-ave:2})
into (\ref{int-thT:4}), the integral reduces to
\begin{eqnarray}
\int_{\bf pq} {n(p) n(q) \over p q} \,
	{p^2(p^2+q^2) \over r^2 \Delta(p,q,r)} \;=\;
{1 \over 2 \epsilon}
\int_{\bf pq} {n(p) n(q) \over p q} \, {p^2 \over r^4}
\,+\, {1 - 2 \epsilon \over 8 \epsilon}
\int_{\bf p} {n(p) \over p}
\int_{\bf q} {n(q) \over q} \, {1 \over  q^2} \,.
\end{eqnarray}
%
The integral in the first term on the right is given in (\ref{int-th:3}),
while the second term can be evaluated using (\ref{int-th:1}).

To evaluate the weighted averages over $c$ of the thermal integrals
in (\ref{intHTL:1})--(\ref{intHTL:3}), we first isolate
the divergent parts, which come from the region $p-q \to 0$.
We write the product of thermal functions
in the form
\begin{equation}
n(p) n(q) =
\left( n(p) n(q) - {s^2 n^2(s) \over p q}  \right)
+  {s^2 n^2(s) \over p q}  \, ,
\label{nnsub-1}
\end{equation}
%
where $s = (p+q)/2$.  In the difference term, the HTL average over
$c$ and the angular average over $x = \hat {\bf p} \cdot \hat {\bf q}$
can be calculated in 3 dimensions:
\begin{eqnarray}
{\rm Re} \left\langle c^2 {r^2 c^2 - p^2 - q^2 \over
	\Delta(p+i\varepsilon,q,r c)}  \right\ranglecx &=&
{1 \over 4 p q} \log {p+q \over |p-q|}
	- {1 \over 2(p^2 - q^2)} \log(p/q) \, ,
\\
{\rm Re} \left\langle c^4 {r^2 c^2 - p^2 - q^2 \over
	\Delta(p+i\varepsilon,q,r c)}  \right\ranglecx &=&
{2(p^2 + q^2) \over 3 (p^2-q^2)^2}
+ {1 \over 12 p q} \log {p+q \over |p-q|}
\nonumber
\\
&&
\;-\; {(3p^2 + q^2)(p^2 + 3 q^2) \over 6 (p^2 - q^2)^3} \log(p/q) \, ,
\\
{\rm Re} \left\langle c^2 {q^2 \over r^2} {r^2 c^2 - p^2 - q^2 \over
	\Delta(p+i\varepsilon,q,r c)}  \right\ranglecx &=&
{q^2 \over 3 (p^2 - q^2)^2}
\left( 2 - {1 \over 2} \log{|p^2-q^2| \over p q} \right.
\nonumber
\\
&&
\left. \;-\; {p^2+q^2 \over 4 p q} \log {p+q \over |p-q|}
	- {p^2 + q^2 \over p^2 - q^2} \log(p/q) \right)\, .
\end{eqnarray}
%
The remaining 2-dimensional integral over $p$ and $q$
can then be evaluated numerically:
\begin{eqnarray}
&& \int_{\bf pq}
\left( {n(p) n(q) \over p q}
	- {s^2 n^2(s) \over p^2 q^2}  \right)
{\rm Re} \left\langle c^2 {r^2 c^2 - p^2 - q^2 \over
	\Delta(p+i\varepsilon,q,r c)}  \right\ranglec \;=\;
\left( 5.292 \times 10^{-3} \right) {T^2 \over (4\pi)^2} \, ,
\label{intHTL:1f}
\\
&& \int_{\bf pq}
\left( {n(p) n(q) \over p q}
	- {s^2 n^2(s) \over p^2 q^2}  \right)
{\rm Re} \left\langle c^4 {r^2 c^2 - p^2 - q^2 \over
	\Delta(p+i\varepsilon,q,r c)}  \right\ranglec \;=\;
\left( 3.292 \times 10^{-3} \right) {T^2 \over (4\pi)^2} \, ,
\label{intHTL:2f}
\\
&& \int_{\bf pq}
\left( {n(p) n(q) \over p q}
	- {s^2 n^2(s) \over p^2 q^2}  \right){q^2 \over r^2}
{\rm Re} \left\langle c^2 {r^2 c^2 - p^2 - q^2 \over
	\Delta(p+i\varepsilon,q,r c)}  \right\ranglec \;=\;
\left( 2.822 \times 10^{-3} \right) {T^2 \over (4\pi)^2} \, .
\label{intHTL:3f}
\end{eqnarray}
%
The integrals involving the $n^2(s)$ term in (\ref{nnsub-1})
are divergent, so the HTL average over
$c$ and the angular average over $x = \hat {\bf p} \cdot \hat {\bf q}$
must be calculated in $3-2\epsilon$ dimensions.
The first step in the calculation of the $n^2(s)$ term
is to change variables from ${\bf p}$ and ${\bf q}$ to
$s = (p+q)/2$, $\beta = 4pq/(p+q)^2$,
and $x= \hat{\bf p} \cdot \hat{\bf q}$:
\begin{eqnarray}
\int_{\bf pq} {s^2 n^2(s) \over p^2 q^2} \, f(p,q,r) &=&
{64 \over (4\pi)^4}
\left[ (e^\gamma \mu^2)^\epsilon
	{\Gamma({3\over2}) \over \Gamma({3\over2}-\epsilon)} \right]^2
	\int_0^\infty ds \, s^{1-4 \epsilon}n^2(s)
\nonumber
\\
&&
\times  s^2 \int_0^1 d \beta \, \beta^{-2 \epsilon} (1-\beta)^{-1/2}
\Big\langle f(s_+,s_-,r) + f(s_-,s_+,r) \Big\rangle_{\!\!x} \, ,
\label{int:n2f}
\end{eqnarray}
%
where $s_\pm = s[1 \pm \sqrt{1-\beta}]$
and $r=s [4 - 2\beta(1-x)]^{1/2}$.
The 2 terms inside the average over $x$
come from the regions $p>q$ and $p<q$, respectively.
The integral over $s$ is easily evaluated:
\begin{equation}
\int_0^\infty ds \, s^{1-4 \epsilon}n^2(s) \;=\;
\Gamma(2-4\epsilon)
\left[ \zeta(1-4\epsilon) - \zeta(2-4\epsilon) \right] T^{2-4\epsilon}\, .
\label{intth:n2}
\end{equation}
%
It remains only to evaluate the averages over $c$ and $x$ and
the integral over $\beta$.

The first step in the calculation of the $n^2(s)$ term
of (\ref{intHTL:1}) is to decompose the integrand into 2 terms:
\begin{equation}
{r^2 c^2 - p^2 - q^2 \over
	\Delta(p+i\varepsilon,q,r c)} \;=\;
-{1 \over 2} \sum_\pm {1 \over (p+i\varepsilon \pm q)^2 - r^2 c^2}\, .
\end{equation}
%
The weighted averages over $c$ give hypergeometric functions:
\begin{eqnarray}
\left\langle {c^2 \over (p+i\varepsilon \pm q)^2 - r^2 c^2}
	\right\ranglec &=&
{1 \over 3 - 2 \epsilon} \,
{1 \over (p+i\varepsilon \pm q)^2} \,
F\left({{3 \over 2},1 \atop {5 \over 2} - \epsilon} \Bigg|
	{r^2 \over (p+i\varepsilon \pm q)^2} \right)\,,
\label{avec:c2}
\\
\left\langle {c^4 \over (p+i\varepsilon \pm q)^2 - r^2 c^2}
	\right\ranglec &=&
{3 \over (3 - 2 \epsilon)(5 - 2 \epsilon)} \,
{1 \over (p+i\varepsilon \pm q)^2} \,
F\left({{5 \over 2},1 \atop {7 \over 2} - \epsilon} \Bigg|
	{r^2 \over (p+i\varepsilon \pm q)^2} \right)\, .
\label{avec:c4}
\end{eqnarray}
%

In the $+q$ case of (\ref{avec:c2}),
the $i\varepsilon$ prescription is unnecessary.
The argument of the hypergeometric function can be written $1 - \beta y$,
where $y = (1-x)/2$.
After using a transformation formula to change the argument to $\beta y$,
we can evaluate the angular average over $x$ to obtain hypergeometric
functions with argument $\beta$.
For example, the average over $x$ of (\ref{avec:c2}) is
\begin{eqnarray}
\left\langle F\left( {{3 \over 2}, 1 \atop {5 \over 2} - \epsilon}
	\Bigg| {r^2 \over (p+q)^2} \right) \right\ranglex &=&
- {3-2\epsilon \over 2 \epsilon}
\left[ F\left( { 1-\epsilon , {3 \over 2} , 1
		\atop 2-2\epsilon , 1+\epsilon } \Bigg| \beta \right)
\right.
\nonumber
\\
&& \hspace{1cm} \left. \;-\;
{ (1)_\epsilon (1)_{-2\epsilon} (2)_{-2 \epsilon} ({3\over2})_{-\epsilon}
	\over (1)_{-\epsilon} (2)_{-3\epsilon} }
\beta^{-\epsilon}
F\left( { 1-2\epsilon , {3\over2}-\epsilon
		\atop 2-3\epsilon } \Bigg| \beta \right)
\right] \, ,
\end{eqnarray}
%
where $(a)_b$ is Pochhammer's symbol which is defined in (\ref{Poch}).
Integrating over $\beta$, we obtain hypergeometric
functions with argument 1:
\begin{eqnarray}
&& s^2 \int_0^1 d\beta \, \beta^{-2\epsilon} (1-\beta)^{-1/2}
\left\langle {c^2 \over (p+q)^2 - r^2 c^2}
	\right\ranglecx \;=\;
- {1 \over 4\epsilon}
{ (1)_\epsilon (2)_{-2\epsilon} \over (1)_{-\epsilon} }
\nonumber
\\
&& \hspace{3cm} \times
\left[ { (1)_{-2\epsilon} (1)_{-\epsilon}
	\over ({3\over2})_{-2\epsilon} (2)_{-2\epsilon} (1)_{\epsilon} }
F\left( { 1-2\epsilon , 1-\epsilon , {3 \over 2} , 1
	\atop {3\over2}-2\epsilon , 2-2\epsilon , 1+\epsilon }
	\Bigg| 1 \right)
\right.
\nonumber
\\
&& \hspace{4cm} \left. \;-\;
{ (1)_{-3\epsilon} (1)_{-2\epsilon} ({3\over2})_{-\epsilon}
	\over ({3\over2})_{-3\epsilon} (2)_{-3\epsilon} }
F\left( { 1-3\epsilon , 1-2\epsilon , {3\over2}-\epsilon
		\atop {3\over2}-3\epsilon,2-3\epsilon } \Bigg| 1 \right)
\right] \, .
\label{FF:1}
\end{eqnarray}
%
The integral weighted by $c^4$ can be evaluated in a similar way.
Expanding in powers of $\epsilon$,
we obtain
\begin{eqnarray}
s^2 \int_0^1 d\beta \, \beta^{-2\epsilon} (1-\beta)^{-1/2}
\left\langle {c^2 \over (p+q)^2 - r^2 c^2}
	\right\ranglecx &=&
{\pi^2 \over 24} (1 + 3.54518 \, \epsilon) \, ,
\label{intavecx:1p}
\\
s^2 \int_0^1 d\beta \, \beta^{-2\epsilon} (1-\beta)^{-1/2}
\left\langle {c^4 \over (p+q)^2 - r^2 c^2}
	\right\ranglecx &=&
{\pi^2 \over 72} (1 + 10.8408 \, \epsilon) \, .
\label{intavecx:2p}
\end{eqnarray}
%

In the $-q$ case of (\ref{avec:c2}),
the argument of the hypergeometric functions can be written
$(1-\beta y)/(1-\beta \pm i \varepsilon)$, where $y=(1-x)/2$
and the prescriptions $+i \varepsilon$ and $-i \varepsilon$
correspond to the regions $p>q$ and $p<q$, respectively.
These regions correspond to the two terms
inside the average over $x$ in (\ref{int:n2f}).
In  order to obtain an analytic result in terms of hypergeometric functions,
it is necessary to integrate over $\beta$ before averaging over $x$.
The integrals over $\beta$ can be evaluated by first using a
transformation formula to change the argument of the hypergeometric
function to $-\beta(1-y)/(1-\beta)$ and then using
the integration formula (\ref{int-2F1}) to obtain hypergeometric
functions with arguments $y$ or $1-y$:
\begin{eqnarray}
&&\int_0^1 d\beta \, \beta^{-2\epsilon} (1-\beta)^{-3/2}
F\left( { {3\over2}, 1 \atop {5\over2}-\epsilon }
	\Bigg| {1-\beta y \over 1-\beta + i \varepsilon} \right)
\nonumber
\\
&&  \hspace{2cm} \;=\;
{3-2\epsilon \over \epsilon} \,
{(1)_{-2\epsilon} \over ({1\over2})_{-2\epsilon}} \,
F\left( { 1-2\epsilon, 1
		\atop 1+\epsilon } \Bigg| 1-y \right)
\nonumber
\\
&& \hspace{2.5cm}
- {3-2\epsilon \over \epsilon} \,
{ (1)_{\epsilon} \over ({1\over2})_{\epsilon} } \,
(1-y)^{-1/2}
F\left( { {1\over2}-2\epsilon, 1 \atop {1\over2}+\epsilon }
	\Bigg| 1-y \right)
\nonumber
\\
&& \hspace{2.5cm}
+ {3 \over 2\epsilon(1-3\epsilon)} e^{i\pi \epsilon} \,
(1)_{\epsilon} (\mbox{${5\over2}$})_{-\epsilon}\,
(1-y)^{-\epsilon}
F\left( { 1-3\epsilon, {3\over2}-\epsilon \atop 2-3\epsilon }
	\Bigg| y \right) \, .
\end{eqnarray}
%
After averaging over $x$, we obtain hypergeometric
functions with argument 1:
\begin{eqnarray}
&& s^2 \int_0^1 d\beta \, \beta^{-2\epsilon} (1-\beta)^{-1/2}
\left\langle {c^2 \over (p+i\varepsilon - q)^2 - r^2 c^2}
	\right\ranglecx
\nonumber
\\
&&  \hspace{1cm} \;=\;
{1 \over 4\epsilon}\,
{(1)_{-2\epsilon} \over ({1\over2})_{-2\epsilon}} \,
F\left( { 1-\epsilon, 1-2\epsilon, 1
		\atop 2-2\epsilon, 1+\epsilon } \Bigg| 1 \right)
\nonumber
\\
&& \hspace{1.5cm}
- {1 \over 2\epsilon}\,
{ (2)_{-2\epsilon} (1)_{\epsilon} ({1\over2})_{-\epsilon}
	\over (1)_{-\epsilon} ({1\over2})_{\epsilon}
			({3\over2})_{-2\epsilon} } \,
F\left( { {1\over2}-\epsilon,{1\over2}-2\epsilon,1
		\atop {3\over2}-2\epsilon,{1\over2}+\epsilon }
	\Bigg| 1 \right)
\nonumber
\\
&& \hspace{1.5cm}
+ {1 \over 8\epsilon(1-3\epsilon)} \, e^{i\pi \epsilon}
{ (2)_{-2\epsilon} (1)_{-2\epsilon} (1)_{\epsilon}({3\over2})_{-\epsilon}
	\over (1)_{-\epsilon} (2)_{-3\epsilon} } \,
F\left( { 1-\epsilon, 1-3\epsilon, {3\over2}-\epsilon
		\atop 2-3\epsilon, 2-3\epsilon }
	\Bigg| 1 \right) \, .
\end{eqnarray}
%
The integral weighted by $c^4$ can be evaluated in a similar way.
Expanding in powers of $\epsilon$ and then taking the real parts,
we obtain
\begin{eqnarray}
{\rm Re} \, s^2 \int_0^1 d\beta \, \beta^{-2\epsilon} (1-\beta)^{-1/2}
\left\langle {c^2 \over (p+i\varepsilon - q)^2 - r^2 c^2}
	\right\ranglecx &=&
- {\pi^2 \over 24} (1 + 0.34275 \, \epsilon) \, ,
\label{intavecx:1m}
\\
{\rm Re} \, s^2 \int_0^1 d\beta \, \beta^{-2\epsilon} (1-\beta)^{-1/2}
\left\langle {c^4 \over (p+i\varepsilon - q)^2 - r^2 c^2}
	\right\ranglecx &=&
- {12+\pi^2 \over 72} (1 + 1.10518 \, \epsilon) \, .
\label{intavecx:2m}
\end{eqnarray}
%
Inserting the sum of the integrals (\ref{intavecx:1p}) and (\ref{intavecx:1m})
into the thermal integral (\ref{int:n2f})
and similarly for the integrals weighted by $c^4$, we obtain
\begin{eqnarray}
&& \int_{\bf pq}
{s^2 n^2(s) \over p^2 q^2}
{\rm Re} \left\langle c^2 {r^2 c^2 - p^2 - q^2 \over
	\Delta(p+i\varepsilon,q,r c)}  \right\ranglec \;=\;
{T^2 \over (4\pi)^2} \left[ \, 0.133434 \, \right] \, , 
\label{intHTL:1d}
\\
&& \int_{\bf pq}
{s^2 n^2(s) \over p^2 q^2}
{\rm Re} \left\langle c^4 {r^2 c^2 - p^2 - q^2 \over
	\Delta(p+i\varepsilon,q,r c)}  \right\ranglec \;=\;
{T^2 \over (4\pi)^2}  \left({\mu\over4\pi T}\right)^{4\epsilon}
\left(-{1\over6 \pi^2} \right)
\left[ {1\over\epsilon} + 7.0292 \right] \, ,
\label{intHTL:2d}
\end{eqnarray}
%
Adding these integrals to the subtracted integrals in
(\ref{intHTL:1f})--(\ref{intHTL:2f}), we obtain the final results in
(\ref{intHTL:1})--(\ref{intHTL:2}).

To evaluate the subtraction in the integral (\ref{intHTL:3f}),
we use the identity
$q^2 = (r^2 + q^2 - p^2 - 2 {\bf p}\cdot{\bf q})/2$.
The integral with $q^2-p^2$ in the numerator is purely imaginary.
Thus the real part of the integral can be expressed as
\begin{eqnarray}
&& \int_{\bf pq} {s^2 n^2(s) \over p^2 q^2} {q^2 \over r^2}
{\rm Re} \left\langle c^2 {r^2 c^2 - p^2 - q^2 \over
	\Delta(p+i\varepsilon,q,r c)}  \right\ranglec
	\nonumber
	\\
&& \hspace{3cm}\;=\;
\int_{\bf pq} {s^2 n^2(s) \over p^2 q^2}
	\left( {1\over2} - {{\bf p}\cdot{\bf q} \over r^2} \right)
{\rm Re} \left\langle c^2 {r^2 c^2 - p^2 - q^2 \over
	\Delta(p+i\varepsilon,q,r c)}  \right\ranglec  \, .
\label{intHTL:3a}
\end{eqnarray}
%
It remains only to evaluate the integral with ${\bf p}\cdot{\bf q}$
in the numerator.
We begin by using the identity
\begin{eqnarray}
\left\langle c^2 \, {{\bf p}\cdot{\bf q} \over r^2} \,
{r^2 c^2 - p^2 - q^2 \over
	\Delta(p+i\varepsilon,q,r c)} \right\ranglecx
&=&
- {p^2+q^2 \over (p^2 - q^2 +i\varepsilon)^2} \langle c^2 \rangle_c
	\left\langle {{\bf p}\cdot{\bf q} \over r^2} \right\ranglex
\nonumber
\\
&&
\;-\; {1 \over 2} \sum_\pm  {1 \over (p+i\varepsilon \pm q)^2} \,
	\left\langle {{\bf p}\cdot{\bf q} \, c^4
			\over (p+i\varepsilon \pm q)^2 - r^2 c^2}
		\right\ranglecx\, .
\label{pqrdelta}
\end{eqnarray}
%
In the first term on the right side,
the average over $c$ is a simple multiplicative factor:
$\langle  c^2\rangle_c = 1/(3-2\epsilon)$.
The average over $x$ gives hypergeometric functions of argument $\beta$:
\begin{equation}
\left\langle { {\bf p}\cdot{\bf q} \over r^2 } \right\ranglex \;=\;
{1 \over 8} \beta
\left[
F\left( { 1-\epsilon, 1 \atop 3-2\epsilon } \Bigg| \beta \right)
- F\left( { 2-\epsilon, 1 \atop 3-2\epsilon } \Bigg| \beta \right)
\right]\, .
\end{equation}
%
The integral over $\beta$ gives hypergeometric functions of argument 1:
\begin{eqnarray}
&&s^2 \int_0^1 d\beta \, \beta^{-2\epsilon} (1-\beta)^{-1/2}
{p^2 + q^2 \over (p^2 - q^2)^2}
\left\langle { {\bf p}\cdot{\bf q} \over r^2 } \right\ranglex
\nonumber
\\
&& \hspace{1cm} \;=\;
- {1 \over 8} \,
{ (2)_{-2\epsilon} \over ({3\over2})_{-2\epsilon} } \,
\left[
F\left( { 2-2\epsilon, 1-\epsilon,1
		\atop {3\over2}-2\epsilon, 3-2\epsilon }
	\Bigg| 1 \right)
- F\left( { 2-2\epsilon, 2-\epsilon,1
		\atop {3\over2}-2\epsilon, 3-2\epsilon }
	\Bigg| 1 \right) \right]
\nonumber
\\
&& \hspace{2cm}
+ {1 \over 12} \,
{ (3)_{-2\epsilon} \over ({5\over2})_{-2\epsilon} } \,
\left[
F\left( { 1-\epsilon, 1 \atop {5\over2}-2\epsilon } \Bigg| 1 \right)
- F\left( { 2-\epsilon, 1 \atop {5\over2}-2\epsilon } \Bigg| 1 \right)
\right]
\, .
\end{eqnarray}
%
Expanding in powers of $\epsilon$, we obtain
\begin{equation}
s^2 \int_0^1 d\beta \, \beta^{-2\epsilon} (1-\beta)^{-1/2}
{p^2 + q^2 \over (p^2 - q^2)^2}
\left\langle { {\bf p}\cdot{\bf q} \over r^2 } \right\ranglex \;=\;
- {\pi^2 \over 16} [ 1 - 1.02148 \, \epsilon ] \, .
\label{intavecx:3}
\end{equation}
%

In the second term of (\ref{pqrdelta}), the average over $c$ is given by
(\ref{avec:c4}). In the $+q$ term, the average over
$x= \hat{\bf p} \cdot \hat{\bf q}$ is
\begin{eqnarray}
&&\left\langle
x F\left( { 1, {5\over2} \atop {7\over2}-\epsilon }
	\Bigg| {r^2 \over (p+q)^2} \right)
\right\ranglex \;=\;
{5-2\epsilon \over 4\epsilon}\,
\left[
F\left( { 2-\epsilon, 1, {5\over2}
	\atop 3-2\epsilon, 1+\epsilon } \Bigg| \beta \right)
- F\left( { 1-\epsilon, 1, {5\over2}
	\atop 3-2\epsilon, 1+\epsilon } \Bigg| \beta \right)
\right]
\nonumber
\\
&& \hspace{0cm}
+ {5 \over 4\epsilon}\,
{ (1)_{\epsilon} (1)_{-2\epsilon} (3)_{-2\epsilon} ({7\over2})_{-\epsilon}
	\over (1)_{-\epsilon} (3)_{-3\epsilon} } \,
\beta^{-\epsilon}
\left[
F\left( { 1-2\epsilon, {5\over2}-\epsilon
	\atop 3-3\epsilon } \Bigg| \beta \right)
- {1-2\epsilon \over 1-\epsilon}
F\left( { 2-2\epsilon, {5\over2}-\epsilon
		\atop 3-3\epsilon } \Bigg| \beta \right)
\right] \, .
\end{eqnarray}
%
Integrating over $\beta$,
we obtain hypergeometric functions of argument 1:
\begin{eqnarray}
&& \int_0^1 d\beta \, \beta^{-2\epsilon} (1-\beta)^{-1/2} \,
\left\langle {{\bf p}\cdot{\bf q} \, c^4 \over (p+q)^2 - r^2 c^2}
	\right\ranglecx
\nonumber
\\
&& \hspace{1cm} \;=\;
{1 \over 4\epsilon(3-2\epsilon)}\,
{(2)_{-2\epsilon} \over ({5\over2})_{-2\epsilon}} \,
\left[
F\left( { 2-2\epsilon, 2-\epsilon, 1, {5\over2}
	\atop {5\over2}-2\epsilon, 3-2\epsilon, 1+\epsilon } \Bigg| 1 \right)
- F\left( { 2-2\epsilon, 1-\epsilon, 1, {5\over2}
	\atop {5\over2}-2\epsilon, 3-2\epsilon, 1+\epsilon } \Bigg| 1 \right)
\right]
\nonumber
\\
&& \hspace{1.5cm}
+ {1 \over 6\epsilon(2-3\epsilon)}\,
{ (1)_{\epsilon} (1)_{-2\epsilon} (3)_{-2\epsilon} ({3\over2})_{-\epsilon}
	\over (1)_{-\epsilon} ({5\over2})_{-3\epsilon} } \,
\nonumber
\\
&& \hspace{2cm} \times \left[
F\left( { 2-3\epsilon, 1-2\epsilon, {5\over2}-\epsilon
		\atop {5\over2}-3\epsilon, 3-3\epsilon }
	\Bigg| 1 \right)
- {1-2\epsilon \over 1-\epsilon}
F\left( { 2-3\epsilon, 2-2\epsilon, {5\over2}-\epsilon
		\atop {5\over2}-3\epsilon, 3-3\epsilon }
	\Bigg| 1 \right)
\right] \, .
\end{eqnarray}
%
Expanding in powers of $\epsilon$, we obtain
\begin{eqnarray}
\int_0^1 d\beta \, \beta^{-2\epsilon} (1-\beta)^{-1/2}  \,
\left\langle {{\bf p}\cdot{\bf q} \, c^4 \over (p+q)^2 - r^2 c^2}
	\right\ranglecx &=&
{\pi^2 - 6 \over 18} (1-0.0728428 \, \epsilon)\, .
\label{intavecx:4p}
\end{eqnarray}
%

In the $-q$ term in the integral of the second term of (\ref{pqrdelta}),
we integrate over $\beta$ before averaging over $x$.
The integral over $\beta$ can be expressed in terms of
hypergeometric functions of type $_2F_1$:
\begin{eqnarray}
&& s^2 \int_0^1 d\beta \, \beta^{-2\epsilon} (1-\beta)^{-1/2}  \,
{4 {\bf p}\cdot{\bf q} \over (p-q)^2}
\left\langle {c^4 \over (p+i\varepsilon - q)^2 - r^2 c^2}
	\right\ranglec
\nonumber
\\
&& \hspace{2cm} \;=\;
- {1 \over 2(3-2\epsilon)\epsilon} \,
{(2)_{-2\epsilon} \over ({1\over2})_{-2\epsilon}} \,
(1-2y) \,
F\left( { 2-2\epsilon, 1 \atop 1+\epsilon } \Bigg| 1-y \right)
\nonumber
\\
&& \hspace{3cm}
- {1 \over 4(3-2\epsilon)\epsilon}  \,
{(1)_{\epsilon} \over (-{1\over2})_{\epsilon}}
(1-2y) \, (1-y)^{-3/2}  \,
F\left( { {1\over2}-2\epsilon, 1 \atop -{1\over2} +\epsilon } \Bigg| 1-y \right)
\nonumber
\\
&& \hspace{3cm}
+ {1 \over 8(2-3\epsilon)\epsilon}
e^{\mp i \pi \epsilon} (1)_\epsilon (\mbox{$3\over2$})_{-\epsilon} \,
(1-2y) \, (1-y)^{-\epsilon} \,
F\left( { 2-3\epsilon, {5\over2}-\epsilon
	\atop 3-3\epsilon } \Bigg| y \right)
\, .
\end{eqnarray}
%
The phase in the last term is $e^{-i \pi \epsilon}$ for the
$f(s_+,s_-,r)$ term of (\ref{int:n2f}), which comes from the $p>q$ region
of the integral, and $e^{i \pi \epsilon}$
for the $f(s_-,s_+,r)$ term, which comes from the $p<q$ region.
The average over $x=\hat{\bf p} \cdot \hat{\bf q}$
can be expressed in terms of
hypergeometric functions of type $_3F_2$ evaluated at 1:
\begin{eqnarray}
&& s^2 \int_0^1 d\beta \, \beta^{-2\epsilon} (1-\beta)^{-1/2}  \,
\left\langle {4 {\bf p} \cdot {\bf q} \over (p-q)^2} \,
	{c^4 \over (p+i\varepsilon - q)^2 - r^2 c^2}
	\right\ranglecx
\nonumber
\\
&& \hspace{1cm} \;=\;
{1 \over 4(3-2\epsilon)\epsilon} \,
{(2)_{-2\epsilon}\over ({1\over2})_{-2\epsilon}}
\left[ F\left( { 1-\epsilon,2-2\epsilon, 1
		\atop 3-2\epsilon,1+\epsilon } \Bigg| 1 \right)
- F\left( { 2-\epsilon,2-2\epsilon, 1
		\atop 3-2\epsilon,1+\epsilon } \Bigg| 1 \right) \right]
\nonumber
\\
&& \hspace{2cm}
- {1 \over (3-2\epsilon)\epsilon} \,
{ (1)_{\epsilon} (3)_{-2\epsilon} (-{1\over2})_{-\epsilon}
\over (1)_{-\epsilon} (-{1\over2})_{\epsilon} ({3\over2})_{-2\epsilon} }
\left[ F\left( { -{1\over2}-\epsilon, {1\over2}-2\epsilon, 1
	\atop {3\over2}-2\epsilon, -{1\over2}+\epsilon } \Bigg| 1 \right)
\right.
\nonumber
\\
&& \hspace{8cm} \left.
+ {1+2\epsilon \over 2(1-\epsilon)}
	F\left( { {1\over2}-\epsilon, {1\over2}-2\epsilon, 1
	\atop {3\over2}-2\epsilon, -{1\over2}+\epsilon } \Bigg| 1 \right)
\right]
\nonumber
\\
&& \hspace{2cm}
+ {1 \over 16(2-3\epsilon)\epsilon} \, e^{\mp i \pi \epsilon} \,
{ (1)_{\epsilon} (2)_{-2\epsilon} (2)_{-2\epsilon} ({3\over2})_{-\epsilon}
	\over (1)_{-\epsilon} (3)_{-3\epsilon} }
\left[ F\left( { 1-\epsilon,  2-3\epsilon, {5\over2}-\epsilon
		\atop 3-3\epsilon, 3-3\epsilon } \Bigg| 1 \right)
\right.
\nonumber
\\
&& \left. \hspace{8cm}
- {1-\epsilon \over 1-2\epsilon}
	F\left( { 2-\epsilon,  2-3\epsilon, {5\over2}-\epsilon
		\atop 3-3\epsilon, 3-3\epsilon } \Bigg| 1 \right)
\right]  \, .
\end{eqnarray}
%
The expansion of the real part of the integral in powers of $\epsilon$ is
\begin{eqnarray}
&&s^2 \int_0^1 d\beta \, \beta^{-2\epsilon} (1-\beta)^{-1/2} \,
{\rm Re} \left\langle {4 {\bf p} \cdot {\bf q} \over (p-q)^2} \,
	{c^4 \over (p+i\varepsilon - q)^2 - r^2 c^2}
	\right\ranglecx
\nonumber
\\
&& \hspace{8cm}
\;=\; {9-\pi^2\over18} \left[ 1 - 0.796579 \, \epsilon \right] \, .
\label{intavecx:4m}
\end{eqnarray}
%
Inserting (\ref{intavecx:3}), (\ref{intavecx:4p}), and (\ref{intavecx:4m})
into the thermal integral of (\ref{pqrdelta}), we obtain
\begin{eqnarray}
&& \int_{\bf pq}
{s^2 n^2(s) \over p^2 q^2} {{\bf p}\cdot{\bf q} \over r^2}
{\rm Re} \left\langle c^2 {r^2 c^2 - p^2 - q^2 \over
	\Delta(p+i\varepsilon,q,r c)}  \right\ranglec \;=\;
{T^2 \over (4\pi)^2}  \left({\mu\over4\pi T}\right)^{4\epsilon}
{1-\pi^2\over24\pi^2} \left[ {1\over\epsilon} + 13.52098 \right] \,.
\end{eqnarray}
%
Inserting this along with (\ref{intHTL:1d}) into (\ref{intHTL:3a}),
we obtain
\begin{eqnarray}
&& \int_{\bf pq}
{s^2 n^2(s) \over p^2 r^2}
{\rm Re} \left\langle c^2 {r^2 c^2 - p^2 - q^2 \over
	\Delta(p+i\varepsilon,q,r c)}  \right\ranglec \;=\;
{T^2 \over (4\pi)^2}  \left({\mu\over4\pi T}\right)^{4\epsilon}
{\pi^2-1\over24\pi^2} \left[ {1\over\epsilon} + 15.302796 \right] \,.
\label{intHTL:3d}
\end{eqnarray}
%
Adding this integral to the subtracted integral in
(\ref{intHTL:3f}), we obtain the final result in (\ref{intHTL:3}).

To evaluate the weighted averages over $c$ of the thermal integrals
in (\ref{intHTL:4})--(\ref{intHTL:7}), we first isolate
the divergent parts, which arise from the region $q \to 0$.
For the integrals (\ref{intHTL:4}) and (\ref{intHTL:5}),
a single subtraction of the thermal distribution $n(q)$
suffices to remove the divergences:
\begin{equation}
n(q) \;=\;
\left( n(q) - {T \over q} \right) \;+\; {T \over q} \, .
\label{nsub-1}
\end{equation}
%
For the integral (\ref{intHTL:6}),
a second subtraction is also needed to remove the divergences:
\begin{equation}
n(q) \;=\;
\left( n(q) - {T \over q} + {1 \over 2} \right)
\;+\; {T \over q} \;-\; {1 \over 2} \, .
\label{nsub-2}
\end{equation}
%
In the last integral (\ref{intHTL:7}), it is convenient to first use the
identity $r_c^2 = p^2 + 2 {\bf p} \cdot {\bf q}/c + q^2/c^2$ to expand
it into 3 integrals, two of which are (\ref{intHTL:4}) and (\ref{intHTL:6}).
In the third integral, the subtraction (\ref{nsub-2})
is needed to remove the divergences.
For the convergent terms, the HTL average over
$c$ and the angular average over $x = \hat {\bf p} \cdot \hat {\bf q}$
can be calculated in 3 dimensions:
\begin{eqnarray}
{\rm Re}\left\langle c^{-1}
	{r_c^2 - p^2 - q^2 \over \Delta(p+i\varepsilon,q,r_c)} \,
	\right\ranglecx &=&
{1 \over 4p^2 - q^2} \log{2p \over q}
\nonumber
\\
&& + {1 \over 4 p q} \left( {p+q \over 2p+q} \log{p+q \over p}
			- {p-q \over 2p-q} \log{|p-q| \over p} \right)
\, ,
\\
{\rm Re} \left\langle c
	{r_c^2 - p^2 - q^2 \over \Delta(p+i\varepsilon,q,r_c)} \,
	\right\ranglecx &=&
{1 \over 6(4p^2 - q^2)}
+ {q^2(4p^2+3q^2) \over 3(4p^2 - q^2)^3} \log{2p \over q}
\nonumber
\\
&& \hspace{-4cm}
+ {(p+q)(4p^2+2p q+q^2) \over 12 p q(2p+q)^3} \log{p+q \over p}
- {(p-q)(4p^2-2p q+q^2) \over 12 p q(2p-q)^3} \log{|p-q| \over p}
\, ,
\\
{\rm Re} \left\langle \hat{\bf p} \cdot \hat{\bf q}
	{r_c^2 - p^2 - q^2 \over \Delta(p+i\varepsilon,q,r_c)} \,
	\right\ranglecx &=&
{1 \over 6pq}
- {q(12p^2-q^2) \over 6p (4p^2 - q^2)^2} \log{4p \over q}
\nonumber
\\
&& \hspace{-4cm}
+ {(p+q)(2p^2-2p q-q^2) \over 12 p^2q(2p+q)^2} \log{p+q \over 4p}
+ {(p-q)(2p^2+2p q-q^2) \over 12 p^2q(2p-q)^2} \log{|p-q| \over 4p} \,.
\end{eqnarray}
%
The remaining 2-dimensional integral over $p$ and $q$
can then be evaluated numerically:
\begin{eqnarray}
&& \int_{\bf pq} {n(p)\over p}
\left( {n(q) \over q} - {T \over q^2} \right)
{\rm Re} \left\langle c^{-1}
	{r_c^2 - p^2 - q^2 \over \Delta(p+i\varepsilon,q,r_c)} \,
	\right\ranglec \;=\;
\left( - 5.113 \times 10^{-1} \right) {T^2 \over (4 \pi)^2} \, ,
\label{intHTL:4f}
\\
&& \int_{\bf pq} {n(p)\over p}
\left( {n(q) \over q} - {T \over q^2} \right)
{\rm Re}\left\langle c
	{r_c^2 - p^2 - q^2 \over \Delta(p+i\varepsilon,q,r_c)} \,
	\right\ranglec \;=\;
\left( - 2.651 \times 10^{-1} \right) {T^2 \over (4 \pi)^2} \, ,
\label{intHTL:5f}
\\
&& \int_{\bf pq} {n(p)\over p}
\left( {n(q) \over q} - {T \over q^2} + {1 \over 2 q} \right)
{p^2 \over q^2} \,
{\rm Re}\left\langle c
	{r_c^2 - p^2 - q^2 \over \Delta(p+i\varepsilon,q,r_c)} \,
	\right\ranglec \;=\;
\left( 2.085 \times 10^{-2} \right) {T^2 \over (4 \pi)^2} \, ,
\label{intHTL:6f}
\\
&& \int_{\bf pq} {n(p)\over p}
\left( {n(q) \over q} - {T \over q^2} + {1 \over 2 q} \right)
{{\bf p} \cdot {\bf q} \over q^2} \,
{\rm Re} \left\langle
	{r_c^2 - p^2 - q^2 \over \Delta(p+i\varepsilon,q,r_c)} \,
	\right\ranglec
\nonumber
\\
&& \hspace{9cm} \;=\;
\left( - 3.729 \times 10^{-3} \right) {T^2 \over (4 \pi)^2} \, .
\label{intHTL:8f}
\end{eqnarray}
%
The integrals involving the terms subtracted from $n(q)$
in (\ref{nsub-1}) and (\ref{nsub-2})
are divergent, so the HTL average over
$c$ and the angular average over $x = \hat {\bf p} \cdot \hat {\bf q}$
must be calculated in $3-2\epsilon$ dimensions.
The first step in the calculation of the subtracted terms
is to replace the average over $c$ of the integral over $q$
by an average over $c$ and $x$:
\begin{eqnarray}
\int_{\bf q} {1 \over q^n} \,
\left\langle f(c)
{r_c^2 - p^2 - q^2 \over \Delta(p+i\varepsilon,q,r_c)} \right\ranglec
& = &
(-1)^{n-1} {1\over 8 \pi^2 \epsilon}
{ (1)_{2 \epsilon} (1)_{-2\epsilon}
	\over ({3\over2})_{-\epsilon} }
(e^\gamma \mu^2)^\epsilon (2p)^{1-n-2\epsilon}
\nonumber
\\
&& \hspace{-1cm}
\times \left\langle f(c) \, c^{3-n-2\epsilon}(1-c^2)^{n-2+2\epsilon}
\sum_\pm (x\mp c - i \varepsilon)^{1-n-2\epsilon} \right\ranglecx \, .
\label{intthq}
\end{eqnarray}
%
The integral over $p$ can now be evaluated easily
using either (\ref{int-th:-1}) or
\begin{equation}
\int_{\bf p} n(p) \, p^{-2-2\epsilon} =
{1\over 2\pi^2}
{(1)_{-4\epsilon} \over ({3\over2})_{-\epsilon}}
\zeta(1-4\epsilon)
(e^\gamma \mu^2)^\epsilon T^{1-4\epsilon} \, .
\label{int-th:-2}
\end{equation}
%
It remains only to calculate the averages
over $c$ and $x$.  The averages over $x$ give $_2F_1$ hypergeometric
functions with argument $[(1 \mp c)/2 - i \varepsilon]^{-1}$:
\begin{eqnarray}
\left\langle  (x\mp c - i \varepsilon)^{-n-2\epsilon}
	\right\ranglex &=&
(1\mp c)^{-n-2\epsilon}
F\left( { 1-\epsilon,n+2\epsilon \atop 2-2\epsilon}
	\Bigg| [(1 \mp c)/2 - i \varepsilon]^{-1} \right) \,,
\label{avex:1}
\\
\left\langle x (x\mp c - i \varepsilon)^{-n-2\epsilon}
	\right\ranglex &=&
{1 \over 2} (1\mp c)^{-n-2\epsilon}
\left[ F\left( { 1-\epsilon,n+2\epsilon \atop 3-2\epsilon}
	\Bigg| [(1 \mp c)/2 - i \varepsilon]^{-1} \right) \right.
\nonumber
\\
&& \hspace{3cm}
\left. -
F\left( { 2-\epsilon,n+2\epsilon \atop 3-2\epsilon}
	\Bigg| [(1 \mp c)/2 - i \varepsilon]^{-1} \right) \right] \, .
\label{avex:2}
\end{eqnarray}
%
Using a transformation formula, the arguments
can be changed to $(1 \mp c)/2 - i \varepsilon$.
If the expressions (\ref{avex:1}) and (\ref{avex:2})
are averaged over $c$ with a weight that is an even
function of $c$, the $+$ and $-$ terms combine to give
$_3F_2$ hypergeometric functions with argument 1.
For example,
\begin{eqnarray}
\left\langle (1-c^2)^{2 \epsilon}
	\sum_\pm (x\mp c - i \varepsilon)^{-1-2\epsilon}
	\right\ranglecx &=&
{1 \over 3\epsilon}
{ (2)_{-2 \epsilon} (1)_\epsilon ({3\over2})_{-\epsilon}
	\over (1)_{-\epsilon} (1)_{-\epsilon} }
\nonumber
\\
&&  \hspace{-1cm} \times
\left\{ - e^{- i \pi \epsilon}
	{ (1)_{3\epsilon} (1)_{-2 \epsilon}
	\over (1)_{2 \epsilon} (2)_{-\epsilon} }
F\left( { 1-2\epsilon,1-\epsilon,\epsilon
	\atop 2-\epsilon,1-3\epsilon} \Bigg| 1 \right)
\right.
\nonumber
\\
&&  \left.
+ e^{i 2\pi \epsilon}
	{ (1)_{-3\epsilon} (1)_\epsilon
	\over  (1)_{-4\epsilon} (2)_{2\epsilon} }
F\left( { 1+\epsilon,1+2\epsilon,4\epsilon
	\atop 2+2\epsilon,1+3\epsilon} \Bigg| 1 \right)
\right\}	\,.
\end{eqnarray}
%
Upon expanding the hypergeometric functions in powers of
$\epsilon$ and taking the real parts, we obtain
\begin{eqnarray}
{\rm Re} \left\langle (1-c^2)^{2 \epsilon}
	\sum_\pm (x\mp c - i \varepsilon)^{-1-2\epsilon}
	\right\ranglecx &=&
\pi^2 \left[ - \epsilon + 2 (1-\log 2) \epsilon^2 \right] \, ,
\label{avecx:1}
\\
{\rm Re} \left\langle c^2 (1-c^2)^{2 \epsilon}
	\sum_\pm (x\mp c - i \varepsilon)^{-1-2\epsilon}
	\right\ranglecx &=&
\pi^2 \left[ - {1 \over 3} \epsilon
	+ {2 \over 9} (2-3\log 2) \epsilon^2 \right] \, ,
\;\;
\label{avecx:2}
\\
{\rm Re} \left\langle (1-c^2)^{2+2 \epsilon}
	\sum_\pm (x\mp c - i \varepsilon)^{-3-2\epsilon}
	\right\ranglecx &=&
\pi^2 \left[ - {8 \over 3} \epsilon^2  \right] \, ,
\label{avecx:3}
\\
{\rm Re} \left\langle x (1-c^2)^{1+2 \epsilon}
	\sum_\pm (x\mp c - i \varepsilon)^{-2-2\epsilon}
	\right\ranglecx &=&
\pi^2 \left[ - {2 \over 3} \epsilon
	+ {2 \over 9} (1-6\log 2) \epsilon^2  \right] \, .
\label{avecx:5}
\end{eqnarray}
%
If the expressions (\ref{avex:2}) and (\ref{avex:2})
are averaged over $c$ with a weight that is an odd
function of $c$, they reduce to integrals of $_2F_1$
hypergeometric functions with argument $y$.
For example,
\begin{eqnarray}
&&\left\langle c (1-c^2)^{1+2 \epsilon}
	\sum_\pm (x\mp c - i \varepsilon)^{-2-2\epsilon}
	\right\ranglecx \;=\;
{(2)_{-2\epsilon} ({3\over2})_{-\epsilon}
\over (1)_{-\epsilon} (1)_{-\epsilon}}
\nonumber
\\
&& \hspace{0.5cm} \times \left\{
- 2 e^{-i\pi \epsilon}
{(1)_{3\epsilon} \over (2)_{2\epsilon}}
\int_0^1 dy \, y^{-2\epsilon} (1-y)^{1+\epsilon} |1-2y|
F\left( { 1-\epsilon,\epsilon \atop -3\epsilon } \Bigg| y \right)
\right.
\nonumber
\\
&& \hspace{1.0cm} \left.
- {8 \over 3(1+3\epsilon)} e^{2i\pi \epsilon}
{(1)_{-3\epsilon} \over (1)_{-4\epsilon}}
\int_0^1 dy \, y^{1+\epsilon} (1-y)^{1+\epsilon} |1-2y|
F\left( { 2+2\epsilon,1+4\epsilon \atop 2+3\epsilon } \Bigg| y \right)
\right\} \, .
\end{eqnarray}
%
The expansions of the integrals of the hypergeometric functions
in powers of $\epsilon$ are given in (\ref{Fabs-1})-(\ref{Fabs-2}).
The resulting expansions for the real parts
of the averages over $c$ and $x$ are
\begin{eqnarray}
{\rm Re} \left\langle c (1-c^2)^{1+2 \epsilon}
	\sum_\pm (x\mp c - i \varepsilon)^{-2-2\epsilon}
	\right\ranglecx &=&
-1 + {14(1-\log2) \over3} \epsilon \,,
\label{avecx:4}
\\
{\rm Re} \left\langle x c(1-c^2)^{2 \epsilon}
	\sum_\pm (x\mp c - i \varepsilon)^{-1-2\epsilon}
	\right\ranglecx &=&
{2 (1 - \log 2) \over3}
\nonumber
\\
&& + \left( {4\over9} + {8\over 9} \log 2
- {4\over3}\log^22 + {\pi^2 \over 18} \right) \epsilon \, .
\label{avecx:6}
\end{eqnarray}
%
Multiplying each of these expansions by the appropriate
factors from the integral over $q$ in (\ref{intthq}) and
the integral over $p$ in (\ref{int-th:-2}) or (\ref{int-th:-1}),
we obtain
\begin{eqnarray}
&& \int_{\bf pq} {n(p)\over p} {1 \over q^2}
{\rm Re} \left\langle c^{-1+2\epsilon}
	{r_c^2 - p^2 -  q^2 \over \Delta(p+i\varepsilon,q,r_c)} \,
	\right\ranglec \;=\;
{T \over (4\pi)^2} \left({\mu\over4\pi T}\right)^{4\epsilon}
\nonumber
\\
&& \hspace{5cm}
\times \left(-{1 \over 8}\right)
\left[ {1\over\epsilon} + 2 + 4 \log(2\pi) \right] \, ,
\label{intHTL:4d}
\\
&& \int_{\bf pq} {n(p)\over p} {1 \over q^2}
{\rm Re}\left\langle c^{1+2\epsilon}
	{r_c^2 - p^2 -  q^2 \over \Delta(p+i\varepsilon,q,r_c)} \,
	\right\ranglec \;=\;
{T \over (4\pi)^2}  \left({\mu\over4\pi T}\right)^{4\epsilon}
\nonumber
\\
&& \hspace{5cm}
\times \left(-{1 \over 24}\right)
\left[ {1\over\epsilon} + {8\over3} + 4 \log(2\pi) \right] \, ,
\label{intHTL:5d}
\\
&& \int_{\bf pq} {n(p)\over p} {p^2 \over q^4}
{\rm Re}\left\langle c^{1+2\epsilon}
	{r_c^2 - p^2 -  q^2 \over \Delta(p+i\varepsilon,q,r_c)} \,
	\right\ranglec \;=\;
{T \over (4\pi)^2}  \left(-{1\over 12}\right)  \, ,
\label{intHTL:6da}
\\
&& \int_{\bf pq} {n(p)\over p}  {{\bf p} \cdot {\bf q} \over q^4}
{\rm Re} \left\langle  c^{2\epsilon}
	{r_c^2 - p^2 -  q^2 \over \Delta(p+i\varepsilon,q,r_c)} \,
	\right\ranglec \;=\;
{T \over (4\pi)^2}  \left({\mu\over4\pi T}\right)^{4\epsilon}
\nonumber
\\
&& \hspace{5cm}
\times {1 \over 24}
\left[ {1\over\epsilon} +  {11\over3} + 4 \log(2\pi) \right] \, ,
\label{intHTL:8da}
\\
&& \int_{\bf pq} {n(p)\over p} {p^2 \over q^3}
{\rm Re}\left\langle c^{1+2\epsilon}
	{r_c^2 - p^2 -  q^2 \over \Delta(p+i\varepsilon,q,r_c)} \,
	\right\ranglec \;=\;
{T^2 \over (4\pi)^2}  \left({\mu\over4\pi T}\right)^{4\epsilon}
\nonumber
\\
&& \hspace{5cm}
\times \left(-{1 \over 24}\right)
\left[ {1\over\epsilon} - {2\over3} + {8\over3} \log2
	+4 {\zeta'(-1) \over \zeta(-1)} \right] \, ,
\label{intHTL:6db}
\\
&& \int_{\bf pq} {n(p)\over p} {{\bf p} \cdot {\bf q} \over q^3}
{\rm Re} \left\langle   c^{2\epsilon}
	{r_c^2 - p^2 -  q^2 \over \Delta(p+i\varepsilon,q,r_c)} \,
	\right\ranglec \;=\;
{T^2 \over (4\pi)^2}  \left({\mu\over4\pi T}\right)^{4\epsilon}
\nonumber
\\
&& \hspace{5cm}
\times \left(-{1 \over 18}\right)
\left[ (1-\log2) \left( {1\over\epsilon} + {14\over3}
		+ 4 {\zeta'(-1) \over \zeta(-1)} \right)
	+ {\pi^2 \over 12} \right] \, .
\label{intHTL:8db}
\end{eqnarray}
%
Adding these integrals to the subtracted integrals in
(\ref{intHTL:4f})--(\ref{intHTL:6f}), we obtain the final results in
(\ref{intHTL:4})--(\ref{intHTL:6}).
Combining (\ref{intHTL:8f}) with (\ref{intHTL:8da}) and
(\ref{intHTL:8db}), we obtain
\begin{eqnarray}
\int_{\bf pq} {n(p) n(q) \over p q} {{\bf p} \cdot {\bf q} \over q^2}
{\rm Re} \left\langle c^{2\epsilon}
	{r_c^2 - p^2 - q^2\over \Delta(p+i\varepsilon,q,r_c)} \,
	\right\ranglec &=&
{T^2 \over (4\pi)^2} \left({\mu\over4\pi T}\right)^{4\epsilon}
\nonumber
\\
&& \hspace{1cm}
\times {5-2\log2 \over 72} \left[ {1\over\epsilon} + 11.6689 \right] \, .
\label{intHTL:8}
\end{eqnarray}
%
The final integral (\ref{intHTL:7}) is obtained from
(\ref{intHTL:4}), (\ref{intHTL:6}), and (\ref{intHTL:8})
by using the identity
$r_c^2 = p^2 + 2 {\bf p} \cdot {\bf q}/c + q^2/c^2$.

\subsection{4-dimensional integrals}

In the sum-integral formula (\ref{int-2loop}),
the second term on the right side involves an integral over
4-dimensional Euclidean momenta. The integrands are functions
of the integration variable $Q$ and $R=-(P+Q)$.
The simplest integrals to evaluate are those whose integrands
are independent of $P_0$:
\begin{eqnarray}
\int_Q {1 \over Q^2 r^2} & = &
{1 \over (4 \pi)^2} \mu^{2 \epsilon} p^{-2\epsilon}
\; 2 \left[{1 \over \epsilon} + 4 - 2 \log 2 \right] \,,
\label{int4:1}
\\
\int_Q {q^2 \over Q^2 r^4} & = &
{1 \over (4 \pi)^2} \mu^{2 \epsilon} p^{-2\epsilon}
\;2 \left[{1 \over \epsilon} + 1 - 2 \log 2  \right]\,,
\\
\int_Q {1 \over Q^2 r^4} & = &
{1 \over (4 \pi)^2} \mu^{2 \epsilon} p^{-2-2\epsilon}
\; ( -2 ) \left[1 + (-2 - 2 \log 2) \epsilon \right] \,.
\end{eqnarray}
%
%
Another simple integral that is needed
depends only on $P^2=P_0^2+p^2$:
\begin{eqnarray}
\int_Q {1 \over Q^2 R^2}
& = &
{1 \over (4 \pi)^2} (e^\gamma \mu^2)^\epsilon (P^2)^{-\epsilon} \;
{1 \over \epsilon} \,
{(1)_\epsilon (1)_{-\epsilon} (1)_{-\epsilon}
	\over (2)_{-2\epsilon}} \, ,
\label{int4:8}
\end{eqnarray}
%
where $(a)_b$ is Pochhammer's symbol which is defined in (\ref{Poch}).
We need the following weighted averages over $c$ of this function
evaluated at $P = (-i p,{\bf p}/c)$:
\begin{eqnarray}
\left\langle c^{-1+2\epsilon}
	\int_Q {1 \over Q^2 R^2} \bigg|_{P \to (-i p,{\bf p}/c)}
	\right\ranglec
& = &
{1 \over (4 \pi)^2} \mu^{2 \epsilon} p^{-2\epsilon}
{1 \over 4}
\left[ {1 \over \epsilon^2} + {2 \log 2 \over \epsilon}
	+ 2 \log^2 2 + {3 \pi^2 \over 4} \right]
\, ,
\label{int4:8.1}
\\
\left\langle c^{1+2\epsilon}
	\int_Q {1 \over Q^2 R^2} \bigg|_{P \to (-i p,{\bf p}/c)}
	\right\ranglec
& = &
{1 \over (4 \pi)^2} \mu^{2 \epsilon} p^{-2\epsilon}
{1 \over 2}
\left[ {1 \over \epsilon} + 2 \log 2  \right]
\, .
\label{int4:8.2}
\end{eqnarray}
%

The remaining integrals are functions of $P_0$ that must
be analytically continued to the point $P_0 = -i p + \varepsilon$.
Several of these integrals are straightforward to evaluate:
\begin{eqnarray}
\int_Q {q^2 \over Q^2 R^2}
	\bigg|_{P_0 = -i p} & = & 0 \,,
\label{int4:4}
\\
\int_Q {q^2 \over Q^2 r^2 R^2}
	\bigg|_{P_0 = -i p} & = &
{1 \over (4 \pi)^2} \mu^{2 \epsilon} p^{-2 \epsilon}
(-1) \left[ {1 \over \epsilon^2} + {1 - 2 \log 2 \over \epsilon}
\right. \nonumber
\\
&& \hspace{4cm} \left.
	+ 10 - 2 \log 2 + 2 \log^2 2 - {7 \pi^2 \over 12} \right] \,,
\label{int4:5}
\\
\int_Q {1 \over Q^2 r^2 R^2}
	\bigg|_{P_0 = -i p} & = &
{1 \over (4 \pi)^2} \mu^{2 \epsilon} p^{-2 -2 \epsilon}
\; \left[ {1 \over \epsilon} - 2 - 2 \log 2 \right] \,.
\label{int4:6}
\end{eqnarray}
%
We also need a weighted average over $c$ of the integral in (\ref{int4:4})
evaluated at $P = (-i p, {\bf p}/c)$.  The integral itself is
\begin{eqnarray}
\int_Q {q^2 \over Q^2 R^2}
	\bigg|_{P \to (-i p, {\bf p}/c)} & = &
{1 \over (4 \pi)^2} (e^\gamma \mu^2)^\epsilon p^{2-2\epsilon}
{(1)_\epsilon \over \epsilon}
\nonumber
\\
&& \times {1\over 4}
{(1)_{-\epsilon} (1)_{-\epsilon}
	\over (2)_{-2\epsilon}}
\left( {1 \over 3 - 2 \epsilon} + c^2 \right)
c^{-2 + 2\epsilon} (1-c^2)^{-\epsilon} \,.
\label{int4:7}
\end{eqnarray}
%
The weighted average is
\begin{eqnarray}
\left\langle c^{1+2\epsilon}
	\int_Q {q^2 \over Q^2 R^2} \bigg|_{P \to (-i p,{\bf p}/c)}
	\right\ranglec
& = &
{1 \over (4 \pi)^2} \mu^{2 \epsilon} p^{2-2\epsilon} \,
{1 \over 48}
\left[ {1 \over \epsilon^2} + {2 (10+3\log 2) \over 3\epsilon}
\right.
\nonumber
\\
&& \hspace{2cm} \left.
	+ {4\over 9} + {40\over 3}\log 2 + 2 \log^2 2
	+ {3 \pi^2 \over 4} \right]
\, .
\label{int4:7a}
\end{eqnarray}
%

The most difficult 4-dimensional integrals to evaluate
involve an HTL average of an integral
with denominator $R_0^2 + r^2 c^2$:
\begin{eqnarray}
{\rm Re} \int_Q {1 \over Q^2}
\left\langle {c^2 \over R_0^2 + r^2 c^2} \right\ranglec
&=&
{1 \over (4 \pi)^2} \mu^{2\epsilon} p^{-2 \epsilon}
\left[ {2 - 2\log 2 \over \epsilon}
\right.
\nonumber
\\
&& \hspace{3cm} \left.
+ 8 - 4 \log 2 + 4 \log^2 2
	- {\pi^2 \over 2} \right] \,,
\label{int4HTL:1}
\\
{\rm Re} \int_Q {1 \over Q^2}
\left\langle {c^2(1-c^2) \over R_0^2 + r^2 c^2} \right\ranglec
&=&
{1 \over (4 \pi)^2} \mu^{2\epsilon} p^{-2 \epsilon}
{1 \over 3}
\left[ {1 \over \epsilon} + {20\over3}  -6 \log 2 \right] \, ,
\\
{\rm Re} \int_Q {1 \over Q^2}
\left\langle {c^4 \over R_0^2 + r^2 c^2} \right\ranglec
&=&
{1 \over (4 \pi)^2} \mu^{2\epsilon} p^{-2 \epsilon}
\left[ {5 - 6\log 2 \over 3 \epsilon}
\right.
\nonumber
\\
&& \hspace{3cm} \left.
+ {52 \over 9} - 2 \log 2 + 4 \log^2 2
	- {\pi^2 \over 2} \right] \,,
\label{int4HTL:2}
\\
{\rm Re} \int_Q {1 \over Q^2 r^2}
\left\langle {c^2 \over R_0^2 + r^2 c^2} \right\ranglec
&=&
{1 \over (4 \pi)^2} \mu^{2\epsilon} p^{-2-2 \epsilon}
\left( - {1 \over 4} \right)
\left[ {1 \over \epsilon} + {4\over 3} + {2\over3} \log 2 \right] \, ,
\label{int4HTL:3}
\\
{\rm Re} \int_Q {q^2 \over Q^2 r^2}
\left\langle {c^2 \over R_0^2 + r^2 c^2} \right\ranglec
&=&
{1 \over (4 \pi)^2} \mu^{2\epsilon} p^{-2 \epsilon}
\left[ {13-16\log2 \over 12 \epsilon}
\right.
\nonumber
\\
&& \hspace{3cm} \left.
	+ {29 \over 9} - {19\over18} \log2
		+ {8\over3}\log^22  - {4\over9} \pi^2 \right] \, .
\label{int4HTL:4}
\end{eqnarray}
%
The analytic continuation to $P_0 =-ip+\varepsilon$
is implied in these integrals and in all the 4-dimensional integrals
in the remainder of this subsection.

We proceed to describe the evaluation of the integrals
(\ref{int4HTL:1}) and (\ref{int4HTL:2}).
The integral over $Q_0$ can be evaluated
by introducing a Feynman parameter to combine $Q^2$
and $R_0^2 + r^2 c^2$ into a single denominator:
\begin{eqnarray}
\int_Q {1 \over Q^2 (R_0^2 + r^2 c^2)}
&=&  {1\over4} \int_0^1 dx
\nonumber
\\
&& \hspace{-1cm} \times
\int_{\bf r}\left[ (1-x+xc^2) r^2 + 2(1-x) {\bf r} \!\cdot\! {\bf p}
	+ (1-x)^2 p^2 - i \varepsilon \right]^{-3/2} ,
\label{fp:1}
\end{eqnarray}
%
where we have carried out the analytic continuation to
$P_0 =-ip+\varepsilon$.
Integrating over ${\bf r}$
and then over the Feynman parameter,
we get a ${}_2F_1$ hypergeometric function with argument $1-c^2$:
\begin{eqnarray}
&& \int_Q {1 \over Q^2 (R_0^2 + r^2 c^2)}
\;=\; {1\over (4\pi)^2} (e^\gamma \mu^2)^\epsilon
	p^{-2\epsilon} {(1)_\epsilon \over \epsilon}
\nonumber
\\
&& \hspace{4cm} \;\times\;
e^{i \pi \epsilon} {(1)_{-2\epsilon} (1)_{-\epsilon} \over (2)_{-3\epsilon}}
	(1-c^2)^{-\epsilon}
	F\left( { {3\over2}-2\epsilon , 1-\epsilon
		\atop 2-3\epsilon } \Bigg| 1-c^2 \right) \,.
\;
\label{int4HTL:12Q}
\end{eqnarray}
%
The subsequent weighted averages over $c$
give ${}_3F_2$ hypergeometric functions
with argument $1$:
\begin{eqnarray}
\int_Q {1 \over Q^2}
\left\langle {c^2 \over R_0^2 + r^2 c^2} \right\ranglec
&=&
{1 \over (4\pi)^2} (e^\gamma \mu^2)^\epsilon
	p^{-2\epsilon} {(1)_\epsilon \over \epsilon}
\nonumber
\\
&& \;\times\:
{1 \over 3} e^{i \pi \epsilon}
{ ({3\over2})_{-\epsilon} (1)_{-2\epsilon} (1)_{-2\epsilon}
	\over ({5\over2})_{- 2\epsilon} (2)_{-3\epsilon} }
F\left({ 1-2\epsilon , {3\over2}-2\epsilon , 1-\epsilon
	\atop {5\over2}-2\epsilon , 2-3\epsilon } \Bigg| 1 \right) \, ,
\\
\int_Q {1 \over Q^2}
\left\langle {c^2 (1-c^2) \over R_0^2 + r^2 c^2} \right\ranglec
&=&
{1 \over (4\pi)^2} (e^\gamma \mu^2)^\epsilon
	p^{-2\epsilon}  {(1)_\epsilon \over \epsilon}
\nonumber
\\
&& \;\times\:
{2 \over 15} e^{i \pi \epsilon}
{ ({3\over2})_{-\epsilon} (1)_{-2\epsilon} (2)_{-2\epsilon}
	\over ({7\over2})_{- 2\epsilon} (2)_{-3\epsilon} }
F\left( { 2-2\epsilon \;\;{3\over2}-2\epsilon , 1-\epsilon
	\atop {7\over2}-2\epsilon , 2-3\epsilon } \Bigg| 1 \right) \, .
\end{eqnarray}
%
After expanding in powers of $\epsilon$, the real part
is (\ref{int4HTL:2}).

The integral (\ref{int4HTL:3})
has a factor of $1/r^2$ in the integrand.
After using (\ref{fp:1}), it is convenient to use a
second Feynman parameter to combine $(1-x+xc^2)r^2$
with the other denominator before integrating over ${\bf r}$:
\begin{eqnarray}
&& \int_Q {1 \over Q^2 r^2 (R_0^2 + r^2 c^2)}
\;= \; {3\over8} \int_0^1 dx \, (1-x+xc^2) \int_0^1 dy \, y^{1/2}
\nonumber
\\
&& \hspace{2cm}
\;\times\;
\int_{\bf r}\left[ (1-x+xc^2) r^2 + 2y(1-x) {\bf r} \!\cdot\! {\bf p}
	+ y(1-x)^2 p^2 - i \varepsilon \right]^{-5/2} \, .
\label{fp-2}
\end{eqnarray}
%
After integrating over ${\bf r}$ and then $y$, we obtain
${}_2F_1$ hypergeometric functions with arguments $x(1-c^2)$.
The integral over $x$ gives a ${}_2F_1$ hypergeometric function
with argument $1-c^2$:
\begin{eqnarray}
\int_Q {1 \over Q^2 r^2 (R_0^2 + r^2 c^2)}
&=&
{1\over (4\pi)^2} (e^\gamma \mu^2)^\epsilon
p^{-2-2\epsilon} {(1)_\epsilon \over \epsilon}
\left\{ {(-{1\over2})_{-\epsilon} (1)_{-\epsilon}
	\over ({1\over2})_{-2\epsilon}}
\right.
\nonumber
\\
&& \hspace{-1cm} \left.
- {3 \over 2(1+2 \epsilon)} e^{i \pi \epsilon}
{(1)_{-2\epsilon} (1)_{-\epsilon} \over (1)_{-3\epsilon}} (1-c^2)^{-\epsilon}
	F\left( { {1\over2}-2\epsilon , -\epsilon
		\atop -3\epsilon } \Bigg| 1-c^2 \right)
	\right\} \, .
\label{int4HTL:3Q}
\end{eqnarray}
%
After averaging over $c$, we get a hypergeometric function with argument 1:
\begin{eqnarray}
\int_Q {1 \over Q^2 r^2}
\left\langle {c^2 \over R_0^2 + r^2 c^2} \right\ranglec
&=&
{1 \over (4\pi)^2} (e^\gamma \mu^2)^\epsilon
p^{-2-2\epsilon}
{(1)_\epsilon \over \epsilon}
\left\{ {1 \over 3-2\epsilon} \,
	{ (-{1\over2})_{-\epsilon} (1)_{-\epsilon}
		\over ({1\over2})_{- 2\epsilon} }
\right.
\nonumber
\\
&& \left.
\;-\; {1 \over 2} e^{i \pi \epsilon} \,
{ (-{1\over2})_{-\epsilon} (1)_{-2\epsilon} (2)_{-2\epsilon}
	\over ({5\over2})_{- 2\epsilon} (1)_{-3\epsilon} }
F \left( { 1-2\epsilon , {1\over2}-2\epsilon , -\epsilon
	\atop {5\over2}-2\epsilon , -3\epsilon } \Bigg| 1 \right)
\right\} \, .
\label{int4HTL:3Qc}
\end{eqnarray}
%
After expanding in powers of $\epsilon$, the real part
is (\ref{int4HTL:3}).

To evaluate the integral (\ref{int4HTL:4}),
it is convenient to first express it as the sum of 3 integrals
by expanding the factor of $q^2$ in the numerator as
$q^2 = p^2 + 2 {\bf p} \cdot {\bf r} + r^2$:
\begin{eqnarray}
\int_Q {q^2 \over Q^2 r^2 (R_0^2 + r^2 c^2)}
\;=\; \int_Q
\left( {p^2 \over r^2} + 2 {{\bf p} \cdot {\bf r} \over r^2} + 1 \right)
{1 \over Q^2 (R_0^2 + r^2 c^2)}
 \, .
\end{eqnarray}
%
To evaluate the integral with ${\bf p} \cdot {\bf r}$ in the numerator,
we first combine the denominators using Feynman
parameters as in (\ref{fp-2}).
After integrating over ${\bf r}$ and then $y$, we obtain
${}_2F_1$ hypergeometric functions with arguments $x(1-c^2)$.
The integral over $x$ gives ${}_2F_1$ hypergeometric functions
with arguments $1-c^2$:
\begin{eqnarray}
\int_Q {{\bf p} \cdot {\bf r}
	\over Q^2 r^2 (R_0^2 + r^2 c^2)}
&=&
{1\over (4\pi)^2} (e^\gamma \mu^2)^\epsilon
p^{-2\epsilon} {(1)_\epsilon \over 2\epsilon^2}
\left\{ - {({3\over2})_{-\epsilon} (1)_{-\epsilon}
	\over ({3\over2})_{-2\epsilon}}
\right.
\nonumber
\\
&& \left.
+ e^{i \pi \epsilon}
{(1)_{-2\epsilon} (1)_{-\epsilon} \over (1)_{-3\epsilon}} (1-c^2)^{-\epsilon}
	F\left( { {3\over2}-2\epsilon , -\epsilon
		\atop 1-3\epsilon } \Bigg| 1-c^2 \right)
	\right\} \, .
\label{int4HTL:5Q}
\end{eqnarray}
%
After averaging over $c$, we get a hypergeometric function with argument 1:
\begin{eqnarray}
\int_Q {{\bf p} \cdot {\bf r} \over Q^2 r^2}
\left\langle {c^2 \over R_0^2 + r^2 c^2} \right\ranglec
&=&
{1 \over (4\pi)^2} (e^\gamma \mu^2)^\epsilon
p^{-2\epsilon}
{(1)_\epsilon \over 2\epsilon^2}
\left\{ - {1 \over 3-2\epsilon} \,
	{ ({3\over2})_{-\epsilon} (1)_{-\epsilon}
		\over ({3\over2})_{- 2\epsilon} }
\right.
\nonumber
\\
&& \left.
\;+\; {1 \over 3} e^{i \pi \epsilon}
{ ({3\over2})_{-\epsilon} (1)_{-2\epsilon} (1)_{-2\epsilon}
	\over ({5\over2})_{- 2\epsilon} (1)_{-3\epsilon} }
F \left( { 1-2\epsilon , {3\over2}-2\epsilon , -\epsilon
	\atop {5\over2}-2\epsilon , 1-3\epsilon } \Bigg| 1 \right)
\right\} \, .
\label{int4HTL:5Qc}
\end{eqnarray}
%
After expanding in powers of $\epsilon$, the real part is
\begin{eqnarray}
{\rm Re} \int_Q {{\bf p} \cdot {\bf r} \over Q^2 r^2}
\left\langle {c^2 \over R_0^2 + r^2 c^2} \right\ranglec
&=&
{1 \over (4 \pi)^2} \mu^{2\epsilon} p^{-2 \epsilon}
\left[ {-1 + \log 2 \over 3\epsilon}
\right.
\nonumber
\\
&& \hspace{3cm} \left.
-{20\over9} + {14\over 9} \log2 -{2\over3} \log^22
	+ {\pi^2\over 36} \right] \, .
\label{int4HTL:5}
\end{eqnarray}
%
Combining this with (\ref{int4HTL:1}) and (\ref{int4HTL:2}),
we obtain the integral (\ref{int4HTL:4}).

\subsection{Hypergeometric functions}
\label{app:hyper}

The generalized hypergeometric function of type $_pF_q$
is an analytic function of one variable with $p+q$ parameters.
In our case, the parameters are functions of $\epsilon$,
so the list of parameters sometimes gets lengthy and the standard notation
for these functions becomes cumbersome.  We therefore introduce a more
concise notation:
\begin{equation}
F\left( { \alpha_1,\alpha_2,\ldots,\alpha_p
	\atop \beta_1,\ldots,\beta_q } \Bigg| z \right)
\;\equiv\;
{}_pF_q(\alpha_1,\alpha_2,\ldots,\alpha_p;\beta_1,\ldots,\beta_q;z) \, .
\end{equation}
%
The generalized hypergeometric function has a power series representation:
\begin{equation}
F\left( { \alpha_1,\alpha_2,\ldots,\alpha_p
	\atop \beta_1,\ldots,\beta_q } \Bigg| z \right)
\;=\; \sum_{n=0}^\infty{ (\alpha_1)_n (\alpha_2)_n \cdots (\alpha_p)_n
	\over (\beta_1)_n \cdots (\beta_q)_n n! } z^n
	\, ,
\label{ps-pFq}
\end{equation}
%
where $(a)_b$ is Pochhammer's symbol:
\begin{equation}
(a)_b = {\Gamma(a+b) \over \Gamma(a)} \,.
\label{Poch}
\end{equation}
%
The power series converges for $|z|<1$.
For $z=1$, it converges if ${\rm Re} s > 0$, where
\begin{equation}
s\;=\; \sum_{i=1}^{p-1} \beta_i -  \sum_{i=1}^p \alpha_i\, .
\label{s-def}
\end{equation}
%
The hypergeometric function of type $_{p+1}F_{q+1}$
has an integral representation in terms of the hypergeometric function
of type $_pF_q$:
\begin{equation}
\int_0^1 dt \, t^{\nu-1} (1-t)^{\mu-1} \,
F\left( { \alpha_1,\alpha_2,\ldots,\alpha_p
	\atop \beta_1,\ldots,\beta_q  } \Bigg| tz \right)
\;=\; { \Gamma(\mu) \Gamma(\nu) \over \Gamma(\mu+\nu)} \,
F\left( { \alpha_1,\alpha_2,\ldots,\alpha_p,\nu
	\atop \beta_1,\ldots,\beta_q,\mu+\nu} \Bigg| z \right) \, .
\label{int-pFq}
\end{equation}
%
If a hypergeometric function has an upper and lower parameter that are
equal, both parameters can be deleted:
\begin{equation}
F\left( { \alpha_1,\alpha_2,\ldots,\alpha_p,\nu
	\atop \beta_1,\ldots,\beta_q, \nu} \Bigg| z \right)
\;=\; F\left( { \alpha_1,\alpha_2,\ldots,\alpha_p
	\atop \beta_1,\ldots,\beta_q } \Bigg| z \right) \, .
\end{equation}
%

The simplest hypergeometric function is the one of type $_1F_0$.
It can be expressed in an analytic form:
\begin{equation}
{}_1F_0(\alpha; \, ;z) \;=\; (1-z)^{-\alpha} \, .
\end{equation}
%
The next simplest hypergeometric functions are those of type $_2F_1$.
They satisfy transformation formulas that allow an $_2F_1$
with argument $z$ to be expressed in terms of an $_2F_1$
with argument $z/(z-1)$ or as a sum of two $_2F_1$'s
with arguments $1-z$ or $1/z$ or $1/(1-z)$.
The hypergeometric functions of type $_2F_1$
with argument $z=1$ can be evaluated analytically in terms of gamma
functions:
\begin{equation}
F\left( { \alpha_1, \alpha_2 \atop \beta_1 } \Bigg| 1 \right)
\;=\; { \Gamma(\beta_1) \Gamma(\beta_1 - \alpha_1 - \alpha_2)
	\over \Gamma(\beta_1 - \alpha_1) \Gamma(\beta_1 - \alpha_2) } \, .
\label{2F1-1}
\end{equation}
%
The hypergeometric function of type $_3F_2$
with argument $z=1$ can be expressed as a $_3F_2$
with argument $z=1$ and different parameters \cite{3F2}:
\begin{equation}
F\left( { \alpha_1, \alpha_2, \alpha_3 \atop \beta_1, \beta_2 } \Bigg| 1 \right)
\;=\; { \Gamma(\beta_1) \Gamma(\beta_2) \Gamma(s)
	\over \Gamma(\alpha_1+s) \Gamma(\alpha_2+s) \Gamma(\alpha_3)} \,
F\left( { \beta_1-\alpha_3, \beta_2-\alpha_3, s
	\atop \alpha_1+s, \alpha_2+s } \Bigg| 1 \right) \,,
\label{3F2-1}
\end{equation}
%
where $s = \beta_1 + \beta_2 - \alpha_1 - \alpha_2 - \alpha_3$.
If all the parameters of a $_3F_2$ are integers and half-odd-integers,
this identity can be used to obtain
equal numbers of half-odd-integers among the upper and lower parameters.
If the parameters of a $_3F_2$
reduce to integers and half-odd-integers in the limit $\epsilon \to 0$ ,
the use of this identity simplifies the expansion of the
hypergeometric functions in powers of $\epsilon$ .

The  most important integration formulas involving $_2F_1$
hypergeometric functions is (\ref{int-pFq}) with $p=2$ and $q=1$.
Another useful integration formula is
\begin{eqnarray}
&&\int_0^1 dt \, t^{\nu-1} (1-t)^{\mu-1} \,
F\left( { \alpha_1,\alpha_2
	\atop \beta_1 } \Bigg| {t \over 1-t} z \right)
\;=\; { \Gamma(\mu) \Gamma(\nu) \over \Gamma(\mu+\nu)} \,
F\left( { \alpha_1,\alpha_2,\nu
	\atop \beta_1,1-\mu } \Bigg| -z \right)
\nonumber
\\
&& \hspace{1cm}
\;+\;
{ \Gamma(\alpha_1+\mu) \Gamma(\alpha_2+\mu) \Gamma(\beta_1) \Gamma(-\mu)
	\over \Gamma(\alpha_1) \Gamma(\alpha_2) \Gamma(\beta_1+\mu) } \,
(-z)^\mu \, F\left( { \alpha_1+\mu,\alpha_2+\mu,\nu+\mu
	\atop \beta_1+\mu,1+\mu} \Bigg| -z \right) \, .
\label{int-2F1}
\end{eqnarray}
%
This is derived by first inserting the integral representation
for $_2F_1$ in (\ref{int-pFq}) with integration variable $t'$ and then
evaluating the integral over $t$ to get a $_2F_1$ with argument
$1+t'z$.  After using a transformation formula to change
the argument to $-t'z$, the remaining integrals over $t'$ are evaluated
using (\ref{int-pFq}) to get $_3F_2$'s with arguments $-z$.

For the calculation of two-loop thermal integrals involving HTL averages,
we require the expansion in powers of $\epsilon$
for hypergeometric functions of type $_pF_{p-1}$ with argument 1
and parameters that are linear in $\epsilon$.
If the power series representation (\ref{ps-pFq})
of the hypergeometric function is convergent at $z=1$ for $\epsilon=0$,
this can be accomplished simply by expanding the summand
in powers of $\epsilon$ and then evaluating the sums.
If the power series is divergent, we must make subtractions
on the sum before expanding in powers of $\epsilon$.
The convergence properties of the power series at $z=1$
is determined by the variable $s$ defined in (\ref{s-def}).
If $s>0$, the power series converges.
If $s\to 0$ in the limit $\epsilon \to 0$,
only one subtraction is necessary to make the sum convergent:
\begin{eqnarray}
&& F\left( { \alpha_1,\alpha_2,\ldots,\alpha_p
	\atop \beta_1,\ldots,\beta_{p-1} } \Bigg| 1 \right)
\;=\;
{ \Gamma(\beta_1) \cdots \Gamma(\beta_{p-1}) \over
	\Gamma(\alpha_1) \Gamma(\alpha_2) \cdots \Gamma(\alpha_p) }
\zeta(s+1)
\nonumber
\\
&&\;+\; \sum_{n=0}^\infty
\left( { (\alpha_1)_n (\alpha_2)_n \cdots (\alpha_p)_n
	\over (\beta_1)_n \cdots (\beta_q)_n n! }
- { \Gamma(\beta_1) \cdots \Gamma(\beta_{p-1}) \over
	\Gamma(\alpha_1) \Gamma(\alpha_2) \cdots \Gamma(\alpha_p)}
	(n+1)^{-s-1} \right)
	\, .
\end{eqnarray}
%
If $s\to -1$ in the limit $\epsilon \to 0$,
two subtractions are necessary to make the sum convergent:
\begin{eqnarray}
F\left( { \alpha_1,\alpha_2,\ldots,\alpha_p
	\atop \beta_1,\ldots,\beta_{p-1} } \Bigg| 1 \right)
&=&
{ \Gamma(\beta_1) \cdots \Gamma(\beta_{p-1}) \over
	\Gamma(\alpha_1) \Gamma(\alpha_2) \cdots \Gamma(\alpha_p)}
\left[ \zeta(s+1) + t \, \zeta(s+2) \right]
\nonumber
\\
&& \hspace{-1cm}
\;+\; \sum_{n=0}^\infty
\left( { (\alpha_1)_n (\alpha_2)_n \cdots (\alpha_p)_n
	\over (\beta_1)_n \cdots (\beta_q)_n n! } \right.
\nonumber
\\
&& \left. - { \Gamma(\beta_1) \cdots \Gamma(\beta_{p-1}) \over
	\Gamma(\alpha_1) \Gamma(\alpha_2) \cdots \Gamma(\alpha_p)}
\left[ 	(n+1)^{-s-1} + t \, (n+1)^{-s-2} \right]
\right) \, ,
\end{eqnarray}
%
where $t$ is given by
\begin{equation}
t \;=\; \sum_{i=1}^p {(\alpha_i-1)(\alpha_i-2)\over2}
- \sum_{i=1}^{p-1} {(\beta_i-1)(\beta_i-2) \over2}  \, .
\end{equation}
%

The expansion of a $_pF_{p-1}$ hypergeometric function
in powers of $\epsilon$ is particularly simple
if in the limit $\epsilon \to 0$
all its parameters are integers or half-odd-integers, with equal numbers
of half-odd-integers among the upper and lower parameters.
If the power series representation for such a hypergeometric function
is expanded in powers of $\epsilon$, the terms in the summand will
be rational functions of $n$, possibly multiplied by
factors of the polylogarithm function $\psi(n+a)$ or its derivatives.
The terms  in the sums can often be simplified by using the obvious identity
\begin{equation}
\sum_{n=0}^\infty \left[ f(n) - f(n+k) \right]
\;=\; \sum_{i=0}^{k-1} f(i) \, .
\end{equation}
%
The sums over $n$ of rational functions of $n$ can be evaluated
by applying the partial fraction decomposition and then
using identities such as
\begin{eqnarray}
\sum_{n=0}^\infty \left({1 \over  n+a} - {1\over n+b} \right)
&=& \psi(b) - \psi(a)  \,,
\\
\sum_{n=0}^\infty {1 \over (n+a)^2} &=& \psi'(a) \, .
\end{eqnarray}
%
The sums of polygamma functions of $n+1$ or $n+{1\over2}$
divided by $n+1$ or $n+{1\over2}$ can be evaluated using
\begin{eqnarray}
\sum_{n=0}^\infty
\left( {\psi(n+1) \over n+1}
	- {\log(n+1) \over n+1} \right)
&=&  - {1\over2} \gamma^2 - {\pi^2 \over 12} - \gamma_1 \,,
\\
\sum_{n=0}^\infty
\left( {\psi(n+1) \over n+{1\over 2}}
	- {\log(n+1) \over n+1} \right)
&=&  - {1\over2} (\gamma + 2 \log 2)^2 + {\pi^2 \over 12} - \gamma_1 \,,
\\
\sum_{n=0}^\infty
\left( {\psi(n+{1\over2}) \over n+1}
	- {\log(n+1) \over n+1} \right)
&=& - {1\over2} \gamma^2 - 4 \log 2 + 2 \log^2 2
	- {\pi^2 \over 12} - \gamma_1 \,,
\\
\sum_{n=0}^\infty
\left( {\psi(n+{1\over2}) \over n+{1\over 2}}
	- {\log(n+1) \over n+1} \right)
&=&  - {1\over2} (\gamma + 2 \log 2)^2 - {\pi^2 \over 4} - \gamma_1 \,,
\end{eqnarray}
%
where $\gamma_1$ is Stieltje's first gamma constant
defined in (\ref{zeta}).
The sums of polygamma functions of $n+1$ or $n+{1\over2}$
can be evaluated using
\begin{eqnarray}
\sum_{n=0}^\infty
\left( \psi(n+1) - \log(n+1)  + {1 \over 2(n+1)} \right)
&=& {1\over2} + {1\over2} \gamma -{1\over 2} \log(2 \pi) \,,
\\
\sum_{n=0}^\infty
\left( \psi(n+\mbox{$1\over2$}) - \log(n+1)  + {1 \over n+1} \right)
&=& {1\over2}\gamma  - \log2 -{1\over 2} \log(2 \pi) \,.
\end{eqnarray}

We also need the expansions in $\epsilon$
of some integrals of $_2F_1$ hypergeometric functions of $y$
that have a factor of $|1-2y|$.  For example,
the following 2 integrals are needed to obtain (\ref{avecx:4}):
\begin{eqnarray}
&& \int_0^1 dy \, y^{-2\epsilon} (1-y)^{1+\epsilon} |1-2y| \,
F\left( { 1-\epsilon,\epsilon
	\atop -3\epsilon} \Bigg| y \right) \;=\;
{1\over 6} + \left( {2\over9} + {4\over9} \log 2 \right) \epsilon  \,,
\label{Fabs-1}
\\
&& \int_0^1 dy \, y^{1+\epsilon} (1-y)^{1+\epsilon} |1-2y| \,
F\left( { 2+2\epsilon,1+\epsilon
	\atop 2+3\epsilon} \Bigg| y \right) \;=\;
{1\over 4} + \left( {7\over12} + {2\over3} \log 2 \right) \epsilon  \,.
\label{Fabs-2}
\end{eqnarray}
These integrals can be evaluated by expressing them in the form
\begin{eqnarray}
\int_0^1 dy \, y^{\nu-1} (1-y)^{\mu-1} |1-2y| \,
F\left( { \alpha_1,\alpha_2
	\atop \beta_1} \Bigg| y \right) &=&
\int_0^1 dy \, y^{\nu-1} (1-y)^{\mu-1}  (2y-1) \,
F\left( { \alpha_1,\alpha_2
	\atop \beta_1} \Bigg| y \right)
\nonumber
\\
&& \hspace{-3cm}
\;+\; 2 \int_0^{1\over2} dy \, y^{\nu-1} (1-y)^{\mu-1}  (1-2y) \,
F\left( { \alpha_1,\alpha_2
	\atop \beta_1} \Bigg| y \right)  \, .
\end{eqnarray}
The evaluation of the first integral on the right side
gives $_3F_2$ hypergeometric functions with argument 1.
The integrals from 0 to $1\over2$ can be evaluated by expanding
the power series representation (\ref{ps-pFq})
of the hypergeometric function in powers of $\epsilon$.
The resulting series can be summed analytically
and then the integral over $y$ can be evaluated.

\end{document}